\documentclass[11pt,a4paper]{article}

\usepackage[a4paper,margin=2.5cm]{geometry}

\usepackage{amsmath,amssymb,amsthm}

\usepackage{natbib}

\usepackage{booktabs}
\usepackage{makecell}
\usepackage{graphicx}

\usepackage[font=small,labelformat=empty]{caption}

\usepackage{bbm}
\usepackage{threeparttable}
\usepackage{comment}

\usepackage{tcolorbox}
\tcbuselibrary{theorems, skins, breakable}

\tcbset{
    propositionstyle/.style={
        enhanced,
        breakable,
        colback=gray!5,
        colframe=gray!50,
        fonttitle=\bfseries,
        separator sign={.},
        description delimiters parenthesis,
        terminator sign={},
        left=6pt,
        right=6pt,
        top=4pt,
        bottom=4pt,
        boxrule=0.4pt,
    }
}

\newtcbtheorem{prop}{Proposition}{propositionstyle}{prop}

\begin{document}

\title{Licensing and Innovation Regimes in Pharmaceutical R\&D}
\author{
Michele Liberatore\thanks{IMT School for Advanced Studies Lucca, Italy and University of Florence, Italy. Email: michele.liberatore@imtlucca.it}
\and
Massimo Riccaboni\thanks{IMT School for Advanced Studies Lucca, Italy and IUSS Pavia, Italy. Email: massimo.riccaboni@imtlucca.it}
}
\date{}
\maketitle

\begin{abstract}
\noindent We study how licensing affects the allocation of innovation in pharmaceutical R\&D. We develop a model in which projects differ in both quality and innovation regime, distinguishing between incremental and novel innovations. Information precision is higher for incremental projects and lower for novel ones, generating different equilibrium dynamics in the market for technology. The model predicts that licensing sustains positive selection and competitive return equalization for incremental innovation, while novel projects may exhibit weaker screening consistent with lemons-type frictions. Using product-level data and Double Machine Learning methods, we test these predictions across success probabilities and monetary returns. We find that licensing increases success probability overall, but return equalization holds primarily for incremental projects. For novel innovation, licensing does not exhibit the same equilibrium adjustment, suggesting residual market imperfections. Instrumenting for licensing using exogenous pipeline shocks confirms this pattern causally: the competitive risk-return trade-off is preserved for incremental ``rushed'' licenses, but it breaks down for novel ones. Our results reconcile evidence on both competitive efficiency and information frictions in markets for technologies, showing that market performance depends systematically on the type of innovation being transacted.
\end{abstract}

\vspace{0.5cm}


\section{Introduction}
\label{sec:intro}

Markets for technology play a central role in modern innovation systems: rather than developing all technologies internally, firms increasingly rely on licensing agreements, alliances, and acquisitions to access external ideas \citep{arora2004markets, arora2010market}. Theoretical and empirical research has examined whether these markets allocate innovative assets efficiently or suffer from informational frictions. A central concern, dating back to \citet{akerlof1970lemons}, is that when originators have superior information about technologies, adverse selection may arise: low-quality technologies are disproportionately brought to market, high-quality technologies are retained, and trade may collapse into a ``lemons'' equilibrium.

This concern is especially relevant in markets for early-stage technologies, where quality is difficult to assess and uncertainty about the potential value of new technological opportunities is high. In the pharmaceutical industry, biotechnology firms often out-license drug candidates to large pharmaceutical companies. Originators usually have a better understanding of a project's scientific foundations, while licensees provide downstream development capabilities, regulatory expertise, and commercialization infrastructure. This asymmetry creates a natural environment for potential adverse selection. Early theoretical work in industrial organization and strategy has emphasized that markets for technology may be prone to selection distortions when evaluation is imperfect \citep{pisano1997,arora2009breath}.

Yet the empirical evidence is mixed: several studies document higher clinical success rates for in-licensed projects, seemingly at odds with adverse selection, while others find evidence of information asymmetries and discounted contract terms for inexperienced originators. The coexistence of these patterns poses a puzzle: how can markets simultaneously show signs of efficiency and informational distortions?
We argue that this coexistence is not a puzzle but an equilibrium property of pharmaceutical market for technologies. Competitive efficiency and lemons-type frictions are not mutually exclusive; they operate simultaneously within the same industry, segmented by innovation regime \citep{arora2022science}. For incremental projects, signal precision is sufficient to sustain competitive screening and equalize returns across organizational forms. For novel projects at the innovation frontier, precision is inherently limited, screening breaks down, and the market fails to fully price innovation risk ex-ante.

Most theoretical models of markets for technology implicitly assume that assets are drawn from a common informational environment: the precision with which buyers can evaluate projects is treated as exogenous and homogeneous. Similarly, empirical analyses typically evaluate licensing outcomes in the aggregate, implicitly pooling projects with fundamentally different risk profiles. In pharmaceutical R\&D, however, projects span a wide continuum from highly predictable incremental innovations (e.g., ``me-too'' drugs, follow-on compounds) to highly novel, first-in-class therapies characterized by fundamental scientific uncertainty. Observed approval rates vary dramatically across these categories, ranging from very high success probabilities for incremental products to single-digit rates for new molecular entities. Such heterogeneity implies that the informational precision with which project quality can be inferred is not constant across licensing deals. Projects that follow well-established development pathways generate relatively informative signals; frontier innovations involve noisier evaluation and greater dispersion in potential value across buyers.

This heterogeneity has important implications for market equilibrium. When signals are sufficiently precise, competitive screening can sustain separating equilibria: only projects above a quality threshold are licensed, and contract terms adjust to reflect expected value. In this regime, markets may display positive selection and return equalization across organizational forms, consistent with competitive efficiency. Conversely, when signals are noisy and evaluation is fundamentally uncertain, pooling equilibria may emerge. Contract terms then reflect average expected quality rather than project-specific information, potentially generating lemons-type distortions.

In this paper, we develop a theoretical and empirical framework that embeds pharmaceutical markets for technologies in an environment with endogenous informational precision across innovation regimes. Originators decide whether to license or develop internally, and licensed development entails coordination costs arising from inter-organizational collaborations.

The model delivers three central predictions. First, selection into licensing depends on the precision of information. For licensed projects, sufficiently precise signals support separating equilibria in which only high-quality assets are licensed, generating positive selection in observed success probabilities. Second, competitive forces equalize expected returns across organizational forms when signals are precise, but return equalization may fail when fundamental uncertainty limits ex-ante pricing. Third, and as a direct consequence of the return structure implied by the second prediction, licensing increases total development costs relative to in-house development when coordination costs arising from inter-organizational collaboration outweigh the effort savings generated by royalty sharing.

We bring these predictions to data using a comprehensive product-level dataset covering pharmaceutical development outcomes, clinical trial costs, and commercial revenues. The richness of the data allows us to move beyond the traditional focus on approval probabilities and to evaluate the full risk–return profile of licensed versus internally developed projects. Crucially, we construct predicted success probabilities using high-dimensional project characteristics and machine learning methods, thereby obtaining a continuous empirical proxy for project risk and informational precision.

Our empirical strategy consists of two complementary steps. The first relies on the Double Machine Learning (DML) framework \citep{chernozhukov2018double}, which flexibly controls for more than one hundred pre-treatment confounders while maintaining valid inference for the parameters of interest. The second step instruments for licensing with exogenous pipeline shocks (Phase III clinical trial failures) experienced by licensee firms, following \citet{hermosilla2021rushed}, to identify a Local Average Treatment Effect for the subpopulation of licenses accelerated by acute pipeline pressure.

The results reconcile the apparently contradictory evidence in the literature. In the aggregate, in-licensed projects exhibit significantly higher success probabilities than internally developed ones, consistent with positive selection and inconsistent with pure adverse selection. However, licensed projects also exhibit lower unconditional net returns - that is, at mean predicted success probability. This pattern is in fact consistent with competitive equilibrium: the pre-market advantage of licensing in success probability is offset by a post-market penalty in net returns, generating a risk-return trade-off in which neither organizational form dominates the other on both dimensions simultaneously. Once we account for heterogeneity along the innovation-risk spectrum, however, a different picture emerges. For low-risk incremental projects, this offsetting mechanism holds in both analyses: in the DML, the success probability advantage of licensing is counterbalanced by lower net returns; in the DML-IV, rushed licenses carry neither a success probability advantage nor a return penalty. In both cases, no strategy dominates the other on both dimensions simultaneously, consistent with competitive return equalization. For high-risk novel projects, by contrast, the compensating structure breaks down. In the descriptive DML results, licensed frontier innovations retain their higher success probabilities without a return penalty. In the DML-IV results, rushed licenses for novel projects carry no success probability advantage but a significant return loss. In both cases the offsetting mechanism is absent and one strategy weakly dominates or is weakly dominated, pointing to an incomplete realization of competitive equilibrium at the innovation frontier. In other words, market forces appear to discipline incremental transactions but are less effective at fully pricing risk and quality at the innovation frontier.

These findings have broader implications for the theory of markets for technology. Rather than asking whether such markets are efficient or plagued by lemons in the aggregate, our results indicate that market performance depends systematically on informational precision and innovation regime. Competitive equilibria and lemons-type distortions can coexist within the same industry, operating over different segments of the risk distribution.

From an industrial organization perspective, the paper contributes to the literature on adverse selection, endogenous screening, and organizational form by highlighting how heterogeneity in signal precision shapes equilibrium outcomes. It also contributes to the growing literature on quasi-experimental identification in innovation economics by providing a product-level application of the \citet{hermosilla2021rushed} instrument that validates the descriptive DML patterns causally. From a policy standpoint, the results raise the possibility that market imperfections may be concentrated in the most innovative segment of pharmaceutical R\&D, precisely where social returns are potentially highest.


The remainder of the paper proceeds as follows. Section~\ref{sec:lit_review} discusses the related literature. Section~\ref{sec:theoretical_model} presents the model and derives the testable predictions. Section~\ref{sec:data} introduces the data and variable construction. Section~\ref{sec:methodology} describes the empirical strategy and identification approach, including the DML and DML-IV specifications. Section~\ref{sec:dml_results} presents the main DML results, including the heterogeneity analysis across the innovation-risk distribution. Section~\ref{sec:dmliv_results} presents the DML-IV causal estimates. 
Section~\ref{sec:conclusion} concludes.

\section{Literature Review}
\label{sec:lit_review}

Pharmaceutical development involves substantial and heterogeneous costs across stages \citep{dimasi2003price}, which motivate firms to seek efficient development strategies. The shift from vertically integrated R\&D models toward distributed innovation systems has made licensing arrangements increasingly central to pharmaceutical innovation \citep{orsenigo2001,arora2004markets}. These licensing agreements allow firms to access specialized knowledge, share development risks, and allocate tasks based on comparative advantages \citep{arora2021knowledge}. Understanding the efficiency and selection properties of these markets for technology has become a critical question in the economics of pharmaceutical innovation.

The efficiency of pharmaceutical technology transfer markets, whether licensing but also acquisition markets work efficiently or exhibit signs of adverse selection, falls under the broader conceptual framework of ``markets for technology'' \citep{arora2001, arora2009breath}. The seminal work by \citet{akerlof1970lemons} on information asymmetry provides a theoretical foundation for understanding potential inefficiencies in these transactions. According to Akerlof's framework, when sellers possess more information about asset quality than buyers, markets may suffer from adverse selection that leads low-quality assets (``lemons'') to be disproportionately brought to market while high-quality assets are retained for internal development. If such dynamics were to characterize pharmaceutical licensing markets, this would suggest fundamental inefficiencies in the allocation of innovative assets and would have significant implications for both firm-level strategy and industry-wide innovation outcomes \citep{pisano1997}.

Empirical evidence on the existence of a ``market for lemons'' problem in pharmaceutical technology markets has produced mixed findings, with different studies identifying both information asymmetries and traces of market efficiency. The resolution of these apparently contrasting findings may lie in recognizing that selection patterns depend on the specific institutional context, contract structures employed, and critically, the risk profile of the projects being transacted.

Several studies document that licensed projects exhibit higher success rates than internally developed ones. \citet{dimasi2010trends} find that drugs in-licensed by the largest pharmaceutical companies demonstrate higher clinical success rates than self-originated drugs, with in-licensed compounds showing cumulative success rates of approximately 27 percent compared to 16 percent for self-originated products during the 1993-2004 period. This finding directly contradicts the prediction of adverse selection theory and suggests that licensed assets may represent lower-risk opportunities efficiently matched with firms possessing comparative advantages in later-stage development. Similarly, \citet{danzon2005productivity} document that products developed through alliances have higher probabilities of success in complex late-stage trials, particularly when the licensee is a large experienced firm, indicating that licensing agreements facilitate efficient allocation of development tasks based on firms' complementary capabilities. These empirical patterns are consistent with a positive selection mechanism in which, for projects with sufficiently high information precision, licensed projects tend to have higher expected success probabilities than internally developed ones, reflecting selection above a quality threshold that varies with project predictability.

While positive selection appears prevalent in the aggregate picture, there are still some pieces of evidence which suggest that information asymmetries nonetheless exist in pharmaceutical licensing markets. \citet{nicholson2005biotech} provide perhaps the most comprehensive examination, concluding that inexperienced biotech companies receive substantially discounted payments when forming their first alliance with pharmaceutical partners, although drugs developed through these alliances demonstrate higher success rates in clinical trials. \citet{palermo2019reliable} extend this line of inquiry by showing that even when deals are closed, licensed patents are more vulnerable to litigation challenges than internally developed ones, particularly when the licensee lacks in-house IP capabilities. These findings suggest asymmetric information and signaling dynamics rather than a classic lemons problem, with inexperienced firms accepting below-market pricing to establish credibility. \citet{reepmeyer2011outlicensing} examine information asymmetries in out-licensing deals and derive recommendations for addressing these through careful structuring of deal terms. These findings indicate that information asymmetries exist in pharmaceutical licensing markets, consistent with the idea that originators possess private information about project quality that developers can only imperfectly observe, but that market participants have developed sophisticated mechanisms to prevent pure adverse selection at least for certain types of projects. This focus on the strategic management of deal terms, however, typically assumes a deliberate and well-planned process on the part of the licensee. \citet{hermosilla2021rushed} challenges this assumption by examining licensing decisions made under pressure, showing that large pharmaceutical firms experiencing a sudden pipeline failure (``rushed innovation'') are significantly more likely to in-license new drugs, but that these rushed licenses subsequently underperform in clinical development.

However, the existing literature has not systematically examined whether this regime dependence influences equilibrium outcomes in licensing markets. Most empirical analyses combine projects with fundamentally different risk profiles, implicitly treating informational precision as homogeneous across transactions. Higher success rates may reflect positive selection of lower-risk projects rather than truly high-quality, innovative but risky assets. The degree of innovation and commercial potential may be inversely related to success probability, with highly innovative, risky projects exhibiting lower success rates but potentially higher returns conditional on success. Moreover, novel inventions are likely to have greater variation across buyers in their ability to extract value, thus resulting in greater transaction costs and increased difficulty in their market evaluation \citep{arora2022science}. Consequently, this distinction between risk and quality motivates our interest in whether licensing markets achieve efficient allocation across the full spectrum of pharmaceutical innovation. Empirical support for the economic relevance of this spectrum comes from \citet{krieger2022missing}, who document a shift in pharmaceutical R\&D toward less novel compounds over time and construct a continuous measure of drug novelty based on chemical similarity. Their evidence that the composition of development pipelines responds to financial incentives and intellectual property conditions implies that novelty is not merely a biological feature of projects but an equilibrium outcome shaped by the economic environment in which firms operate. Our analysis complements these findings by examining whether the market for technology allocates novel and incremental projects differently.

Some institutional features of pharmaceutical licensing markets help mitigate information asymmetries and support positive selection equilibria. First, the extensive use of milestone payments, contingent royalties, and performance-based deal structures allows buyers and sellers to share risks and align incentives in ways that mitigate adverse selection \citep{nicholson2005biotech}.
Second, scientific and regulatory transparency - including published preclinical data, clinical trial registrations, and peer-reviewed publications - substantially reduces hidden information about asset quality. Third, the closed network between biotech and pharmaceutical firms, combined with reputational concerns in a relatively concentrated industry, creates strong incentives for honest disclosure \citep{danzon2005productivity}. Fourth, sophisticated due diligence processes reduce information gaps before deal completion. These transparency mechanisms effectively reduce information asymmetry and increase the precision of available signals, moving the market closer to a full-information benchmark where selection would be more efficient. Nevertheless, the effectiveness of these mechanisms is unlikely to be uniform across the innovation-risk spectrum. Where development pathways are well established and outcomes are relatively predictable, transparency and reputational incentives may be sufficient to eliminate adverse selection. Where outcomes are fundamentally uncertain, as in frontier innovations with novel mechanisms of action or unproven biological targets, these same mechanisms may prove insufficient, leaving information asymmetries unresolved despite their presence.

Although licensing is often presumed to improve efficiency through specialization and scale, the total costs of licensing, including coordination costs arising from inter-organizational interfaces, must be considered. The broader context of declining pharmaceutical R\&D efficiency \citep{pammolli2011productivity} makes the question of how licensing affects resource allocation especially consequential. The empirical heterogeneity in licensing outcomes across therapeutic areas, development stages, and firm types \citep{hay2014clinical, kola2004can} suggests that licensing efficiency may vary systematically based on the underlying cost structure and, importantly, the risk profile of the projects being licensed.

\section{R\&D Licensing with Innovation Types}
\label{sec:theoretical_model}

In this section we define a conceptual framework for our analysis of pharmaceutical markets for technologies.
We consider a market for R\&D projects populated by biotechnology firms (originators) and pharmaceutical companies (developers). Projects differ by quality and innovation regime, and developers choose how to access them among three development strategies. The primary object of analysis is the comparison between \emph{in-licensing} and \emph{in-house} product development: these two strategies are the central focus of our analysis, and the testable predictions of the model are formulated around them. Nonetheless, \textit{company acquisition} is included as a third strategy — in which a developer obtains access to a project by acquiring the originating firm outright, thus integrating its full pipeline and scientific capabilities — to check whether the equilibrium properties we derive for licensing are specific to that contractual governance mode or extend to alternative forms of external sourcing. 

\subsection{Projects, Information, and Development Strategies}
\label{subsec:formalization_general}

Each project is characterized by a latent quality parameter $\theta \in [0,1]$, drawn from distribution $F$ with density $f$, which determines its expected development and commercialization value. In addition, projects differ in their innovation regime. Let
\[
\tau \in \{I,N\}
\]
denote the technological type, where $\tau=I$ represents \emph{incremental innovation} and $\tau=N$ \emph{novel innovation}\footnote{The binary formulation is adopted for clarity; the framework extends naturally to a continuous innovation-type space $\tau \in [0,1]$, with information precision $\lambda(\tau)$ strictly decreasing in novelty.}. Incremental projects follow established development pathways with well-characterized mechanisms, whereas novel projects involve fundamentally uncertain biological targets and/or therapeutic approaches. The regime $\tau$ is observable to both originators and developers, while quality $\theta$ is privately known to the originator.

Information precision depends on the innovation regime. Let $\lambda = \lambda(\tau)$, with $\lambda(I) > \lambda(N) > 0$. Developers observe a noisy signal
\[
\hat{\theta} = \theta + \varepsilon, \qquad \varepsilon \sim \mathcal{N}\!\left(0, \sigma^2(\tau)\right), \quad \sigma^2(\tau) = \frac{1}{\lambda(\tau)},
\]
so that signal precision is higher for incremental projects and substantially noisier for novel ones.

Updating follows standard Bayesian inference: combining a Gaussian prior on $\theta$ with precision $\lambda_0$ and mean $\mu_0$ with the Gaussian likelihood of the signal yields a Gaussian posterior whose mean is a precision-weighted average of the prior mean and the observed signal:
\[
\mathbb{E}[\theta \mid \hat{\theta}, \tau] = \frac{\lambda(\tau)\,\hat{\theta} + \lambda_0\,\mu_0}{\lambda(\tau) + \lambda_0},
\]
where $\lambda_0$ and $\mu_0$ denote the precision and mean of the prior respectively, the weight assigned to the signal is $\frac{\lambda(\tau)}{\lambda(\tau)+\lambda_0}$ and the weight assigned to the prior mean is $\frac{\lambda_0}{\lambda(\tau)+\lambda_0}$; as $\lambda(\tau) \to \infty$ (incremental projects) the posterior collapses to $\hat{\theta}$, while as $\lambda(\tau) \to 0$ (novel projects) it collapses to $\mu_0$, leaving the developer unable to discriminate across projects regardless of the signal received.

A developer can access a project through three alternative strategies $s \in \{H, L, A\}$: \textit{in-house development}, \textit{licensing}, or \textit{company acquisition}. Each strategy differs along three dimensions: (i) the degree of organizational integration, (ii) the structure of incentives, and (iii) the magnitude of associated frictions. Under \textit{in-house} \textit{development} ($s = H$), the developer retains full control of the project and its associated cash flows. Under \textit{licensing} ($s = L$), the originator transfers development rights in exchange for a royalty rate $r \in (0,1)$ on realized revenues, along with upfront milestone payments. Under \textit{company acquisition} ($s = A$), the developer acquires the originating firm outright at a negotiated price $P_A$, obtaining full organizational integration at the cost of a fixed integration expenditure $\phi > 0$ that entails coordination frictions through structural consolidation.

\subsection{Effort, Success, and Costs}
\label{subsec:formalization_strategy}

If a developer supplies effort $e\geq0$ on a project of quality $\theta$, the probability of regulatory approval (i.e., success) is $p(\theta,e)=\theta\cdot e$, with convex effort cost $C(e)=\frac{k}{2}e^{2}$, $k>0$. Revenue conditional on success is normalized to $V>0$ and the structure of payoffs differs across strategies.

\paragraph{In-house development.} The developer solves
\[
\max_{e} \;\theta e V - \frac{k}{2}e^2,
\]
yielding optimal effort $e^*_H = \frac{\theta V}{k}$ and expected profits
\[
\pi_H(\theta) = \frac{\theta^2 V^2}{2k}.
\]
Profits are strictly convex in quality, so high-quality projects generate disproportionately larger returns under in-house development.

\paragraph{Licensing.} Under a licensing contract with royalty rate $r$, the developer retains share $(1 - r)$ of realized revenues. The developer's problem becomes
\[
\max_{e} \;(1-r)\,\theta e V - \frac{k}{2}e^2,
\]
yielding $e^*_L = \frac{(1-r)\theta V}{k} < e^*_H$ for $r > 0$. Royalty sharing thus introduces a moral hazard wedge: effort is distorted downward relative to the first-best. Expected developer profits under licensing are
\[
\pi_L(\theta) = \frac{(1-r)^2\,\theta^2 V^2}{2k} - C_c(\tau),
\]
where $C_c(\tau) \geq 0$ denotes additional strategy-specific coordination costs arising from inter-organizational collaboration — technology transfer frictions, contractual monitoring, and knowledge integration costs. These costs must satisfy $C_c(N) > C_c(I) > 0$, reflecting the greater organizational complexity of licensing frontier innovations with less codified knowledge.

\paragraph{Company acquisition.} Under acquisition, the developer integrates the originating entity and recovers full revenue rights, so that effort incentives replicate the in-house development case
\[
e^*_A = \frac{\theta V}{k}, \qquad \pi_A(\theta) = \frac{\theta^2 V^2}{2k} - \phi,
\]
where $\phi > 0$ is a fixed integration cost - comprising organizational restructuring, cultural alignment, and the elimination of duplicated functions - that is invariant to project quality and innovation type. Importantly, acquisition eliminates the royalty-induced moral hazard wedge and the type-dependent coordination costs $C_c(\tau)$, replacing them with a type-invariant fixed cost $\phi$.

\subsection{Strategy Choice, Equilibrium, and Testable Predictions}
\label{subsec:propositions}

The originator licenses when the expected licensing payoff, comprising royalty income and upfront payments, exceeds internal commercialization returns. Let $U_H(\theta)$ denote the originator's payoff under in-house development, and $U_L(\theta, r)$ the payoff under licensing. The originator licenses if and only if $U_L(\theta, r) \geq U_H(\theta)$. Because $U_L$ is increasing in $\theta$ through royalty income $r \cdot p(\theta, e^*_L) \cdot V$, and $U_H$ grows convexly in $\theta$, the comparison depends on the curvature of both payoff functions and on the precision of information available to developers. Under acquisition, the developer makes a take-it-or-leave-it offer $P_A$ to the originator based on expected project quality $\mathbb{E}[\theta \mid \hat{\theta}, \tau]$. Because the price reflects only the developer's posterior rather than the true $\theta$, the quality of the pricing mechanism depends critically on $\lambda(\tau)$.

\begin{prop}{Positive Selection}{selection}
\noindent Licensed projects exhibit higher success probabilities than in-house developed ones, consistent with positive selection into licensing. When signal precision is sufficiently high, there exists a cutoff $\theta^*(I) \in (0,1)$ such that originators license projects above this threshold and retain the remainder for in-house development. Because contract terms $r(\hat{\theta})$ are responsive to the observed signal, the licensing market sustains a separating equilibrium: licensed projects are positively selected along the innovation-risk spectrum, with higher expected success rates than the population average.
\end{prop}

With precise signals ($\lambda$ large), the developer's posterior is closely aligned with true $\theta$. Contract terms $r(\hat{\theta})$ are responsive to the innovation-risk profile of the project, making licensing attractive for originators above the threshold. A single-crossing condition on $U_L(\theta, r) - U_H(\theta)$ yields the interior cutoff $\theta^*$.

\begin{prop}{Competitive Equilibrium and Return Equalization}{equalization}
\noindent\textit{Incremental projects.} For incremental projects $(\tau = I)$, where information precision is sufficient to sustain separating equilibria, competition equalizes expected developer risk-adjusted returns across in-house and licensed development:
\[
\mathbb{E}[\pi_H(\theta)] = \mathbb{E}[\pi_L(\theta) \mid \theta \geq \theta^*(I)],
\]
so that neither strategy hierarchically dominates. The higher success probability of licensed incremental projects is offset by lower conditional returns, generating a competitive risk-return trade-off.

\tcblower
\noindent\textit{Novel projects.} For novel projects $(\tau = N)$, fundamental uncertainty limits the ability to price innovation risk \textit{ex-ante}. Return equalization need not hold: licensed novel projects may exhibit higher success probabilities without a corresponding reduction in returns, constituting a violation of competitive equilibrium consistent with lemons-type frictions.
\end{prop}

\noindent Under company acquisition, the offer price $P_A = P_A(\hat{\theta}, \tau)$ reflects the developer's posterior, and since acquisition eliminates royalty distortions the acquisition premium is higher than the licensing value for high-$\theta$ projects; however, the type-invariant fixed cost $\phi$ makes acquisition relatively less attractive for weaker projects and, under competitive bidding, is priced into the acquisition offer itself, yielding expected returns that are largely invariant to $\tau$ — so that acquisitions should exhibit neither the equilibrium-adjustment mechanism of licensing nor its distortions across innovation types.

\begin{prop}{Cost Comparison}{costs}
\noindent\textit{Licensing vs. in-house development.} Licensing increases total development cost relative to in-house development if and only if coordination costs dominate effort savings
\[
C_c(\tau) > C(e^*_H) - C(e^*_L) = \frac{\theta^2 V^2 r(2-r)}{2k}.
\]
This condition provides a cost-side mechanism that reinforces the return disadvantage of licensing documented in Proposition 2
\end{prop}

The conceptual framework delivers three central testable predictions. First (\emph{positive selection}), licensed projects exhibit higher success probabilities than in-house developed ones, with the effect driven primarily by incremental projects where signal precision supports separating equilibria. Second (\emph{return equalization}), the success-probability advantage of licensing is offset by lower conditional returns for incremental projects, consistent with competitive equilibrium; this offset is absent for novel projects, where low information precision prevents the market from fully pricing innovation risk \textit{ex-ante}, generating lemons-type frictions. Third (\emph{cost premia}), licensing entails higher total development costs when coordination costs outweigh effort savings, a mechanism that supports the return patterns predicted in Proposition 2 and whose empirical content is examined in the appendix. These predictions motivate our empirical distinction between incremental and novel pharmaceutical projects and guide the empirical strategy described in Section~\ref{sec:methodology}.

\section{Data and Descriptive Statistics}
\label{sec:data}

\subsection{Data Sources}
\label{subsec:data_sources}

Data are from Evaluate Pharma\textsuperscript{\textcircled{R}} and relate to pharmaceutical products across multiple dimensions of drug development. The dataset gathers information about development strategies, therapeutic characteristics, technological features, regulatory outcomes, and cost and sales information, offering a representative sample of industry standards and practices across various layers of dimension. Data are collected from multiple sources to ensure comprehensive coverage of the pharmaceutical development landscape, including both approved drugs and failed product development projects. This dual coverage allows us to analyze both success probabilities for completed projects and realized costs and sales for approved drugs, providing a comprehensive view of how development strategies relate to different pharmaceutical outcomes. This empirical setting includes rich project-level characteristics that allow us to construct predicted success probabilities (serving as empirical proxies for project novelty) and enables us to test the theoretical predictions derived from our model regarding selection patterns (Proposition 1), equilibrium returns (Proposition 2), and development cost efficiency (Proposition 3).

\subsection{Classification of Development Status and Strategies}
\label{subsec:dev_status_and_strat}

The primary distinguishing feature for each pharmaceutical product in the dataset is its current market status worldwide, operationalized as a two-tiered classification system that captures the position within the pharmaceutical product lifecycle. This taxonomic framework is defined by two hierarchical dimensions: (1) Market Status, which includes categories such as Marketed, Research \& Development (R\&D), Suspended in R\&D, Abandoned in R\&D, Disposed in R\&D, and Transferred through Mergers \& Acquisitions (M\&A); and (2) Development Phase, which ranges from various clinical trial phases (Phase I, Phase II, Phase III, and preclinical stages) to filed and approved products. For analytical purposes, observations are subsequently stratified into three mutually exclusive categories: abandoned products, approved products, and ongoing development projects. This classification scheme maps the chronological and regulatory timeline of each product's developmental trajectory, allowing us to observe both the probability of success $p(\theta,e)$ through approval outcomes and realized development costs for completed projects.

The core strategy variable in the empirical analysis distinguishes between  alternative development modes, with our primary focus on the comparison between in-licensed and internally developed products. Products developed in-house — which constitute the baseline category labeled as ``Organic development'' in our data — are defined as products developed internally, relying entirely on a firm's in-house R\&D capabilities. In-licensed products are those sourced externally through licensing agreements, typically involving milestone payments, royalties, or other forms of contingent compensation. As anticipated in Section~\ref{sec:theoretical_model}, our data also include a residual category of products sourced through company acquisition — that is, products obtained through the acquisition of an entire firm or a specific business division. This category enters our DML analysis exclusively as a governance-mode benchmark, as its role is to verify that the equilibrium patterns we document for licensing are specific to that contractual form and do not generalize to other forms of external sourcing. Notably, such acquisitions may not always be motivated by the intention to further develop the acquired products; as shown by \citet{cunningham2021killer}, some incumbent firms acquire innovative targets with the strategic aim of discontinuing their innovation projects to preempt future competition. Although additional development strategies (such as standalone product acquisitions and co-development partnerships like joint-ventures) can be observed in the raw data, these are excluded from the main analysis due to their low rate of occurrence and lack of statistical relevance.

\subsection{Confounding Features and Selection on Observables}

The choices of firms among these alternative development strategies are inherently endogenous \citep{liu2024pharmaceutical} and determined by a complex set of observed confounding factors. This endogeneity reflects the selection mechanism formalized in our theoretical model: firms choose to license projects based on the expected net benefit from licensing relative to internal development, which depends on unobserved project quality $\theta$ as well as on observable characteristics. Among the observable confounders, these include therapeutic area and subcategory of the new product, its technological classification, the degree of product proprietary control and patenting decisions — consistent with evidence that intellectual property rights can shape subsequent innovation \citep{williams2013intellectual, kogan2017technological, azoulay2019public}. Importantly, these characteristics also capture dimensions related to project risk: more innovative therapeutic approaches, novel mechanisms of action, and earlier-stage technologies typically exhibit lower predicted success probabilities but potentially higher value conditional on success. All such characteristics are explicitly modeled into the empirical framework through a comprehensive matrix of confounding variables influencing both strategic choices and developmental outcomes. The DML framework allows us to flexibly control for these high-dimensional confounders while estimating the parameters of interest, effectively conditioning on observables to approximate the selection process.

In Evaluate Pharma\textsuperscript{\textcircled{R}} data, these confounding factors are coded as categorical variables and are subsequently one-hot encoded to construct a comprehensive covariate matrix comprising more than 140 binary features, enabling the model to control for the full heterogeneity in project characteristics that may influence both strategy assignment and outcomes. This rich set of controls allows us to approximate the distribution of project quality $F(\theta)$ conditional on observables, reducing concerns about selection on unobservables that drive the information asymmetry in our theoretical framework. In addition, these controls are used to construct predicted success probabilities that approximate project risk, i.e. our key dimension of heterogeneity for understanding how licensing markets function differently across different degrees of pharmaceutical project innovation.

In order to validate the assumption that our confounding matrix captures genuine structural determinants of pharmaceutical outcomes, we conduct bootstrap variable inclusion analysis for all outcomes, complemented by feature importance decomposition for the success probability outcome. Across 100 bootstrap resamples, proprietary level exhibits the highest level of selection stability for all outcomes, accounting for approximately two-thirds of predictable variance in success probability according to both Gini and permutation importance measures. This dominance mirrors the consolidated pharmaceutical reality in which the distinction between first-in-class innovation (new molecular entities) and proven mechanisms (generics, established compounds) determines development risk more than any other observable characteristic. Therapeutic category and subcategory, technology category, and technology type explain progressively smaller shares of predictive variance, forming a clear hierarchy that aligns with domain knowledge about pharmaceutical development. This consistency across outcomes and importance metrics, combined with strong replication stability, supports the interpretation of our high-dimensional control matrix as a reliable proxy for the genuine confounding structure that influences both outcomes and strategy choices.

Table~\ref{tab:success_by_prop_level} provides a preliminary illustration of how development risk varies with project novelty by reporting realized success probabilities across proprietary level categories using the sample employed for selection testing. The data reveal a defined monotonic gradient, with approval rates declining from 90.7\% for generics to only 9.1\% for new molecular entities, offering a first-order confirmation that innovation novelty inversely correlates with development success. Notably, although proprietary level serves as a useful initial proxy for the innovation dimension, our subsequent heterogeneity analysis employs predicted success probability, constructed via machine learning from the full set of observable project characteristics, as a more granular and continuous measure of project risk and related novelty that better captures the information precision parameter $\lambda(\tau)$ in our theoretical framework.

\begin{table}[h!]
\centering
\caption{\textbf{Table 1} - Actual Success Probability by Proprietary Level Category}
\label{tab:success_by_prop_level}
\small
\setlength{\tabcolsep}{12pt}
\renewcommand{\arraystretch}{1.20}
\begin{tabular}{lrrrcc}
\toprule\toprule
 & & & & \multicolumn{2}{c}{\textit{Success Probability}} \\
\cmidrule(lr){5-6}
\textbf{Proprietary Level} & \textbf{N} & \textbf{Approved} & \textbf{Failed} & \textbf{Mean} & \textbf{Std. Dev.} \\
\midrule
Generics             & 22,926 & 20,803 &  2,123 & 0.907 & 0.290 \\
Other                &  4,246 &  3,664 &    582 & 0.863 & 0.344 \\
Biosimilars          &  1,283 &    734 &    549 & 0.572 & 0.495 \\
New Drug Application                  & 13,834 &  6,107 &  7,727 & 0.441 & 0.497 \\
New Molecular Entity & 47,680 &  4,350 & 43,330 & 0.091 & 0.288 \\
\bottomrule\bottomrule
\end{tabular}
\end{table}

\subsection{Outcomes}

Outcome variables are constructed to capture the key quantities in our theoretical model. The two primary outcomes examined in the body of the paper are success probability ($Y_1$) and net return ($Y_2$), which correspond directly to the central testable predictions of Propositions 1 and 2 respectively. The two components of $Y_2$ — clinical trial costs ($Y_3$) and lifetime sales ($Y_4$) — are also separately used as outcomes in the appendix, providing supplementary evidence on Proposition 3 and on the commercial dimension of licensing performance. Finally, an alternative measure of commercial performance is computed as the ratio of lifetime sales to clinical trial costs ($Y_5$) and is also used as outcome in the appendix as a robustness check on the net return findings.

\paragraph{Success Probability ($Y_1$).} First, the probability of success feature captures the likelihood of successful completion and regulatory approval for completed projects, and it is computed as a binary feature which takes value 1 for approved products and 0 for abandoned projects. This measure directly corresponds to the success probability $p(\theta, e)$ in our theoretical model, allowing us to test the explicit prediction in Proposition 1 regarding positive selection.  However, it is crucial to interpret this measure correctly: higher success probabilities often reflect low-uncertainty, more incremental projects following well-established development pathways, while lower success rates characterize higher-risk, more innovative projects pursuing novel mechanisms or unproven therapeutic approaches. If licensed projects systematically exhibit higher success rates, this may indicate selection of lower-risk rather than higher-quality projects — a distinction that becomes central to understanding whether licensing markets efficiently allocate the most innovative pharmaceutical assets.

Ultimately, this measure allows comparison between the effectiveness of different strategic approaches in relation to the substantial attrition that characterizes pharmaceutical development, in which most compounds entering the clinical trial phase eventually fail to reach the market. This focus on approval outcomes connects to a broader literature documenting how short-term incentives, patent term length, and the organizational structure of R\&D can systematically skew private investment away from longer-horizon projects \citep{budish2015firms,de2022biopharmaceutical}.

To ensure the temporal validity of strategy assignment and avoid endogeneity concerns, particular care is taken to ensure that all externally developed products — whether in-licensed or acquired — are considered if and only if the corresponding acquisition or licensing deal was signed before market approval. Consequently, while all in-house developed products are investigated without temporal restrictions, products sourced through licensing or company acquisitions are filtered accordingly to avoid post-approval selection bias. This temporal restriction ensures that the development strategy choice precedes the outcome (approval decision), maintaining the ordering implicit in our theoretical model in which firms choose licensing based on pre-development assessment of project quality $\theta$.

\paragraph{Net Return ($Y_2$).} Second, net return is defined as lifetime sales minus clinical trial costs (in millions of U.S. dollars) and constitutes our primary measure of economic performance across development strategies. This outcome allows us to assess the implications of Proposition 2 regarding equilibrium returns: for incremental projects, competitive markets should equalize risk-adjusted returns across licensed and internally developed products, while for novel projects this equalization may not hold. Abandoned projects are assigned a lifetime sales value of zero, resulting in negative net return values that reflect unrecovered development expenditures. The inclusion of failed projects ensures a more realistic assessment of strategic performance, consistent with our model's emphasis on total economic surplus rather than conditional-on-success outcomes.

\subsection{Sample Construction and Filtering}

The construction of the sample required several important decisions regarding inclusion criteria and data quality requirements, with specific filtering procedures applied depending on the outcome variable under analysis. The starting point is a dataset of 192,611 raw product-level observations.

For the primary outcome of success probability ($Y_1$), the sample includes both approved and abandoned products, yielding 98,712 observations, of which 36,778 relate to approved products (37.3\%) and 61,934 to abandoned projects (62.7\%). Excluding company acquisition observations (8,376 products) for the DML analysis yields a final sample of 90,336 observations comprising 79,098 in-house developed products and 11,238 in-licensed products.

For the net return outcome ($Y_2$), the sample is constructed from products with non-missing information on trial costs, restricted to products with resolved outcomes (approved or abandoned). After applying strategy-specific temporal restrictions, the final sample contains 13,595 observations: 2,722 approved products (approximately 20\%) and 10,873 abandoned projects (approximately 80\%). Abandoned projects are assigned lifetime sales of zero, generating negative net return values that reflect unrecovered development expenditures.

\begin{figure}[tp]
    \centering
    \includegraphics[width=1\linewidth]{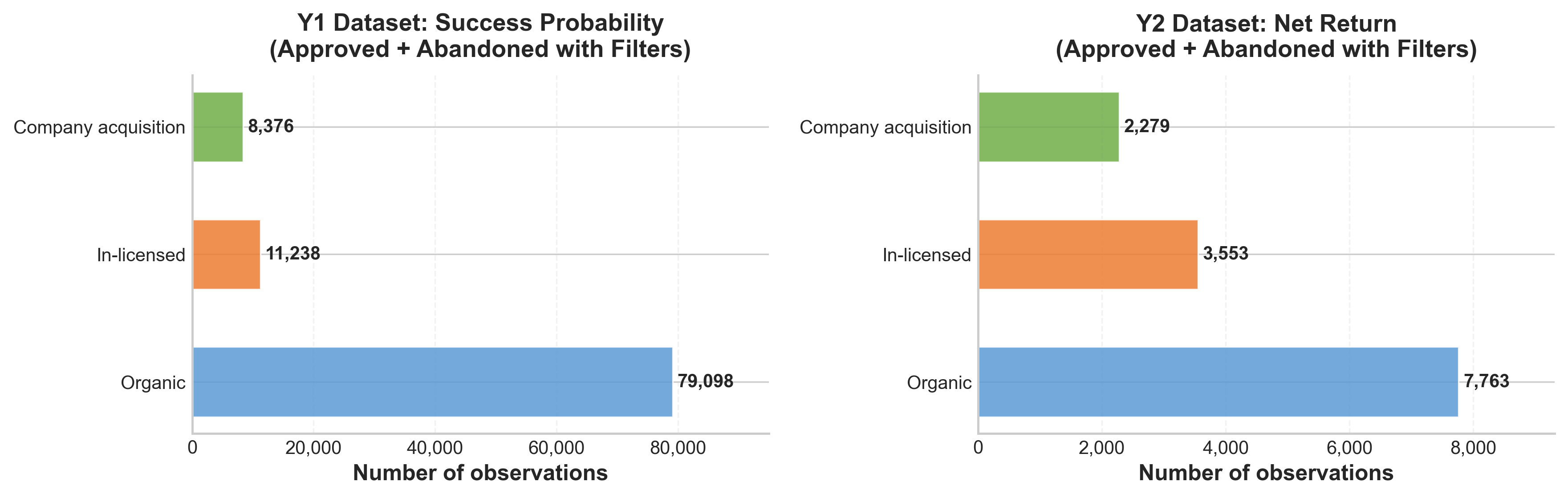}
    \caption{\textbf{Figure 1} - Distribution of Products by Strategy across Datasets}
    \label{fig:fig1}
\end{figure}

\begin{figure}[tp]
    \centering
    \includegraphics[width=0.8\linewidth]{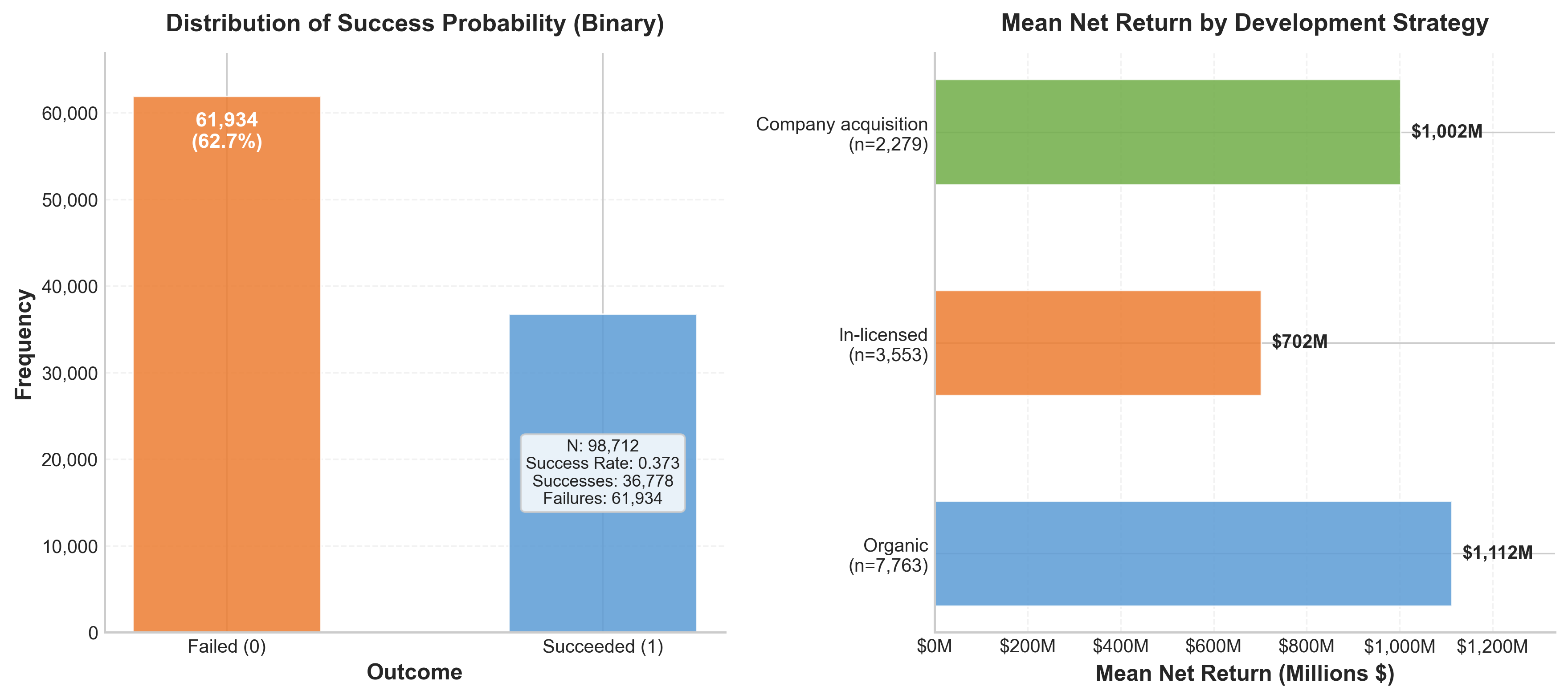}
    \caption{\textbf{Figure 2} - Summary Statistics for $Y_1$ and $Y_2$}
    \label{fig:fig2}
\end{figure}

Throughout our empirical analysis, we focus on the comparison between in-licensed and in-house developed products, as this comparison directly tests our theoretical model's predictions regarding licensing markets. Products sourced through company acquisition are excluded from DML estimation concerning positive selection to maintain analytical focus, though they are included in descriptive statistics for completeness — see Figure~\ref{fig:fig1} and Figure~\ref{fig:fig2} — and in the later DML heterogeneity analysis as a benchmark for licensing-specific results.

\section{Methodology}
\label{sec:methodology}

The propositions derived in the previous section yield empirical implications regarding selection patterns, return equalization, and development cost efficiency across innovation regimes. To bring these predictions to the data, we rely on the Double Machine Learning (DML) framework \citep{chernozhukov2018double}, which is particularly well suited to the analysis of high-dimensional observational data characterized by complex and potentially non-linear relationships. The DML approach allows us to assess the predictions derived from our theoretical model while flexibly controlling for the high-dimensional confounding structure that characterizes pharmaceutical product development. More in detail, the reasons behind the choice of DML can be found in several considerations that are central to the empirical setting of pharmaceutical product development.

First, clinical trial development data is inherently high-dimensional. Due to extensive preprocessing and one-hot encoding of categorical confounders, the set of pre-treatment factors includes more than 140 distinct features referring to product-level attributes that jointly influence both firms' strategic development choices and downstream outcomes. This dimensionality substantially exceeds what traditional parametric methods can adequately accommodate while maintaining sufficient degrees of freedom for meaningful statistical inference. Moreover, classical regression approaches are prone to overfitting, multicollinearity, and unstable coefficient estimates, while \textit{ad hoc} variable selection may lead to omitted-variable bias. On the other hand, DML is specifically designed to handle settings of this kind with many controls relative to sample size, allowing for flexible regularization without compromising valid inference on parameters of interest.

Second, the relationships between confounders and both the choice of development strategy and outcomes are likely to be highly non-linear and characterized by complex interactions. Pharmaceutical firms' strategic decisions regarding whether to develop products in-house or obtain them via licensing agreements result from a complex optimization process that accounts for product-specific characteristics, project risk profiles, and organizational capabilities. Similarly, the mapping from confounders to development outcomes is mediated by regulatory requirements, clinical trial designs, patient recruitment challenges, and market competition, all of which are likely to operate through non-linear dynamics. In this context, a key advantage of the DML framework is that it places no \textit{a priori} restrictions on the functional form of these relationships, succeeding in capturing non-linearities and complex interaction patterns that would be difficult to model explicitly.

Third, manually specifying the appropriate functional form for these relationships would require extensive model selection procedures with a high risk of specification error. The traditional approach of testing various polynomial terms, interaction effects, and non-linear transformations would have been not only computationally intensive but also subject to multiple testing problems and researcher discretion in model specification that can lead to spurious findings. In contrast, the DML framework offers a unique way to combine the predictive power of machine learning methods to flexibly model nuisance relationships with the inferential rigor necessary for identification \citep{belloni2014inference}. The first key methodological insight underlying DML is that, by employing cross-fitting procedures, researchers can avoid the overfitting problems that would otherwise arise from using the same data to both train machine learning models and estimate coefficients of interest. The second relevant point concerns orthogonalization (achieved through the partialling-out approach) that removes the direct dependence of the estimator on the nuisance parameters \citep{athey2019machine}, thus yielding estimates that are robust to small misspecifications in the machine learning models and achieve $\sqrt{n}$-consistency and asymptotic normality under regularity conditions.

The adoption of machine learning methods is further supported by a growing body of evidence demonstrating that non-linear algorithms substantially outperform traditional linear models in predicting trial outcomes and drug approval probabilities \citep{siah2021predicting, lo2019machine}. Because the relationship between product characteristics and development outcomes is better captured by flexible non-linear models, incorporating such learners in both the outcome and strategy-selection equations of the DML procedure yields more accurate nuisance approximations and, consequently, more reliable inference on the parameters of interest.

\subsection{DML Specification and Estimation}
\label{subsec:dml_specification}

The DML estimation procedure considers the following partially linear model
\begin{equation*}
Y_i = \theta_0 + \theta_{\text{lic}} D_{\text{lic},i} + g(X_i) + \varepsilon_i,
\end{equation*}
where $i$ indexes products and $D_{\text{lic},i}$ is an indicator for in-licensed products. The outcome variable $Y_i$ represents the two primary measures of development performance examined in the body of the paper: the binary approval indicator ($Y_1$, corresponding to success probability $p(\theta,e)$ in Proposition 1) and net return, defined as lifetime sales minus clinical trial costs ($Y_2$, related to the equilibrium return predictions in Proposition 2). Additional outcomes are examined in the appendix as robustness checks. In-house development serves as the baseline category. The high-dimensional vector $X_i$ collects all observed confounders determined before the strategy choice, approximating the observable correlates of project risk and other factors influencing the strategy selection decision. In the heterogeneity analysis that follows, these confounders enable us to construct predicted success probabilities that proxy project risk, empirically corresponding to the information precision parameter $\lambda(\tau)$ in our theoretical framework (with higher precision associated with incremental projects and lower precision with novel ones). The function $g(\cdot)$ represents an unknown, potentially highly non-linear mapping from confounders to outcomes and is treated as a nuisance component. The parameter $\theta_{\text{lic}}$ captures the average estimated difference in outcomes between in-licensed and internally developed products, conditional on the confounders $X_i$.

A crucial feature of the DML framework is that the nuisance functions are not interpreted structurally. Their unique purpose is to partial out the influence of confounders as accurately as possible. Accordingly, estimation prioritizes predictive performance rather than interpretability when modeling $g(X_i)$ and the strategy selection mechanisms. This distinction is essential: while the parameter $\theta_{\text{lic}}$ is the object of economic interest, the nuisance components function only as statistical means to remove confounding bias.

Estimation proceeds via the partialling-out approach \citep{chernozhukov2018double}. The sample is repeatedly partitioned into $K$ folds: $\{I_1, I_2, \ldots, I_K\}$ such that $I_k \cap I_j = \emptyset$ for $k \neq j$. For each split the conditional expectations of the outcome and strategy variables given $X_i$ are estimated on auxiliary subsample $I_k^c = \{1,2,\ldots,n\} \setminus I_k$ using machine learning methods
\begin{align*}
\text{Outcome estimation:} \quad &\hat{m}_{k}(X) = \arg\min_{m \in \mathcal{M}} \sum_{i \in I_k^c} L(Y_i, m(X_i)) \\
\text{Strategy estimation:} \quad &\hat{\pi}_{k}(X) = \arg\min_{\pi \in \Pi}\sum_{i \in I_k^c} \ell(\mathbbm{1}\{D_{\text{lic},i} = 1\}, \pi(X_i))
\end{align*}
where $\mathcal{M}$ and $\Pi$ represent appropriate function classes, $L(\cdot,\cdot)$ denotes the loss function for outcome prediction, and $\ell(\cdot,\cdot)$ the loss function for strategy estimation. The fitted values $\hat{m}_k(X_i)$ estimate $\mathbb{E}[Y_i|X_i]$ and $\hat{\pi}_k(X_i)$ estimate the propensity score $P(D_{\text{lic},i}=1|X_i)$.

These estimates are used to construct orthogonalized residuals
\begin{align*}
\tilde{Y}_i &= Y_i - \hat{m}_{k}(X_i) \\
\tilde{D}_{\text{lic},i} &= \mathbbm{1}\{D_{\text{lic},i} = 1\} - \hat{\pi}_{k}(X_i)
\end{align*}
which capture the variation in outcomes and strategies that remains after removing the predictable component associated with the confounders, effectively isolating variation in development strategies and outcomes that is orthogonal to observable risk proxies.

In the final stage, the parameter of interest is recovered via OLS of the orthogonalized outcome residuals on the orthogonalized strategy residuals:
\begin{equation*}
\hat{\theta}_{\text{lic}} = \left[\sum_{i=1}^{n} \tilde{D}_{\text{lic},i}^2\right]^{-1} \left[\sum_{i=1}^{n} \tilde{D}_{\text{lic},i} \tilde{Y}_i\right]
\end{equation*}
For certain outcomes reported in the appendix, this final step is modified to employ weighted least squares (WLS) after applying inverse probability of selection weighting (IPSW) to correct for the differential probability of observing approved versus failed projects; the full specification is described in Appendix~\ref{app:results}.

\paragraph{Machine Learning Algorithms}
\label{subsec:dml_learners}

Different machine learning algorithms are employed depending on the nature of the outcome variable. For the binary approval indicator ($Y_1$), conditional expectation functions are estimated using Logistic Regression, Lasso Logistic, Random Forest classifier, and Gradient Boosting classifier. For $Y_2$ (net return), only Random Forest is employed, as it consistently delivers the best predictive performance among the learners evaluated for $Y_1$ and carries over its superiority to the continuous net return setting. For the remaining continuous outcomes reported in the appendix - trial costs ($Y_3$), lifetime sales ($Y_4$), and the ROI ratio ($Y_5$) - the estimation strategy implements Lasso, Ridge, and Elastic Net regressions with penalty parameters selected via cross-validation, as well as Random Forest and Gradient Boosting regressors. This combination of estimators enhances robustness by allowing the DML procedure to exploit different modeling strengths across outcomes. The use of multiple learners also provides a form of sensitivity analysis: stability of estimates across machine learning algorithms strengthens confidence that results reflect genuine patterns in the data rather than specific modeling assumptions.

Standard errors and confidence intervals for $\hat{\theta}_{\text{lic}}$ are constructed using the asymptotic distribution theory of \citet{chernozhukov2018double}. Under regularity conditions — including sufficient sample size, appropriate convergence rates for the nuisance parameter estimates, and overlap in the propensity score distribution — the DML estimator achieves $\sqrt{n}$-consistency and asymptotic normality, enabling valid inference despite the use of machine learning in the first stage.

\subsection{DML-IV Specification and Estimation}
\label{subsec:dmliv_specification}

Although the DML framework delivers well-controlled estimates of the average return difference between licensed and organically developed products, a fundamental identification challenge remains: firms' licensing decisions are endogenous. Even with a rich control matrix, a skeptical reader could attribute the observed return differences to unobserved quality differences between licensed and organic projects rather than to the causal effect of the licensing process itself. To address this concern, we complement the DML analysis with a Instrumental Variable identification strategy (from now on, DML-IV) that exploits exogenous variation in licensing propensity.

Following \citet{hermosilla2021rushed}, we instrument for licensing using exogenous pipeline shocks experienced by licensee firms. The identification logic rests on a well-documented feature of pharmaceutical R\&D: Phase III clinical trial failures (P3Fs) represent large, sudden, and largely unanticipated negative shocks to a firm's pipeline. Since Phase III trials are both the longest and most expensive stage of clinical development — and because projects at this point have already cleared most attrition hurdles, making them more likely to succeed than fail — their sudden termination constitutes a major negative shock to a company’s pipeline. This, in turn, creates immediate pressure from stakeholders — i.e. investors, analysts, and patient organizations — on management to replenish the pipeline quickly. Firms respond to these shocks by significantly increasing their licensing activity within a short window of approximately one year, a timeline that is compressed relative to the usual licensing process, which under normal conditions requires extensive planning, candidate screening, due diligence, and contract negotiation. The quasi-experimental nature of P3Fs arises from the fact that, conditional on a firm's pipeline composition and Phase III exposure (both of which can be controlled for) the precise timing of a trial failure is driven by clinical and biological factors largely beyond managerial control. It follows that once a firm's Phase III risk score is accounted for, no observable contextual factor systematically predicts P3F incidence, supporting the interpretation that P3Fs are as-good-as-random conditional on controls. This makes P3Fs a credible source of exogenous variation in licensing propensity, as they shift the probability that a firm engages in licensing without directly affecting the quality of the products available for licensing in the market.

Formally, the instrument is defined at the product level as:
\begin{equation*}
Z_i = \begin{cases} 1 & \text{if product $i$ has been developed under licensing agreement and at least one of } \\ & \text{its licensees experienced a P3F in the 365 days strictly before the deal date} \\ 0 & \text{otherwise} \end{cases}
\end{equation*}
The 365-day window follows \citet{hermosilla2021rushed} and reflects the empirical regularity that P3F-fueled licensing activity is concentrated within one year of the shock and attenuates at longer horizons, consistent with the ``rushed'' rather than strategic nature of these transactions. By construction, in-house developed products receive $Z_i=0$, as they are not associated with any licensing transaction. The instrument therefore isolates a specific subset of in-licensed products — those whose deal was plausibly accelerated by acute pipeline pressure on the licensee — from both in-house and non-rushed licensed products, generating exogenous variation in licensing propensity that is orthogonal to unobserved product quality.

In our structural model, the causal object of interest is the Local Average Treatment Effect (LATE) of in-licensing on outcome $Y_i$, defined as the average effect among complier projects. This complier population corresponds to \emph{rushed} licenses in the sense of \citet{hermosilla2021rushed}: products drawn into licensing transactions under time pressure rather than through the deliberate quality-screening process that characterizes non-rushed deals. The LATE is therefore a local parameter, pertaining to this specific subpopulation, and should be interpreted as the causal effect of licensing under compressed due diligence conditions and time for project evaluation. The equation estimated is equivalent to the DML case, with the only difference that the control vector $X_i$ now comprises the same one-hot encoded product-level characteristics used in the DML analysis, augmented by a continuous firm-level risk score — defined as the ratio of Phase III failures to total Phase III-exposed projects ever attributed to any firm associated with product $i$. Including the risk score as a control ensures that the instrument $Z_i$ isolates variation in licensing propensity driven by the \emph{timing} of pipeline shocks instead of firms' structural exposure to Phase III risk. Both these refinements mirror the identification strategy of \citet{hermosilla2021rushed}. The exclusion restriction requires that $Z_i$ affects outcomes $Y_i$ only through its effect on the licensing decision $D_{\text{lic},i}$: conditional on $X_i$, a licensee's P3F affects product-level outcomes uniquely by shifting the probability of the licensing transaction, not through any direct channel. This restriction is plausible in our setting because P3Fs are failures of \emph{other} projects in the licensee's pipeline, unrelated to the scientific or commercial characteristics of the product being licensed. Finally, one additional concern arising from the product-level application of the instrument is that firms experiencing a P3F are likely to rush-license a replacement project in the same therapeutic area as the failure. This expected compositional shift is directly absorbed by the therapeutic area dummies included in $X_i$, ensuring that residual variation in $Z_i$ reflects only the exogenous timing of the pipeline shock rather than disease-area targeting, and preserving the exclusion restriction conditional on controls.

The first stage is
\begin{equation*}
D_{\text{lic},i} = \pi_0 + \pi Z_i + h(X_i) + \eta_i,
\end{equation*}
where $h(X_i)$ flexibly controls for the full set of one-hot encoded project characteristics and the continuous firm-level risk score $R_i$ - defined as the maximum historical P3F rate across all developers of the project - both of which are included in the augmented control matrix $X_i$. 
The coefficient $\pi$ captures the residual effect of the instrument after removing all variation explained by these controls, thus ensuring that the instrument truly reflects pipeline shocks rather than heterogeneous firm-level failure propensity.

\paragraph{Estimation procedure.} The DML-IV estimator follows the partially linear IV model of \citet{chernozhukov2018double}. Three nuisance functions are estimated via cross-fitted Random Forest models with $K=5$ folds: $\hat{\ell}(X_i) = \hat{\mathbb{E}}[Y_i \mid X_i]$, $\hat{m}(X_i) = \hat{\mathbb{E}}[D_{\text{lic},i} \mid X_i]$, and $\hat{r}(X_i) = \hat{\mathbb{E}}[Z_i \mid X_i]$. Subtracting these fitted values from the observed outcome, treatment, and instrument yields the residuals $\tilde{Y}_i$, $\tilde{D}_{\text{lic},i}$, and $\tilde{Z}_i$, which capture the variation in each variable that is orthogonal to the full set of controls. The parameter of interest is then recovered as the Wald ratio on residualized variables:
\begin{equation*}
\hat{\theta}_{\text{lic}}^{\text{IV}} = \frac{\sum_i \tilde{Z}_i \tilde{Y}_i}{\sum_i \tilde{Z}_i \tilde{D}_{\text{lic},i}},
\end{equation*}
which scales the residual covariance between instrument and outcome by the residual covariance between instrument and treatment. Heteroskedasticity-robust standard errors are constructed following Theorem 4.2 of \citet{chernozhukov2018double}, and instrument relevance is assessed via the first-stage $F$-statistic from regressing $\tilde{D}_{\text{lic},i}$ on $\tilde{Z}_i$. The DML-IV identifies a LATE for the complier subpopulation — products whose licensing was triggered by a pipeline shock — and is therefore local rather than average. Its role in the analysis is to provide causal evidence that the equilibrium patterns documented by the DML are not driven by unobserved quality selection, as discussed in Section~\ref{sec:dmliv_results}.

\section{DML Results}
\label{sec:dml_results}

This section presents the empirical estimates for in-licensing relative to in-house development across the two dimensions of pharmaceutical outcomes identified (success probability and monetary returns). The analysis employs the Double Machine Learning framework described in Section~\ref{subsec:dml_specification}, assessing the theoretical predictions regarding selection patterns (Proposition 1) and equilibrium returns (Proposition 2). We begin by examining selection into licensing through success probability estimates, then proceed to the heterogeneity analysis for the net returns that reveals how estimated differences vary with predicted success probability (our proxy for project risk) across the innovation-risk distribution. Results for trial costs ($Y_3$), lifetime sales ($Y_4$), and the ROI ratio ($Y_5$) are reported in Appendix~\ref{app:results} and are consistent with the patterns discussed below.

\subsection{Selection Patterns: Success Probability}
\label{subsec:dml_y1_results}

Our key reference question is whether licensing exhibits positive or negative selection — that is, whether projects with higher predicted success probability are more likely to be licensed (as predicted by Proposition 1 for incremental projects with sufficiently high information precision) or whether adverse selection dominates. The success probability outcome ($Y_1$) is employed to directly assess whether licensing exhibits positive selection in terms of approval likelihood. According to our theoretical model, for incremental projects with sufficiently predictable outcomes, we should observe that only projects exceeding a quality threshold are licensed, implying higher success probabilities for in-licensed products after controlling for observable characteristics.

\begin{figure}[tp]
    \centering
    \includegraphics[width=.8\linewidth]{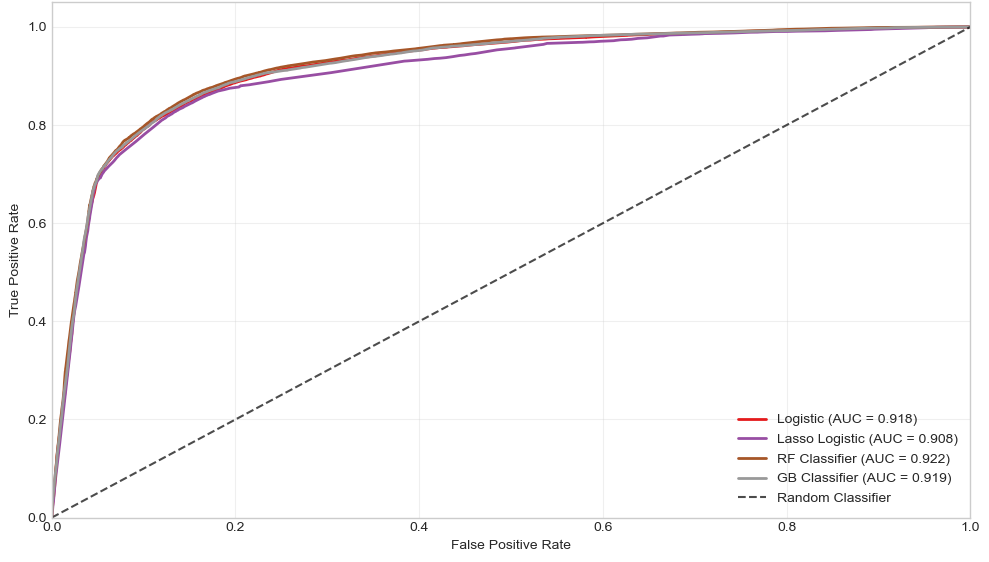}
    \caption{\textbf{Figure 3} - ROC Curves for $Y_1$ Prediction}
    \label{fig:roc_y1}
\end{figure}

\begin{figure}[tp]
    \centering
    \includegraphics[width=.8\linewidth]{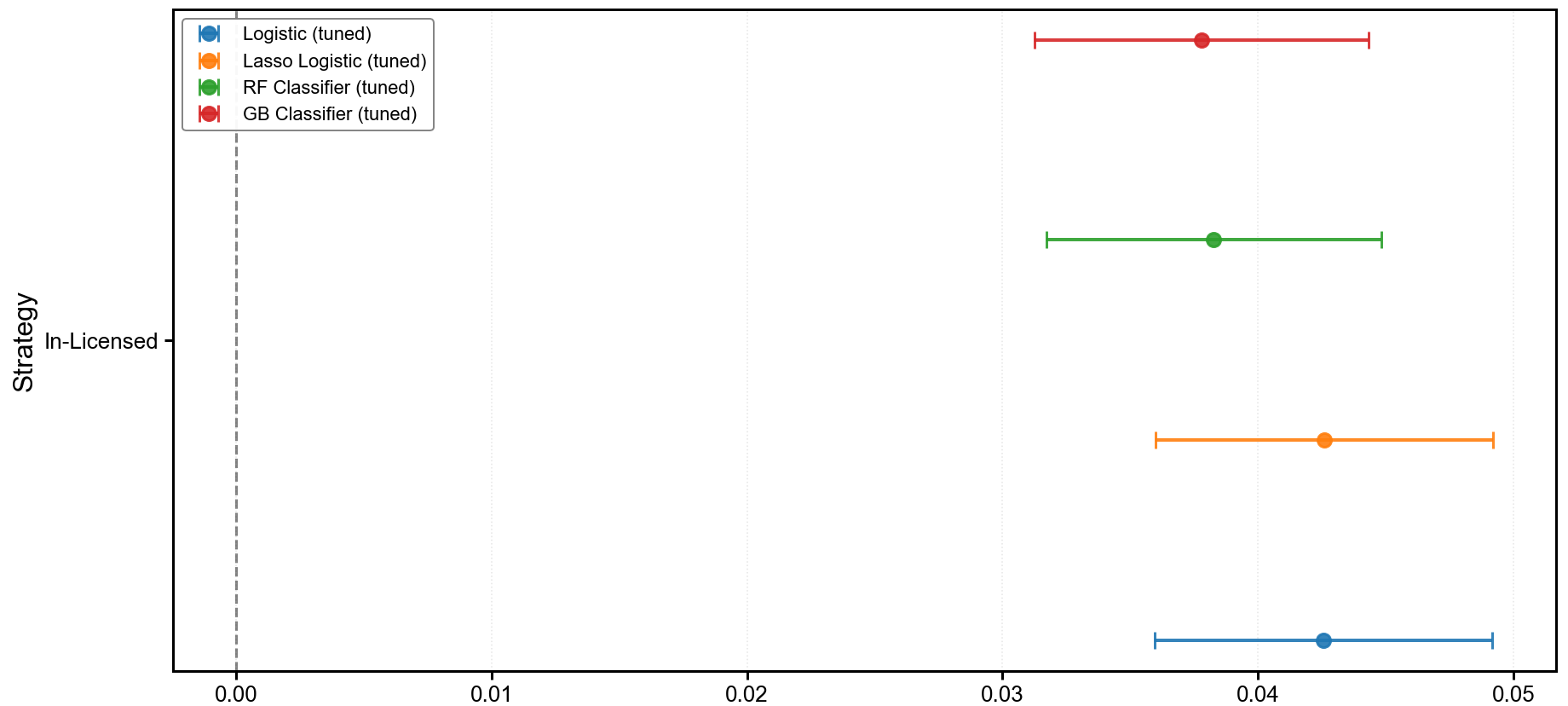}
    \caption{\textbf{Figure 4} - DML Estimate: Success Probability ($Y_1$) with 95\% Confidence Intervals}
    \label{fig:dml_y1}
\end{figure}

The sample includes 90,336 products (both approved and abandoned), of which 39.6\% (35,774 projects) achieved market approval. In this binary outcome setting (1 for approved products, 0 for failed projects), machine learning algorithms demonstrate strong predictive performance, with Random Forest classifier achieving AUC values exceeding 0.92, indicating excellent discrimination between approved and abandoned products (Fig.~\ref{fig:roc_y1}). On the other hand, the DML estimates provide clear evidence of positive selection into licensing, strongly supporting Proposition 1 (Fig.~\ref{fig:dml_y1}). In-licensed products exhibit significantly higher success probabilities than internally developed products, with point estimates ranging from 0.038 (Random Forest) to 0.043 (Logistic) across different learners. Given the baseline approval rate of 39.6\%, these estimates represent 3.8-4.3 percentage point increases in absolute terms, corresponding to approximately 9.6--10.9\% relative increases in success probability. All estimates are highly statistically significant ($p<0.001$) and remarkably stable across different machine learning estimators\footnote{Hyperparameters for all learners are selected via Bayesian optimization minimizing cross-validated log-loss. The selected hyperparameters are then applied to the full 5-fold DML cross-fitting procedure.}.

These findings carry important implications regarding selection patterns in pharmaceutical technology markets. For in-licensing, the results directly contradict the pure adverse selection hypothesis \citep{akerlof1970lemons} that would predict lower success rates for externally sourced projects if biotechnology firms systematically bring their lowest-quality assets to market. Instead, the results align with the positive selection equilibrium characterized in Proposition 1: licensed projects systematically exhibit higher predicted success probability, consistent with selection above a quality threshold for projects with sufficiently high information precision. These findings also confirm prior empirical evidence from \citet{dimasi2010trends} documenting higher success rates for in-licensed compounds, though our analysis controls for a much richer set of observable confounders. However, as noted earlier, ``positive selection'' in terms of success probability may represent selection of lower-risk rather than higher-value projects. Higher success rates could reflect licensing of more incremental innovations following established development pathways (corresponding to the incremental regime $\tau = I$ with high information precision $\lambda(I)$ in our framework), rather than novel, high-risk projects with fundamental uncertainty (corresponding to the novel regime $\tau = N$ with low information precision $\lambda(N)$) that represent the frontier of pharmaceutical innovation.

Mechanisms underlying this positive selection for in-licensing can become clearer through the lens of our theoretical model. Biotechnology firms choose to license projects when the expected payoff from licensing exceeds that from internal commercialization. For projects with higher quality $\theta$ in predictable categories (incremental projects), the licensing option becomes more appealing because the probability of success increases with project quality, making royalty income more valuable, and because the commercialization advantage of large pharmaceutical firms is more valuable for products with higher success probabilities. As a consequence, biotech firms with higher-quality projects in predictable categories show stronger incentives to license than those with marginal projects, generating positive selection at the equilibrium threshold \citep{gans2003product}.

The positive selection result for in-licensing establishes a critical benchmark: since licensed projects have higher success probabilities, reflecting lower \textit{ex-ante} risk, how do they perform on the monetary net return dimension? Do the higher approval rates translate into superior economic performance for pharmaceutical firms, or are they offset by other factors? Is the effect homogeneous across project degree of innovation? These questions motivate the heterogeneity analysis that follows.

\subsection{Robustness Checks}
\label{subsec:dml_robustness}

\paragraph{Bad Controls.} Following \citet{hunermund2023double}, we verify that our control variables satisfy the backdoor criterion and do not introduce collider or mediator bias driven by the presence of ``bad controls'' — variables that violate the conditional independence assumption required for causal identification. These scholars demonstrate that even a small number of bad controls can produce substantial bias in DML estimates, as these variables tend to be highly correlated with treatment (through unobserved confounders) and are likely to be selected by the algorithm.

Although most of the features in vector $X_i$ represent confounders that determine both licensing propensity and development outcomes — such as disease area determining success rates or technology platform determining development requirements — patent status warrants particular scrutiny. Because patenting decisions may be strategic and time-dependent, patents filed during or after licensing negotiations may act as mediators or colliders rather than as valid confounders. 

In order to assess whether patent status actually constitutes a bad control, we implement two complementary diagnostic tests proposed by \citet{hunermund2023double}. First, we employ sequential control addition, estimating our DML model with and without patent status in the confounding matrix using Random Forest as the only learner. This check reveals substantial sensitivity in the success probability specification: the estimated difference for in-licensing changes from 0.0250 with patent status included to 0.0379 without it, a significant increase of 51.6\%. This significant difference in the success probability specification suggests that patent status may be endogenous to strategic product development choices in ways that violate the conditional independence assumption. Therefore, we conduct a second diagnostic test comparing DML estimates to na\"ive LASSO (post-LASSO) coefficients for this outcome. The key insight from their framework is that when controls are exogenous, DML double selection procedure (which identifies variables correlated with both strategy and outcome) should substantially outperform na\"ive LASSO (which selects only outcome-relevant controls). However, when bad controls are present, both methods become similarly biased, and the advantage of DML disappears because bad controls are correlated with both strategy and outcome through unobserved confounders. Comparing DML to na\"ive LASSO estimates confirms the diagnosis: excluding patent status increases the absolute difference between the two estimators from 0.2874 to 0.3475 — a gain of 0.0601 (or 20.9\%) - indicating that DML double selection performs better once the endogenous variable is removed and can properly identify true confounders.

As a consequence, all success probability results reported in Section~\ref{subsec:dml_y1_results} are obtained from the specification that excludes patent status from the confounding matrix. Importantly, the qualitative conclusion of positive selection into licensing remains unchanged when patent status is included (the estimated difference is positive and statistically significant in both specifications), though the magnitude increases substantially when the bad control is removed. Overall, these robustness checks provide confidence that our main qualitative result regarding positive selection is not driven by bad control bias.

\paragraph{Late-stage Licensing Bias.} A related concern is that our temporal filter — retaining only licenses signed before market approval — may not fully address a late-stage licensing bias. Compounds licensed in late clinical stages (Phase II or Phase III) already carry substantially higher information about their likely success, as they have cleared earlier attrition hurdles. The positive effect of licensing on success probability could therefore be partly driven by a subset of licenses occurring when development risk is already low. First, this concern is mitigated by the composition of our sample: only approximately one in four licensed products is transacted during the clinical phase, confirming that the majority of licensing activity occurs at the pre-clinical or research project stage, in which outcome uncertainty remains high. Nonetheless, to address this concern directly, we strengthen the temporal filter by retaining only licenses with a deal date occurring at the pre-clinical stage or Phase I at most, thereby excluding Phase II and Phase III transactions. Under this restricted specification, the estimated advantage of licensing over in-house development declines to approximately 2 percentage points increase in success probability in absolute terms, corresponding to a relative increase of approximately 5.4\% — thus smaller than the baseline increase of 9.6-10.9\% and consistent with the presence of a modest late-stage licensing effect. Importantly, however, the direction and statistical significance of the result are preserved, confirming that positive selection into licensing is robust to late-stage licensing bias and that the qualitative interpretation of our findings remains valid.

\subsection{Heterogeneity Analysis: Net Returns across Innovation Regimes}
\label{sec:heterogeneity}

Having established positive selection into licensing, we now examine whether the economic returns from licensing versus in-house development vary systematically with project risk. This analysis constitutes the central empirical contribution of the paper as it reveals under which conditions a competitive equilibrium holds in pharmaceutical markets for technology, and where lemons-type frictions persist. Building on \citet{dimasi2007cost}, we note preliminarily that estimated differences do not vary significantly between biotechnology and pharmaceutical companies (interaction terms: $p=0.335$ for in-licensing), motivating focus on project-level risk characteristics rather than firm type. Our primary focus remains on the comparison between in-licensed and internally developed products, but we also examine company acquisition as an alternative governance-mode benchmark because, as argued in Section~\ref{sec:theoretical_model}, its inclusion allows us to assess whether the return patterns we document for licensing are specific to that contractual form or reflect more general features of external sourcing.

\subsubsection{Formal Specification of the Continuous Interaction Model}
\label{subsec:dml_specification_interaction}

To examine how estimated differences vary with predicted success probability, we adapt the partially linear model to include interactions within the DML framework:
\begin{equation*}
Y_i = \theta_0 + \sum_{j} \left[\theta_j D_{ji} + \gamma_j D_{ji} \times \tilde{S}_i\right] + g(X_i) + \varepsilon_i,
\end{equation*}
where $Y_i$ denotes net returns (in millions of U.S. dollars or logs), $D_{ji}$ are strategy indicators for in-licensing and company acquisition relative to the in-house development baseline, and $\tilde{S}_i = S_i - \bar{S}$ represents centered predicted success probability with $\bar{S}$ being the sample mean. The parameters of interest are: (i) $\theta_j$, the main effect of strategy $j$, capturing the estimated difference at mean success probability $\bar{S}$; (ii) $\gamma_j$, the interaction effect, capturing how the estimated difference changes with success probability (and, by extension, with innovation regime); (iii) the total effect at any success probability $S_i$: $\theta_j + \gamma_j(S_i - \bar{S})$. As in previous specifications, the function $g(X_i)$ captures the potentially non-linear relationship between the full set of confounders and outcomes, estimated via machine learning as described in Section~\ref{sec:methodology}. As before, the DML framework ensures $\sqrt{n}$-consistent estimation of $\theta_j$ and $\gamma_j$ despite the high-dimensional nuisance function $g(\cdot)$, with standard errors robust to heteroskedasticity and computed using the asymptotic distribution theory in \citet{chernozhukov2018double}\footnote{Because $\hat{S}_i$ enters the interaction term $D_{ji} \times \tilde{S}_i$ as a generated regressor, analytical standard errors treat $\hat{S}_i$ as fixed and may understate total sampling uncertainty. We validate the analytical inference via a two-stage bootstrap ($B = 500$ replications) that jointly resamples the $Y_1$ classifier training sample and the $Y_2$ estimation sample in each replication. Bootstrap standard errors for $\hat{\theta}_j$ and $\hat{\gamma}_j$ are within 3--6\% of their analytical counterparts (SE ratios of 1.024 and 0.940 respectively), and the bootstrap 95\% confidence intervals for the total marginal effects replicate the analytical significance pattern exactly. Full results are reported in Appendix~\ref{app:dml_interaction_validation}.}.

\subsubsection{Data Construction and Success Probability Prediction}
\label{subsec:dml_data_&_succ_prob}

The analysis employs net returns ($Y_2$, lifetime sales minus clinical trial costs in millions of U.S. dollars) as the outcome variable, incorporating both approved and abandoned products to provide an inclusive evaluation that accounts for failure costs. After applying strategy-specific temporal restrictions and filtering for non-missing cost and sales data, the final sample contains 13,595 observations: 2,722 approved products ($\approx 20\%$) and 10,873 abandoned projects ($\approx 80\%$). For the latter, lifetime sales are coded as zero, generating negative net returns that reflect unrecovered development expenditures.

In order to examine heterogeneity across the risk distribution, we construct predicted success probabilities by training a Random Forest classifier on the full $Y_1$ estimation sample (N=98,712 projects) and applying it out-of-sample to the net return sample, yielding predictions with mean 0.220 and median 0.116. This two-sample approach ensures that $\hat{S}_i$ is not mechanically correlated with the $Y_2$ outcome. Importantly, since this sample comprises only products with realized outcomes (either approved or abandoned), the predicted probabilities should be interpreted as an \textit{ex-post} assessment of project risk. These probabilities are constructed using only the product characteristics known at the outset of development and serve as empirical proxies for information precision, $\lambda(\tau)$, in our conceptual framework \citep{akcigit2018growth}. Higher predicted probabilities indicate lower-risk projects corresponding to incremental projects with high $\lambda(I)$, while lower predicted probabilities indicate higher-risk projects corresponding to novel projects with low $\lambda(N)$. As a validation, approved products show substantially higher predicted probabilities (mean: 0.437, median: 0.426) than abandoned projects (mean: 0.166, median: 0.078), confirming that predictions correlate strongly with realized outcomes. Across development strategies, in-house projects exhibit the highest predicted probability of success (N: 7,763; mean: 0.232; median: 0.111), followed by licensed projects (N: 3,553; mean: 0.215; median: 0.120), while projects originating from company acquisitions display the lowest predicted probabilities (N: 2,279; mean: 0.190; median: 0.116).

\subsubsection{The Risk-Return Trade-off across Innovation Regimes}
\label{subsec:dml_main_results}

The main effect estimates capture differences for products with average predicted success probability. Company acquisition shows a point estimate of $-\$30.70$M (SE: \$183.54M, $p=0.867$), indicating no significant difference at the mean. In-licensing exhibits a large and significant negative main effect of $-\$746.29$M (SE: \$152.82M, $p<0.001$), indicating that in-licensed projects with average success probability generate substantially lower net returns than comparable in-house developed products. Crucially, however, the interaction terms reveal that this aggregate pattern masks important heterogeneity. On one hand, for company acquisition the interaction coefficient is $-\$839.11$M (SE: \$829.06M, $p=0.312$), again not statistically significant as was the main effect. On the other hand, for in-licensing the interaction is $-\$2{,}443.39$M (SE: \$659.03M, $p<0.001$), highly significant and negative, indicating that the return disadvantage of licensing intensifies monotonically as success probability increases — equivalently, as project risk decreases.

\begin{table}[tp]
\centering
\caption{\textbf{Table 2} - Heterogeneous Effect on Net Returns ($Y_2$) by Innovation-Risk Percentile: Company Acquisition}
\label{tab:company_acquisition_effects}
\small
\renewcommand{\arraystretch}{1.20}
\begin{tabular}{lrrrrc}
\toprule\toprule
 & & & \multicolumn{3}{c}{\textit{Total Effect: $\theta_j + \gamma_j \tilde{S}_i$}} \\
\cmidrule(lr){4-6}
\textbf{Percentile of $\tilde{S}_i$} & \textbf{Main Effect ($\theta_j$)} & \textbf{Interaction ($\gamma_j \tilde{S}_i$)} & \textbf{Estimate} & \textbf{SE} & \textbf{$p$-value} \\
\midrule
10th (most novel)  & $-30.70$ & $+162.41$ & $+131.71$  & 243.79 & 0.589 \\
25th  & $-30.70$ & $+153.91$ & $+123.21$  & 238.35 & 0.605 \\
Median (50th)    & $-30.70$ & $+87.25$  & $+56.54$   & 202.77 & 0.780 \\
75th & $-30.70$ & $-83.84$  & $-114.55$  & 201.37 & 0.570 \\
90th (most incremental)  & $-30.70$ & $-360.86$ & $-391.57$  & 401.01 & 0.329 \\
\bottomrule\bottomrule
\end{tabular}
\vspace{0.1cm}
\end{table}

\begin{table}[tp]
\centering
\caption{\textbf{Table 3} - Heterogeneous Effect on Net Returns ($Y_2$) by Innovation-Risk Percentile: In-Licensing}
\label{tab:inlicensed_effects}
\small
\renewcommand{\arraystretch}{1.20}
\begin{tabular}{lrrrrc}
\toprule\toprule
 & & & \multicolumn{3}{c}{\textit{Total Effect: $\theta_j + \gamma_j \tilde{S}_i$}} \\
\cmidrule(lr){4-6}
\textbf{Percentile of $\tilde{S}_i$} & \textbf{Main Effect ($\theta_j$)} & \textbf{Interaction ($\gamma_j \tilde{S}_i$)} & \textbf{Estimate} & \textbf{SE} & \textbf{$p$-value} \\
\midrule
10th (most novel) & $-746.29$ & $+472.91$     & $-273.38$             & 199.06 & 0.170 \\
25th  & $-746.29$ & $+448.17$     & $-298.13$             & 194.85 & 0.126 \\
Median (50th)    & $-746.29$ & $+254.05$     & $-492.24^{***}$       & 167.48 & 0.003 \\
75th  & $-746.29$ & $-244.14$     & $-990.43^{***}$       & 166.41 & $<$0.001 \\
90th (most incremental) & $-746.29$ & $-1{,}050.79$ & $-1{,}797.08^{***}$   & 321.99 & $<$0.001 \\
\bottomrule\bottomrule
\end{tabular}
\begin{minipage}{\textwidth}
\vspace{4pt}
\centering
\footnotesize\textbf{Notes:} $^{*}p<0.10$, $^{**}p<0.05$, $^{***}p<0.01$.
\end{minipage}
\end{table}

\begin{figure}[tp]
    \centering
    \includegraphics[width=1\linewidth]{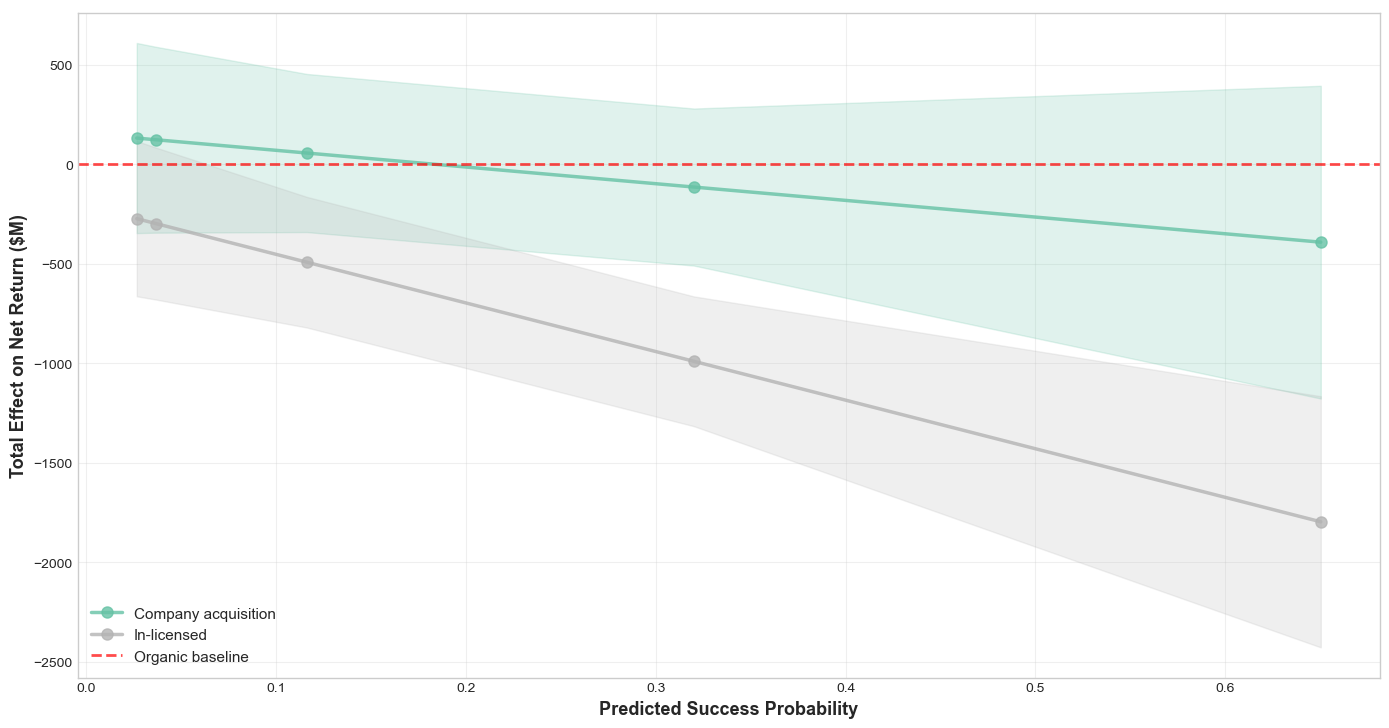}
    \captionsetup{justification=centering}
    \caption{\textbf{Figure 5} - Marginal Effect on Net Returns ($Y_2$): How Strategy Impact Varies with Innovation Regimes \\
    (Shaded areas = 95\% Confidence Intervals)}
    \label{fig:dml_marginal_effects_shaded}
\end{figure}

\begin{figure}[tp]
    \centering
    \includegraphics[width=1\linewidth]{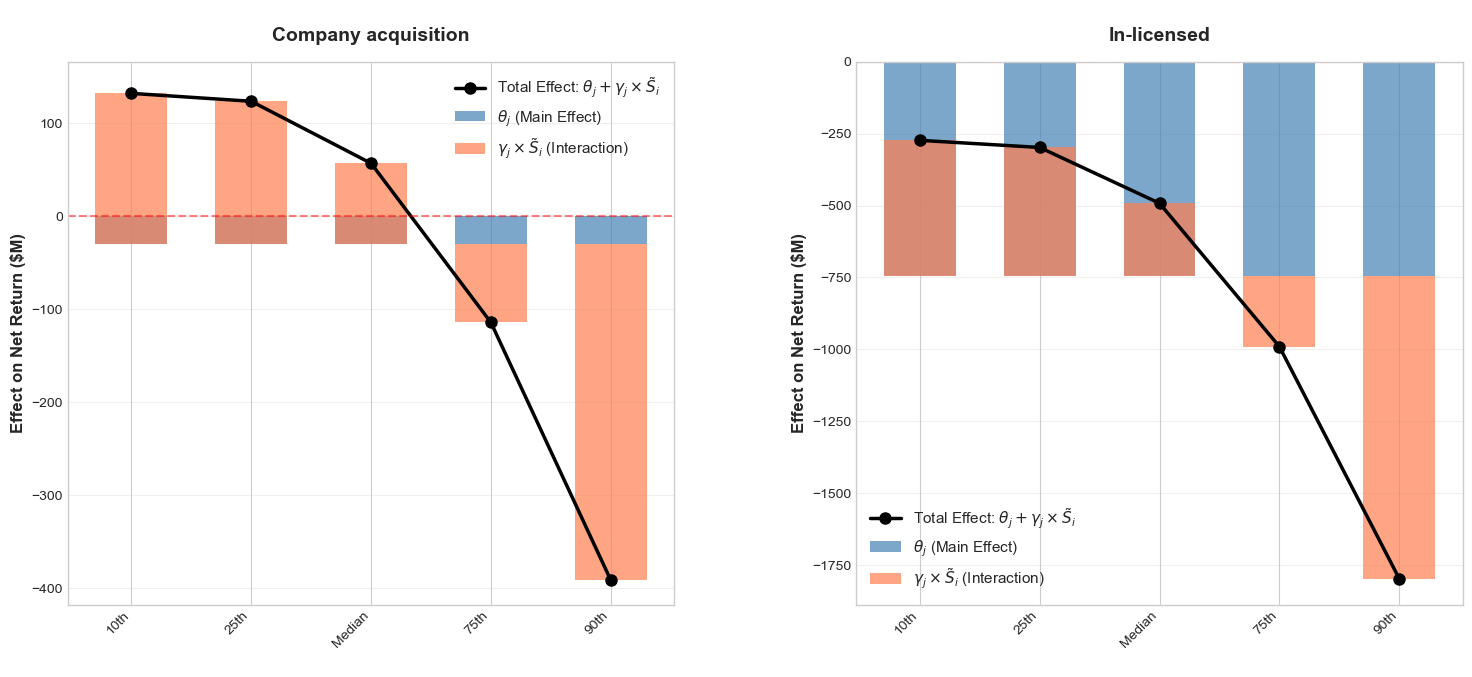}
    \caption{\textbf{Figure 6} - Decomposition of Total Effect on Net Returns ($Y_2$): $\theta_j + \gamma_j \tilde{S}_i \;\big|\; \tilde{S}_i = Q_{\tau}(\tilde{S})$}
    \label{fig:dml_marginal_effects_decomposition}
\end{figure}

Computing marginal effects at different percentiles reveals the full pattern (Tab.~\ref{tab:company_acquisition_effects} and Tab.~\ref{tab:inlicensed_effects}). For company acquisition, effects range from $+\$131.71$M at the 10th percentile to $-\$391.57$M at the 90th percentile, but none achieve statistical significance, confirming the acquisition neutrality with respect to organic development across the entire risk distribution. By contrast, for in-licensing the pattern reveals a monotonic relationship between project risk and estimated differences which turns significant only for low-risk projects (i.e. incremental innovations): at the 10th percentile of the success probability distribution (0.027), the estimated difference is $-\$273.38$M ($p=0.170$), indistinguishable from zero; at the 25th percentile (0.037), the effect remains insignificant at $-\$298.13$M ($p=0.126$); at the median (0.116), the effect becomes both economically substantial and statistically significant ($-\$492.24$M, $p=0.003$); at the 75th percentile (0.320), it intensifies to $-\$990.43$M ($p<0.001$); and at the 90th percentile (0.650), it reaches $-\$1{,}797.08$M ($p<0.001$). These results are visualized in Fig.~\ref{fig:dml_marginal_effects_shaded} and Fig.~\ref{fig:dml_marginal_effects_decomposition}.

\begin{figure}[tp]
    \centering
    \includegraphics[width=1\linewidth]{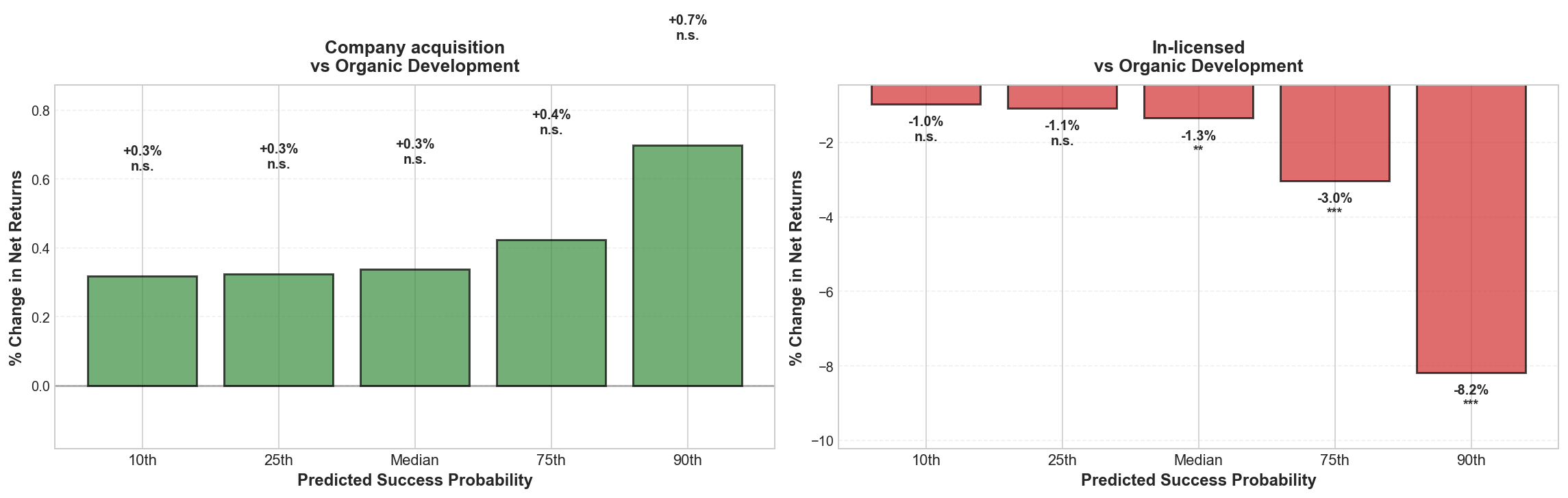}
    \caption{\textbf{Figure 7} - Strategy Effects on Net Returns ($Y_2$, Log-Transformed): \% Change across Innovation Regimes}
    \label{fig:dml_marginal_effects_percentage}
\end{figure}

To provide an alternative specification, we re-estimate using log-transformed net returns (applying a shift transformation to handle negative values), but also in this case the log model yields substantively identical conclusions with more interpretable magnitudes. The main effect for company acquisition is +0.0040 (SE: 0.0066, $p=0.5388$), whereas that of in-licensing is -0.0272 (SE: 0.0058, $p<0.001$), negative and significant. Once again, the interaction for company acquisition is $+0.0048$ (SE: 0.0273, $p=0.861$), not significant, whereas that for in-licensing is $-0.0958$ (SE: 0.0191, $p<0.001$). Marginal effects by percentile (Tab.~\ref{tab:marginal_effects_log} and Fig.~\ref{fig:dml_marginal_effects_percentage}) confirm the pattern: for company acquisition, effects range from $+0.32\%$ to $+0.70\%$ across the risk distribution, none significant; in contrast, for in-licensing, effects range from statistical neutrality at lower percentiles (higher-risk projects) to significant negative estimated differences for projects at or above the median probability level, with the highest negative effect of -8.17\% at the 90th percentile (lowest-risk projects). Overall, the pattern is unambiguous: licensing has negligible effects on returns for novel projects but substantial negative effects on returns for incremental projects, with the effect intensifying monotonically as predicted success probability increases (equivalently, as project risk decreases). 

\subsubsection{Competitive Equilibrium and Lemons-Type Frictions}
\label{subsec:dml_main_results_interpretation}

The heterogeneity results reveal a coherent picture that reconciles the aggregate return disadvantage of licensing with the competitive equilibrium predicted by Proposition 2. Three findings stand out.

First, for \emph{low-risk incremental projects} (those with above-median predicted success probability), the trade-off inherent in licensing becomes fully apparent. In-licensing new projects under development provides higher success rates ($+3.8$--$4.3$ percentage points), reflecting positive selection as predicted by Proposition 1. However, this advantage in the development phase is substantially offset in the post-approval phase through lower net returns — from $-1.3\%$ to $-8.7\%$ at the 50th and 90th percentiles respectively — consistent with risk-adjusted return equalization predicted by Proposition 2 for incremental projects. This pattern reveals a fundamental trade-off that pharmaceutical firms face when choosing development strategies for projects with less intrinsic uncertainty. When selecting licensing over in-house development, firms can obtain projects with higher predicted probability of approval, but these projects subsequently generate lower net returns conditional on approval. In other words, the higher success probability provides insurance against development failure, but this insurance comes at the cost of reduced profitability when the product eventually reaches the market. In the Appendix~\ref{app:results} we demonstrate that the cost premium associated with licensing and the lower commercial revenues of licensed products compound to generate the net return disadvantage, consistent with Proposition 3. The critical implication — and the strongest evidence for competitive equilibrium in the lower tail of risk distribution (equivalently, in the market segment of incremental innovations) — is that \emph{no strategy hierarchically dominates}. The risk-adjusted trade-off structure is in fact the equilibrium outcome that economic theory predicts in competitive markets, reflecting enhanced competition due to the existence of markets for technologies \citep{arora2010ideas,akcigit2016buy}. If licensing systematically generated both higher success rates \textit{and} higher net returns, competitive forces would drive all high-quality projects toward licensing, bidding up costs until the return advantage disappeared. Conversely, if in-house development dominated on both dimensions, firms would avoid licensing, driving down licensing costs until the equilibrium is restored. In sum, the observed coexistence of both strategies, with licensing advantage in success probability offset by its profitability disadvantage for low-risk projects, indicates that the market reaches a competitive equilibrium in which pharmaceutical firms are indifferent at the margin between the two development modes \citep{arora2001, robinson2007financial}.

Second, for \emph{high-risk novel projects} (bottom quartile of the success probability distribution), the pattern differs: in-licensing still provides higher success rates but has no significant negative impact on net returns — effects are statistically indistinguishable from zero at the 10th and 25th percentiles. This asymmetry suggests that market mechanisms ensuring return equalization across development strategies operate less forcefully for projects with less peers in the market that carry fewer information about potential future trajectories (low $\lambda(N)$) compared to incremental projects with high information precision (high $\lambda(I)$). Same to say, these riskier projects benefit from pharmaceutical firms' superior development capabilities without incurring the severe commercialization disadvantages that afflict licensed incremental products, as they can generate exceptional commercial value through premium pricing and first-mover advantages in novel therapeutic categories. This finding aligns with \citet{arora2022science}, who show that novel inventions face higher transaction costs in technology markets but may offer greater gains from trade due to increased buyer heterogeneity. Overall, the favorable licensing outcome for high-risk novel projects — higher success probability without profitability costs — is consistent with elements of ``lemons-type'' frictions that appear to persist in this market segment.

\begin{table}[tp]
\centering
\caption{\textbf{Table 4} - Marginal Effect on Net Returns ($Y_2$, Log-Transformed) by Success Probability Percentile}
\label{tab:marginal_effects_log}
\small
\renewcommand{\arraystretch}{1.20}
\resizebox{\textwidth}{!}{%
\begin{tabular}{lrccc}
\toprule\toprule
 & & \multicolumn{2}{c}{\textit{Total Effect ($\theta_j + \gamma_j \tilde{S}_i$)}} \\
\cmidrule(lr){3-4}
\textbf{Percentile of $\tilde{S}_i$} & \textbf{Success Probability} & \textbf{Company Acquisition} & \textbf{In-Licensing} \\
\midrule
10th (most novel)  & 0.018 & $+0.32\%$ $(0.668)$ & $-0.96\%$ $(0.174)$            \\
25th & 0.030 & $+0.32\%$ $(0.656)$ & $-1.07\%$ $(0.121)$            \\
Median (50th) & 0.058 & $+0.34\%$ $(0.627)$ & $-1.34\%^{**}$ $(0.042)$       \\
75th & 0.237 & $+0.42\%$ $(0.538)$ & $-3.02\%^{***}$ $(<0.001)$     \\
90th (most incremental) & 0.807 & $+0.70\%$ $(0.715)$ & $-8.17\%^{***}$ $(<0.001)$     \\
\bottomrule\bottomrule
\end{tabular}}
\begin{minipage}{\textwidth}
\vspace{4pt}
\centering
\footnotesize\textbf{Notes:} p-values in parentheses: $^{*}p<0.10$, $^{**}p<0.05$, $^{***}p<0.01$.
\end{minipage}
\end{table}

Third, \emph{company acquisition} exhibits a uniformly neutral pattern in returns relative to in-house development across both innovation regimes, with estimated differences consistently small and statistically insignificant. This confirms the acquisition neutrality predicted alongside Proposition 2: full integration eliminates the inter-organizational frictions that drive licensing coordination costs, but also eliminates the risk-sharing and screening mechanisms that licensing contracts provide. This finding indicates that governance mode matters: the risk-return trade-off we document for licensing is specific to that contractual form rather than representing a general feature of any market for technologies.

In sum, we find that mechanisms ensuring risk-adjusted equilibrium in licensing markets operate effectively where development pathways are well-established and outcomes more predictable, but the key contribution of our risk-stratified analysis is revealing that these equilibrium patterns are not stable across innovation regimes and vary systematically with project novelty. The absence of a return penalty for licensed high-risk novel projects, despite their higher success rates, suggests that different market dynamics operate for frontier innovation compared to incremental projects, and that competitive equilibrium may not fully characterize this market segment in the same way it does for incremental projects. Originators of frontier innovations may accept licensing terms that undervalue their assets, potentially leading to an over-representation of high-risk novel projects in licensing transactions relative to what competitive pricing would imply, and this asymmetry is consistent with a ``lemons-type equilibrium'' in which the market fails to fully capture the commercial potential of frontier innovation through licensing arrangements.

\section{DML-IV Results}
\label{sec:dmliv_results}

This section presents the causal estimates from the DML-IV analysis. As discussed in Section~\ref{sec:methodology}, the DML framework delivers well-controlled descriptive estimates of equilibrium conditions in markets for technology conditional on observable confounders, but cannot rule out that unobserved quality differences drive the estimated patterns. The DML-IV addresses this concern by isolating exogenous variation in licensing decisions — those driven by acute pipeline pressure following a Phase~III failure — and estimating a LATE for the complier subpopulation of rushed licenses. The two analyses are complementary as on one hand the DML establishes the equilibrium pattern across the full distribution of licenses, while on the other hand the DML-IV provides causal identification for a specific subset of licenses. The bridge between them therefore lies in the mechanism: if the same innovation-regime heterogeneity documented in Section~\ref{sec:dml_results} survives instrumentation then the DML pattern cannot be attributed to selection on unobservables, and the $\lambda(\tau)$ mechanism identified in the theoretical framework receives direct causal support at least for the specific subset of licenses for which we can clearly identify an exogenous source of variation.

\subsection{First Stage}
\label{subsec:dmliv_firststage}

The instrument passes the relevance condition with large margins. In the primary specification, we restrict both in-licensed and organic products to firms with at least one Phase III–exposed project, i.e. firms that during their existence had at least one project in which they were the developer that reached Phase III, ensuring that they were truly exposed to the risk of Phase III failure. The first-stage coefficient ($\hat{\pi} = 0.650$, SE: 0.012, $p < 0.001$) confirms strong instrument relevance: after removing the variation in licensing explained by confounders and the continuous firm-level risk score, products associated with a Phase III pipeline failure by the licensee in the 365 days preceding the deal date are 65 percentage points more likely to be in-licensed relative to the comparison group. Notably, since the instrument is defined at the product level as the joint occurrence of a licensing deal and a preceding P3F, identification rests on the exogeneity of P3F timing conditional on controls. While \citet{hermosilla2021rushed} documents the same qualitative mechanism at the firm level - Phase III failures generate pipeline pressure that significantly increases subsequent licensing activity - his analysis operates at a different unit of observation, measuring the effect of a failure on the count of new licensing transactions per firm, while ours captures the effect on the probability that a specific product is licensed rather than organically developed. The strong first stage is nonetheless consistent with his findings and confirms that the rushed-licensing mechanism identified in his framework operates at the product level in our more heterogeneous sample, which spans thousands of firms of varying sizes compared to the 20 largest global pharmaceutical companies in his study.

\subsection{Selection Patterns: Success Probability under Pipeline Pressure}
\label{subsec:dmliv_y1}

Before examining net returns, we assess whether the positive selection result documented in Section~\ref{subsec:dml_y1_results} can be extended to the complier subpopulation of rushed licenses with a causal interpretation. The estimation sample contains 34,722 completed projects - 6,717 in-licensed (with 802 rushed license), 28,005 in-house developed projects - and a mean approval rate of 0.257. The Random Forest outcome classifier achieves cross-validated AUC = 0.899, confirming strong nuisance performance in Table~\ref{tab:dmliv_y1}.

\begin{table}[tp]
\centering
\caption{\textbf{Table 5} - DML-IV Estimate: Success Probability ($Y_1$)}
\label{tab:dmliv_y1}
\small
\renewcommand{\arraystretch}{1.20}
\resizebox{\textwidth}{!}{%
\begin{tabular}{lcccc}
\toprule\toprule
\multicolumn{5}{c}{\textit{Panel A: LATE Estimate}} \\
\cmidrule(lr){1-5}
\textbf{Outcome} & $\hat{\theta}^{IV}_{\text{lic}}$ & \textbf{SE} & \textbf{$p$-value} & \textbf{95\% CI} \\
\midrule
Success Probability ($Y_1$) & $+0.025$ & 0.018 & 0.167 & $[-0.011,\; +0.061]$ \\
\addlinespace[6pt]
\toprule\toprule
\multicolumn{5}{c}{\textit{Panel B: First Stage}} \\
\cmidrule(lr){1-5}
 & $\hat{\pi}$ & \textbf{SE} & \textbf{$p$-value} & \textbf{95\% CI} \\
\midrule
Pipeline failure $\to$ In-licensing & $0.650^{***}$ & 0.012 & $<0.001$ & $[0.626,\; 0.674]$ \\
\addlinespace[6pt]
\toprule\toprule
\multicolumn{5}{c}{\textit{Panel C: Nuisance Performance}} \\
\cmidrule(lr){1-5}
 & \textbf{AUC($Y$)} & \textbf{AUC($D$)} & & \\
\midrule
Random Forest (5-fold cross-fit) & 0.899 & 0.842 & & \\
\bottomrule\bottomrule
\end{tabular}}
\begin{minipage}{\textwidth}
\vspace{4pt}
\centering
\footnotesize\textbf{Notes:} First-stage $F = 2{,}967.2$, testing the null of instrument irrelevance ($\hat{\pi} = 0$). AUC($Y$) and AUC($D$) denote the cross-validated AUC of the Random Forest nuisance models for the outcome and treatment respectively. $^{*}p<0.10$, $^{**}p<0.05$, $^{***}p<0.01$.
\end{minipage}
\end{table}

The LATE of $+0.025$ ($p = 0.167$) is not statistically significant and stands in contrast to the DML result of Section~\ref{subsec:dml_y1_results}, where the full population of in-licensed products exhibits a highly significant positive selection advantage of 3.8--4.3 percentage points. However, the disappearance of this advantage for licenses is both expected and informative when referred to only rushed ones. As \citet{hermosilla2021rushed} concludes, rushed licenses represent a subset of transactions in which the standard screening and due diligence process is compressed by exogenous pipeline pressure, reducing the ability of licensees to select high-quality projects. The positive selection result that characterises the full licensing population therefore does not extend to the complier subpopulation: rushed licenses are not positively selected in terms of approval likelihood. This finding establishes an important baseline for interpreting the net return results that follow.

\subsection{Heterogeneity Analysis: Net Returns across Innovation Regimes for Rushed Licenses}
\label{subsec:dmliv_y2}

\begin{figure}[tp]
    \centering
    \includegraphics[width=1\linewidth]{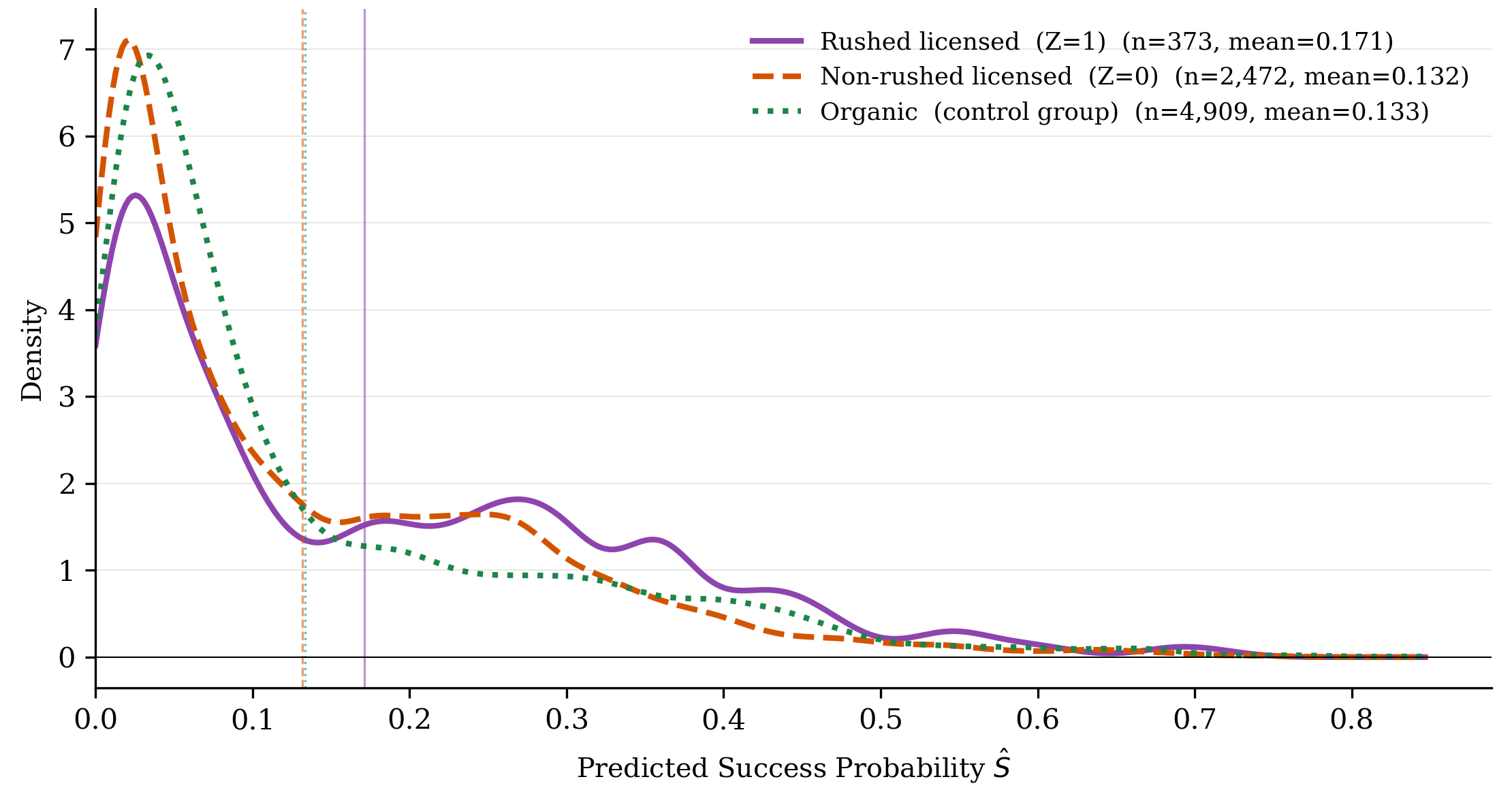}
    \caption{\textbf{Figure 8} - Distribution of Innovation-Risk Proxy $\hat{S}$ by Group \\ (Vertical Lines = Group Means)}
    \label{fig:dmliv_succ_prob_distribution}
\end{figure}

To examine how the causal effect of rushed licensing on net returns varies across alternative innovation regimes, we estimate the interacted DML-IV model introduced in Section~\ref{subsec:dmliv_specification}, extended to allow heterogeneous treatment effects:
\begin{equation*}
Y_i = \theta_0 + \theta^{IV}_{\text{lic}}\, D_{\text{lic},i} + \gamma^{IV}_{\text{lic}}\,(D_{\text{lic},i} \times \tilde{S}_i) + g(X_i) + \varepsilon_i,
\end{equation*}
where $Y_i$ denotes the net return outcome, $\tilde{S}_i = \hat{S}_i - \bar{S}$ is the demeaned predicted success probability, constructed by applying out-of-sample the Random Forest classifier trained on the $Y_1$ estimation sample to the net return sample, yielding a mean predicted probability of $\bar{S} = 0.135$. Figure~\ref{fig:dmliv_succ_prob_distribution} plots the distribution of $\hat{S}_i$ across the three groups. The three distributions largely overlap, confirming broad comparability in innovation-risk composition across groups, though rushed licenses display a slightly increased mass at higher $\hat{S}_i$ relative to organic in-house products and non-rushed licenses, consistent with firms under pipeline pressure selectively seeking lower-risk projects to fill their development gap as quickly as possible. The two endogenous regressors $\{D_{\text{lic},i},\, D_{\text{lic},i} \times \tilde{S}_i\}$ are instrumented by the expanded instrument set $\{Z_i,\, Z_i \times \tilde{S}_i\}$, yielding a just-identified two-stage system with exactly as many instruments as endogenous regressors. The parameter $\theta^{IV}_{\text{lic}}$ captures the LATE at mean innovation type, $\gamma^{IV}_{\text{lic}}$ captures how the LATE varies across the innovation spectrum, and the LATE at any percentile $p$ is $\theta^{IV}_{\text{lic}} + \gamma^{IV}_{\text{lic}}\, \tilde{S}(p)$. The joint first-stage $F$-statistic of 340.0 confirms instrument strength well above conventional thresholds.

\begin{figure}[!t]
    \centering
    \includegraphics[width=.7\linewidth]{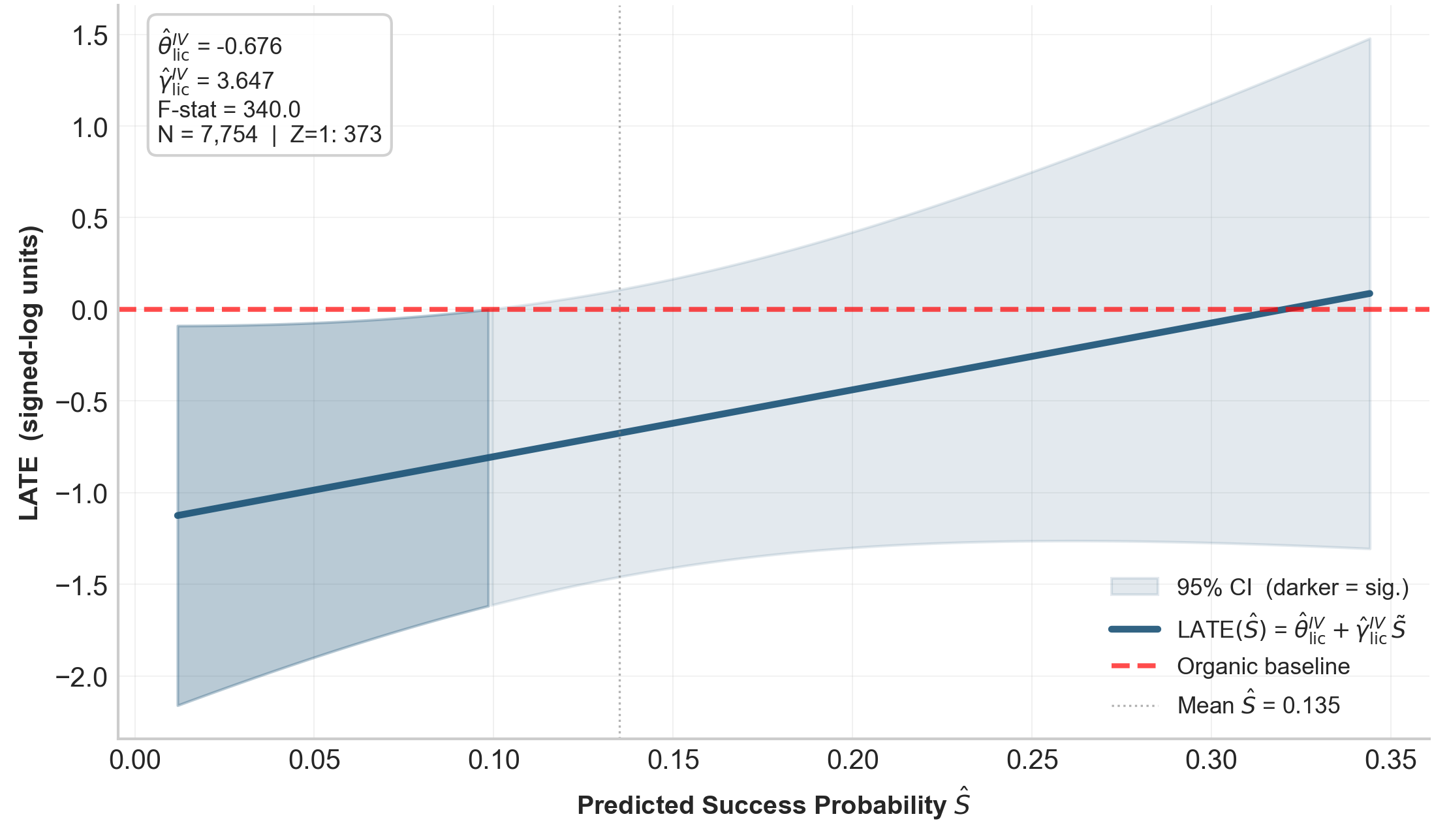}
    \caption{\textbf{Figure 9} - Marginal Effect of Rushed Licensing on Net Return by Innovation Regimes}
    \label{fig:dmliv_marginal_effect_shaded}
\end{figure}

Table~\ref{tab:dmliv_interacted_log} reports the main coefficients and LATEs for each percentile for the signed-log net return outcome, defined as $f(x) = \text{sign}(x)\cdot\log(1+|x|)$ where $x$ denotes raw net return in millions of U.S. dollars. The table is complemented by Figure~\ref{fig:dmliv_marginal_effect_shaded}. This transformation is applied because raw net return spans both large positive values (blockbuster approved products) and large negative values (expensive abandoned projects), generating substantial variance that inflates standard errors in the second stage; the signed-log compresses both tails symmetrically but preserves the sign and the ordering of observations, thus reducing the outcome standard deviation.

For the most novel projects (10th percentile, $\hat{S} = 0.012$), rushed licensing reduces net return by 0.355 standard deviations relative to organic development ($p = 0.012$). The effect remains large and significant at the 25th percentile ($-0.336$ SD, $p = 0.012$) and at the median ($-0.282$ SD, $p = 0.021$). Moving toward more incremental projects, the effect fades monotonically: at the 75th percentile the LATE shrinks to $-0.128$ SD and loses significance ($p = 0.418$), and at the 90th percentile it is essentially zero ($+0.028$ SD, $p = 0.913$). In order to provide a concrete dollar-scale complement, Table~\ref{tab:dmliv_interacted_trim} reports the corresponding LATEs from a trimmed specification, which restricts the net return outcome to the 5th-95th percentile of its distribution, dropping 638 observations with extreme values and retaining 5,732 products.\footnote{Trimming bounds: [$-\$372.8$M, $+\$1{,}119.0$M]. Trimming is applied within-sample at the 5th and 95th percentiles of the net return distribution prior to estimation.} The trimmed results confirm and quantify the signed-log findings on the original USD million scale. For the most novel projects (10th percentile), rushed licensing reduces trimmed net return by \$43.9M relative to organic development ($p = 0.042$), and this effect remains significant at the 25th percentile ($-\$42.3$M, $p = 0.042$) and marginally significant at the median ($-\$37.9$M, $p = 0.059$). The effect fades monotonically and loses significance at the 75th percentile ($-\$25.2$M, $p = 0.345$) and the 90th percentile ($-\$13.3$M, $p = 0.737$), neither of which is statistically distinguishable from zero. The monotone pattern is consistent across both outcome transformations.

\begin{table}[!t]
\centering
\caption{\textbf{Table 6} - Heterogeneous LATE on Net Returns ($Y_2$) by Innovation-Risk Percentile: Rushed Licensing}
\label{tab:dmliv_interacted_log}
\small
\renewcommand{\arraystretch}{1.20}
\setlength{\tabcolsep}{2pt}
\begin{tabular}{lrrrrc}
\toprule\toprule
 & & \multicolumn{4}{c}{\textit{Total Effect: $\theta^{IV}_{\text{lic}} + \gamma^{IV}_{\text{lic}}\, \tilde{S}_i$}} \\
\cmidrule(lr){3-6}
\textbf{Percentile of $\tilde{S}_i$} & \textbf{Main Effect ($\theta^{IV}_{\text{lic}}$)} & \textbf{LATE (signed-log)} & \textbf{Effect (SD units)} & \textbf{SE} & \textbf{$p$-value} \\
\midrule
10th (most novel)       & $-0.676$ & $-1.124^{**}$ & $-0.355$ & 0.447 & 0.012 \\
25th                    & $-0.676$ & $-1.065^{**}$ & $-0.336$ & 0.425 & 0.012 \\
Median (50th)           & $-0.676$ & $-0.893^{**}$ & $-0.282$ & 0.387 & 0.021 \\
75th                    & $-0.676$ & $-0.405$      & $-0.128$ & 0.501 & 0.418 \\
90th (most incremental) & $-0.676$ & $+0.088$      & $+0.028$ & 0.800 & 0.913 \\
\bottomrule\bottomrule
\end{tabular}
\begin{minipage}{\textwidth}
\vspace{4pt}
\centering
\footnotesize\textbf{Notes:} SD units = LATE divided by the outcome standard deviation of 3.167 signed-log units. Joint $F = 340.0$. $^{*}p<0.10$, $^{**}p<0.05$, $^{***}p<0.01$.
\end{minipage}
\end{table}

\begin{table}[!t]
\centering
\caption{\textbf{Table 7} - Heterogeneous LATE on Net Returns ($Y_2$) by Innovation-Risk Percentile: Rushed Licensing (in \$M)}
\label{tab:dmliv_interacted_trim}
\small
\renewcommand{\arraystretch}{1.20}
\setlength{\tabcolsep}{8pt}
\begin{tabular}{lrrrc}
\toprule\toprule
 & & \multicolumn{3}{c}{\textit{Total Effect: $\theta^{IV}_{\text{lic}} + \gamma^{IV}_{\text{lic}}\, \tilde{S}_i$}} \\
\cmidrule(lr){3-5}
\textbf{Percentile of $\tilde{S}_i$} & \textbf{Main Effect ($\theta^{IV}_{\text{lic}}$)} & \textbf{LATE (\$M)} & \textbf{SE} & \textbf{$p$-value} \\
\midrule
10th (most novel)       & $-31.4$ & $-43.9^{**}$ & $21.6$ & $0.042$ \\
25th                    & $-31.4$ & $-42.3^{**}$ & $20.8$ & $0.042$ \\
Median (50th)           & $-31.4$ & $-37.9^{*}$  & $20.0$ & $0.059$ \\
75th                    & $-31.4$ & $-25.2$      & $26.7$ & $0.345$ \\
90th (most incremental) & $-31.4$ & $-13.3$      & $39.6$ & $0.737$ \\
\bottomrule\bottomrule
\end{tabular}
\begin{minipage}{\textwidth}
\vspace{4pt}
\centering
\footnotesize\textbf{Notes:} Coefficients interpretable in millions of U.S. dollars. Net return trimmed within-sample at the 5th and 95th percentiles (bounds: $[-\$372.8\text{M},\; +\$1{,}119.0\text{M}]$; 638 observations dropped). Joint $F = 213.3$. $^{*}p<0.10$, $^{**}p<0.05$, $^{***}p<0.01$.
\end{minipage}
\end{table}
\medskip

\subsection{Connecting DML and DML-IV: Causal Validation of the Equilibrium Pattern across Innovation Regimes}
\label{subsec:dmliv_bridge}

Table \ref{tab:summary} summarizes the sign and significance of the estimated effects on success probability and net returns across analyses and innovation regimes.

\begin{table}[tp]
\centering
\caption{\textbf{Table 8} - Summary of Estimated Effects on Success Probability and Net Returns}
\label{tab:summary}
\small
\renewcommand{\arraystretch}{1.20}
\begin{tabular}{lcccc}
\toprule\toprule
& \multicolumn{2}{c}{\textit{DML (full population)}} & \multicolumn{2}{c}{\textit{DML-IV (rushed licenses)}} \\
\cmidrule(lr){2-3} \cmidrule(lr){4-5}
& \textbf{Success Prob.} & \textbf{Net Return} & \textbf{Success Prob.} & \textbf{Net Return} \\
\midrule
Incremental & $\uparrow^{***}$ & $\downarrow^{***}$ & $=$ & $=$ \\
Novel       & $\uparrow^{***}$ & $=$ & $=$ & $\downarrow^{**}$ \\
\midrule
Equilibrium (incremental) & \multicolumn{2}{c}{Neither strategy dominates} & \multicolumn{2}{c}{Neither strategy dominates} \\
Equilibrium (novel)       & \multicolumn{2}{c}{Licensing weakly dominates} & \multicolumn{2}{c}{In-house weakly dominates} \\
\bottomrule\bottomrule
\end{tabular}
\begin{minipage}{\textwidth}
\vspace{4pt}
\centering
\footnotesize\textbf{Notes:} $\uparrow$ ($\downarrow$) denotes a positive (negative) and statistically significant effect of in-licensing relative to organic development; $=$ denotes an effect indistinguishable from zero. For DML, success probability effects are from Section~\ref{subsec:dml_y1_results} and net return effects from Section~\ref{subsec:dml_main_results} (10th percentile for novel, 90th percentile for incremental). For DML-IV, effects are from Sections~\ref{subsec:dmliv_y1} and~\ref{subsec:dmliv_y2}. $^{**}p<0.05$, $^{***}p<0.01$.
\end{minipage}
\end{table}

The DML-IV results provide causal validation of the equilibrium pattern documented in Section~\ref{sec:dml_results}, while simultaneously hinting at the mechanisms through which innovation regime moderates the consequences of licensing. To properly interpret the net return results, they must be read jointly with the success probability findings of Section~\ref{subsec:dmliv_y1}: the two outcomes together determine whether a competitive equilibrium holds for each innovation regime. Three points of contact between the DML and DML-IV analyses deserve emphasis.

First, the existence of a competitive equilibrium is confirmed for incremental projects, but through a different channel. In the DML analysis, incremental licensed projects exhibit higher success probabilities than in-house developed ones (+3.8-4.3 percentage points) but lower net returns, so that neither strategy dominates the other on both dimensions simultaneously — a pattern consistent with competitive equilibrium. The DML-IV results reproduce this non-dominance structure for the complier subpopulation of rushed licenses, but through a different combination: rushed licenses for incremental projects carry no significant success probability advantage (LATE $= +0.025$, $p = 0.167$) and no significant return penalty (LATE $= +0.028$ SD at the 90th percentile, $p = 0.913$). Competitive equilibrium therefore holds in both analyses — in the DML because the success probability advantage of licensing is offset by a return disadvantage, and in the DML-IV because neither advantage nor disadvantage is present on either dimension. In both cases, no strategy hierarchically dominates the other, which is precisely the equilibrium condition that competitive markets predict \citep{arora2001, robinson2007financial}.

Second, lemons-type frictions persist for novel projects even after an exogenous source of variation in licensing propensity (P3F) is considered in the model. In the DML analysis, the competitive equilibrium that holds for incremental projects rests on an offsetting mechanism: the success probability advantage of licensing is counterbalanced by a return penalty, so that neither strategy dominates on both dimensions simultaneously. For novel projects, this offsetting mechanism breaks down - higher success probabilities are not accompanied by any return penalty - and competitive equilibrium fails to hold. The DML-IV results confirm this failure causally. For the complier subpopulation of rushed novel licenses, the success probability advantage disappears entirely (LATE $= +0.025$, $p = 0.167$) while a significant return penalty emerges ($-0.355$ SD at the 10th percentile, $p = 0.012$). The offsetting mechanism is again absent, though the asymmetry of advantages and disadvantages is now inverted - the two dimensions of product development move together rather than against each other, preventing the market from reaching the balanced risk-return trade-off that characterizes competitive equilibrium. This causal confirmation is important, as it demonstrates that the failure of competitive equilibrium for novel projects is a genuine consequence of the difficulty in evaluating frontier innovations, which is further exacerbated by the time constraint intrinsic in the rushed licensing process following a Phase III failure.

Third, the DML-IV identifies the channel through which information precision shapes the consequences of licensing. In our model, $\lambda(\tau)$ denotes the signal precision for innovation type $\tau$: high for incremental projects ($\lambda(I)$), low for novel ones ($\lambda(N)$). Rushed licensing does not alter $\lambda(\tau)$ --- it does not change the intrinsic information structure of the innovation. What it does is compress the time available to process whatever signal is available. For incremental projects with high $\lambda(I)$, the signal is precise enough that even a compressed time window is enough for firms to capture most of the relevant quality information, such that competitive pricing is maintained and neither success probability nor returns are significantly affected relative to in-house development. For novel projects with low $\lambda(N)$, the signal is inherently diffuse; a rushed process fails to compensate for this fundamental uncertainty, the latent lemons problem becomes manifest, and returns fall significantly below organic development even as the success probability advantage disappears. Together, the DML and DML-IV analyses tell a coherent two-layer story: the DML establishes the descriptive equilibrium pattern across the full distribution of licensing agreements, documenting how competitive forces operate differently across innovation regimes; the DML-IV confirms that this pattern holds causally, at least for the subset of licenses for which a credible source of exogenous variation - Phase III pipeline failures - can be identified and exploited for identification.

\section{Conclusion}
\label{sec:conclusion}

This study examines how product development strategy shapes pharmaceutical innovation outcomes across innovation regimes, combining a descriptive DML analysis with a causal DML-IV analysis that exploits exogenous pipeline shocks as a source of variation in licensing propensity. 

The main empirical findings can be summarized along two dimensions. First, in-licensed products exhibit significantly higher success probabilities than internally developed ones (+3.8--4.3 percentage points, all $p < 0.001$), consistent with positive selection and inconsistent with pure adverse selection. This advantage generalizes to all licensing transactions but disappears specifically for rushed licenses (LATE $= +0.025$, $p = 0.167$). Second, the aggregate success probability advantage of licensing is offset by lower net returns at mean innovation risk ($\hat{\theta}_{\text{lic}} = -\$746.29$M, $p < 0.001$), generating a risk-return trade-off in which neither strategy dominates. However, this pattern masks fundamental heterogeneity. For incremental projects, the offsetting mechanism holds precisely as the model predicts: competitive return equalization sustains a trade-off consistent with market efficiency. The DML-IV confirms this causally for rushed licenses, which generate neither a success probability advantage nor a return penalty for incremental projects, preserving the non-dominance structure. By contrast, for novel projects the offsetting mechanism breaks down in both analyses. In the DML, licensed frontier innovations retain higher success probabilities without a return penalty. In the DML-IV, rushed licenses for novel projects carry no success probability advantage but a significant and monotonically increasing return loss ($-0.355$ SD at the 10th percentile, $p = 0.012$, corresponding to approximately \$43.9M on the trimmed scale). In both cases, the compensating structure that characterizes competitive equilibrium is absent, pointing to an incomplete realization of competitive equilibrium at the innovation frontier. The potential underlying mechanism behind this asymmetry lies in the assumption that rushed licensing compresses the time available for firms to process project-specific signals, and for novel projects - where signals are inherently diffuse - this compression may contribute to making the latent lemons problem manifest, driving returns below the in-house development baseline without any offsetting improvement in approval rates.

Overall, our empirical findings reinforce the core insight of the existing literature on markets for technology: pharmaceutical licensing markets do not suffer from systematic adverse selection in the aggregate. Market forces appear to function efficiently for incremental projects and the non-dominance result among strategies confirms this interpretation. However, the central insight is that this competitive equilibrium in pharmaceutical licensing markets is not homogeneous: competitive forces operate effectively where development pathways are well-established and project characteristics are predictable (and that is what drives also the aggregate effect), but prove insufficient at the frontier of pharmaceutical innovation, where information asymmetries and the difficulty of evaluating frontier innovations generate persistent frictions. The governance mode through which external innovation is accessed also matters: contractual licensing arrangements carry coordination costs and information frictions that full organizational integration through acquisition does not, as confirmed by the uniformly neutral acquisition effects documented across all innovation regimes. 

Some limitations of the present analysis deserve acknowledgment. Our empirical proxy for innovation regime - predicted success probability constructed from observable product characteristics - captures the average risk profile of projects within each category but may not fully distinguish between high-quality novel projects and low-quality incremental ones. The fundamental challenge is that latent project quality ($\theta$ in our model) is unobservable by assumption, which is indeed what generates the information asymmetry explored throughout the study. One promising approach for fine-tuning the proxy is the use of text analysis of licensing contracts and patent documents to extract project-specific features that correlate more directly with the quality dimension. As a reference, related work by \citet{krieger2022missing} demonstrates that chemical similarity measures can serve as informative proxies for the degree of novelty of pharmaceutical projects. A second limitation is that the DML-IV instrument is constrained to the subpopulation of rushed licenses triggered by Phase III pipeline failures and cannot be generalized to the full set of licensing transactions. Future research should seek valid instruments covering a broader range of licensing motivations, which would allow causal identification for the entire licensing population instead of addressing main results only to the complier subpopulation. Finally, our analysis focuses on the pharmaceutical industry, chosen for its high degree of standardization in development pathways, regulatory requirements, and outcome measurement. Whether the regime-differentiated equilibrium patterns we document extend to other technology markets in which licensing is prevalent but development pathways are less standardized remains to be established and constitutes a natural direction for future research.

From a policy perspective, the evidence of market inefficiencies concentrated in the novel innovation segment raises concerns about whether current institutional arrangements adequately support the efficient allocation of pharmaceutical assets with the greatest potential social value. As information asymmetries are most severe precisely where novelty is highest, transparency may emerge as a natural policy lever. Reducing information asymmetries at the evaluation stage - through standardized disclosure of preclinical and early-clinical data, third-party scientific assessments, or milestone-based contract structures that reveal quality information progressively - could enhance the ability of licensees to price novel projects accurately, thus restoring the competitive return adjustment that our results suggest is currently absent at the innovation frontier. However, the effectiveness and design of such interventions require careful consideration, as reported by \citet{tyagi2025does} who show that increased clinical trial transparency, while reducing information asymmetries, may induce firms to shift toward more incremental research to avoid public scrutiny of novel but risky initiatives. Their findings confirm our conclusion that information frictions are strongest at the innovation frontier, while cautioning that policies designed to improve information disclosure must not undermine the novel R\&D they are meant to encourage. Whether such interventions are feasible and welfare-improving in
practice remains an important open question for future research.

\clearpage
\newpage

\appendix
\renewcommand{\thesection}{\Alph{section}}
\renewcommand{\thetable}{\Alph{section}.\arabic{table}}
\renewcommand{\thefigure}{\Alph{section}.\arabic{figure}}
\setcounter{table}{0}
\setcounter{figure}{0}

\section*{Appendices}
\addcontentsline{toc}{section}{Appendices}
\section{Additional Outcomes and Supplementary Analyses}
\label{app:first}

\subsection{Additional Outcomes}
\label{app:appendix_outcomes}

The body of the paper focuses on two primary outcomes: success probability ($Y_1$) and net return ($Y_2$). To shed light on each dimension of the net return outcome, here we report separate DML analyses for the two components of net return — trial costs ($Y_3$) and lifetime sales ($Y_4$) — allowing us to assess whether the aggregate net return response to licensing is driven primarily by the cost side, the revenue side, or both simultaneously. In addition, we define and analyze a ratio-based alternative to net return ($Y_5$, the \emph{ROI} ratio), which is employed in the break-even descriptive analysis of Appendix~\ref{app:ipsw_selection_bias} and~\ref{app:descriptive_results}. Throughout our empirical analysis, we focus on the comparison between in-licensed and in-house developed products, as this comparison directly tests our theoretical model's predictions regarding licensing markets. Products sourced through company acquisition are excluded from the DML estimation in Appendix~\ref{app:results} to maintain analytical focus, though they are included in Appendix~\ref{app:ipsw_selection_bias} and~\ref{app:descriptive_results} for completeness.

\paragraph{Trial Costs ($Y_3$).} The trial cost variable represents the logarithm of total trial costs for approved drugs, providing a measure of the resource intensity of successful development projects. This empirical measure proxies total social development costs in Proposition 3, for both licensing and in-house development strategies. Costs include all direct expenditures associated with clinical trials, including patient recruitment, site management, regulatory compliance, and monitoring activities. Importantly, the adoption of the Evaluate cost model represents a substantial methodological advancement over existing literature in this domain, which has always relied on survey-based data or publicly disclosed aggregate figures to estimate development costs — see for example \citet{dimasi2016innovation} or \citet{adams2010spending}. On the other hand, the Evaluate methodology employs a bottom-up modeling approach that integrates trial-specific parameters, therapeutic area benchmarks, and regulatory complexity factors to generate product-level cost estimates with considerably greater precision and granularity. This substantial improvement in the measurement approach facilitates more accurate assessments of the cost implications associated with different development strategies and substantially reduces measurement error that could bias the estimates.

For the DML analysis of trial costs, the sample is restricted to approved products with non-missing trial cost data, yielding 15,890 observations. Excluding projects developed as a result of a company acquisition deal for the primary in-licensed versus in-house development comparison results in a final estimation sample of 11,220 observations.

\paragraph{Sales ($Y_4$).} Commercial performance is measured using product-level lifetime sales (in logs), defined as cumulative worldwide revenues generated over the product market life. This variable is computed exclusively for approved products to ensure reliability of sales data and captures commercial success accounting for both peak sales performance and market longevity. While not explicitly modeled in our theoretical framework, commercial sales provide an \textit{ex-post} measure of realized project value and market potential, offering an empirical proxy for the revenue from successful development.

For the DML analysis of lifetime sales, the sample is again restricted to approved products with non-missing lifetime sales data, yielding 9,346 observations. Excluding company acquisition products results in a final estimation sample of 6,451 observations comprising 4,117 in-house developed products and 2,334 in-licensed products.

\paragraph{ROI Ratio ($Y_5$).} A systematic measure of clinical development efficiency is constructed as the ratio of lifetime sales to clinical trial costs, capturing the revenue generated per unit of development expenditure. This specification is necessarily restricted to approved products with observable commercial revenues, thus excluding abandoned projects from the analysis. From the initial universe of products with non-missing information on both trial costs and lifetime sales (7,474 products, including approved, abandoned, and ongoing projects), the sample is restricted to approved products, resulting in 4,749 observations. 

While referred to as ROI for simplicity, this measure should be interpreted as a gross revenue-cost ratio rather than a full return on investment in the corporate finance sense, as it excludes pre-clinical costs, post-marketing expenditures, and opportunity costs of capital. Alternative and equally appropriate labels include \textit{break-even ratio} or \textit{commercial recovery ratio}. The terminology adopted in this Appendix reflects this narrower interpretation. The break-even descriptive analysis in Appendix~\ref{app:descriptive_results} employs this measure alongside the net return specification, which addresses the limitation of conditioning on approval by assigning a lifetime sales value of zero to abandoned products, thus incorporating development failures into the analysis.

\subsection{Break-Even Analysis and Selection Bias Correction}
\label{app:ipsw_selection_bias}

Before implementing the DML framework, here we evaluate break-even performance across alternative development strategies through a preliminary statistical comparison. This exercise focuses on products for which both clinical development costs and commercial outcomes are jointly observable. From the initial universe of products with non-missing information on both trial costs and lifetime sales (7,474 products), the sample is restricted to approved products, resulting in 4,749 observations. This restriction allows the construction of a break-even indicator conditional on market approval — the ROI ratio defined in Appendix~\ref{app:appendix_outcomes} — where a value greater than one indicates that cumulative revenues exceed clinical development expenditures, i.e. that project reaches a break-even point after commercialization.

\begin{table}[tp]
\centering
\caption{\textbf{Table A.1} - Top 10 Features with Largest Balance Improvement after IPSW}
\label{tab:balance_improvement}
\small
\renewcommand{\arraystretch}{1.20}
\resizebox{\textwidth}{!}{%
\begin{tabular}{llrrr}
\toprule\toprule
\textbf{Feature} & \textbf{Category} & \textbf{SMD Before IPSW} & \textbf{SMD After IPSW} & \textbf{Improvement} \\
\midrule
Patent Status        & Patented                  & $-0.420$ & $-0.041$ & $0.380$ \\
Technology           & Monoclonal antibody       & $-0.248$ & $-0.026$ & $0.222$ \\
Therapeutic Subcat.  & Immuno-oncology           & $-0.268$ & $-0.068$ & $0.200$ \\
Proprietary Level    & New molecular entity      & $-0.225$ & $-0.032$ & $0.193$ \\
Therapeutic Cat.     & Oncology                  & $-0.217$ & $-0.033$ & $0.184$ \\
Technological Cat.   & Biotechnology             & $-0.230$ & $-0.055$ & $0.175$ \\
Technological Cat.   & Conventional              & $+0.230$ & $+0.055$ & $0.175$ \\
Technology           & Small molecule chemistry  & $+0.196$ & $+0.048$ & $0.147$ \\
Proprietary Level    & NDA                       & $+0.148$ & $+0.026$ & $0.121$ \\
Therapeutic Subcat.  & Other cancer treatments   & $-0.123$ & $+0.001$ & $0.121$ \\
\bottomrule\bottomrule
\end{tabular}}
\begin{minipage}{\textwidth}
\vspace{4pt}
\centering
\footnotesize\textbf{Notes:} Standardized Mean Differences (SMDs) computed between the universe of products with non-missing trial costs and lifetime sales and the approved-products subsample, before and after IPSW reweighting. Improvement = $|$SMD Before$|$ $-$ $|$SMD After$|$. Following \citet{rosenbaum1985constructing} and \citet{austin2009balance}, absolute SMD values below 0.1 indicate negligible imbalance.
\end{minipage}
\end{table}

Restricting attention to approved products raises potential selection concerns, as the resulting subsample may not be representative of the broader population of products for which both metrics are observed. In particular, conditioning on approval may drive systematic differences in the distribution of pre-strategy covariates across the universe and the analytical subsample. To assess the severity of this issue, a formal covariate balance check is conducted comparing the distribution of pre-strategy characteristics between the universe of products with both metrics and the approved-products subsample. Two complementary diagnostics are employed: first, Pearson chi-square tests assess whether the distributions of each confounder differ substantially across the two groups; second, Standardized Mean Differences (SMDs) are computed for all one-hot encoded covariates to quantify the magnitude of imbalance in a scale-free way. Following standard practice in the program evaluation literature, absolute SMD values below 0.1 are interpreted as negligible imbalance, values between 0.1 and 0.2 as moderate imbalance, and values above 0.2 as substantial imbalance \citep{rosenbaum1985constructing, austin2009balance}. The unweighted comparison reveals several covariates exhibiting moderate to substantial imbalance, indicating that the approved-products subsample differs systematically from the broader universe along economically meaningful dimensions.

In order to address this concern, Inverse Probability of Selection Weighting (IPSW) is applied so that the approved-products subsample mirrors the universe of products with complete information in terms of confounding features. Selection probabilities are estimated using a logistic regression model in which the dependent variable is a binary indicator for inclusion in the approved-products subsample and the covariates correspond to the full matrix of confounders. Each approved product is then weighted by the inverse of its estimated selection probability, assigning higher weights to products that were \textit{ex-ante} less likely to be approved given their characteristics. This procedure preserves the sample size while correcting for differential selection on observables, effectively reweighting the approved sample to resemble the \textit{ex-ante} population of all products entering development. The effectiveness of IPSW is evaluated by re-computing SMD statistics after weighting; results summarized in Tab.~\ref{tab:balance_improvement} show that all previously identified cases of moderate and substantial imbalance are eliminated, with the vast majority of covariates falling well below conventional imbalance thresholds. This selection correction is particularly important because approval likelihood varies with project risk: lower-risk projects with higher predicted success probabilities are more likely to appear in the approved subsample, while higher-risk innovative projects are disproportionately represented among failures. Without IPSW correction, comparisons of returns conditional on approval would over-represent lower-risk projects and underweight higher-risk projects at the frontier of pharmaceutical innovation.

\subsection{Descriptive Results on Break-Even Performance}
\label{app:descriptive_results}

Descriptive findings provide preliminary evidence on whether the empirical patterns align with our theoretical predictions about selection and returns.

\begin{figure}[tp]
    \centering
    \includegraphics[width=1\linewidth]{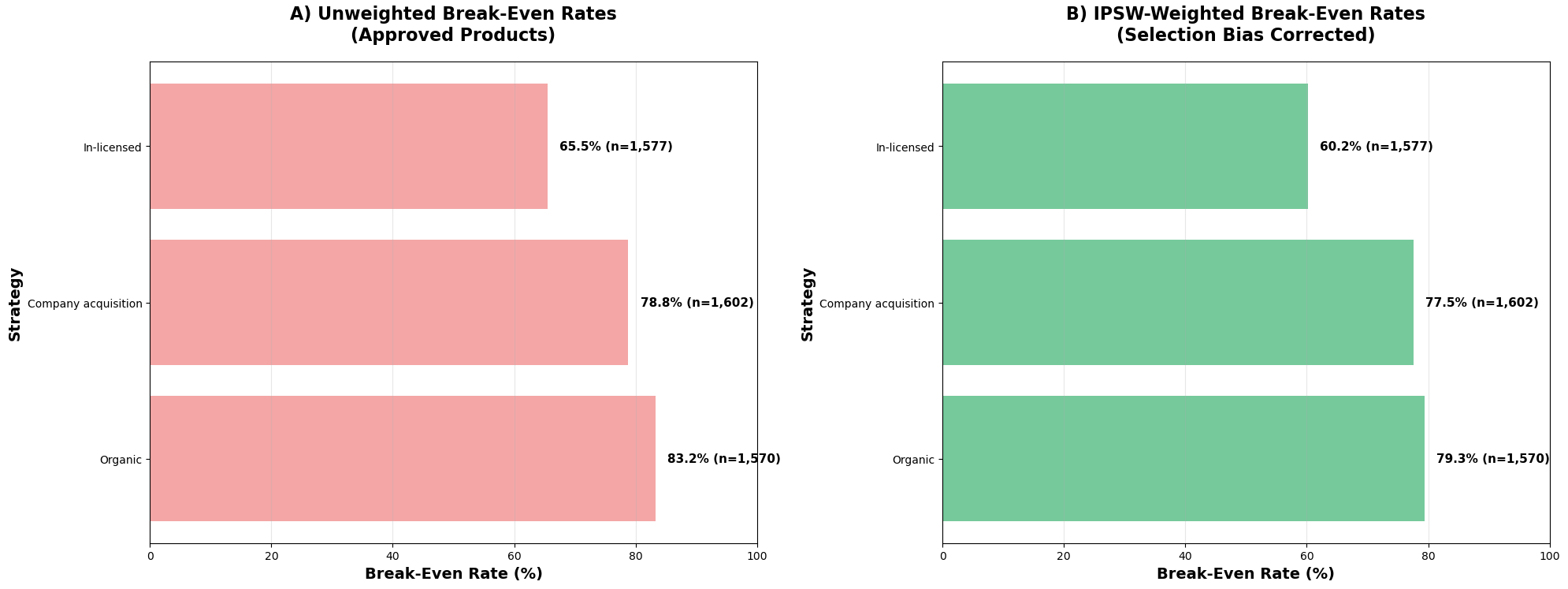}
    \caption{\textbf{Figure A.1} - Break-Even Rate Comparison: Unweighted vs IPSW-weighted}
    \label{fig:breakeven_rates}
\end{figure}

Fig.~\ref{fig:breakeven_rates} and Tab.~\ref{tab:metric_comparison} show that the impact of reweighting, while necessary to address the aforementioned balancing issues, is relatively modest in magnitude. After applying IPSW, break-even rates - defined as the fraction of approved products whose lifetime sales exceed clinical trial costs, hence with a $ROI > 1$ - are revised downward across all strategies, yielding rates of 79.3\% for organic development and 60.2\% for in-licensed products, suggesting that the initial subsample of approved products systematically overrepresents characteristics associated with higher commercial success and that correcting for this selection effect reduces estimated break-even rates by approximately 3-5 percentage points across strategies.

\begin{table}[tp]
\centering
\caption{\textbf{Table A.2} - Comparison of Metrics: Unweighted vs.\ IPSW-Weighted}
\label{tab:metric_comparison}
\small
\renewcommand{\arraystretch}{1.20}
\begin{tabular}{lrrr}
\toprule\toprule
\textbf{Metric} & \textbf{Unweighted} & \textbf{IPSW-Weighted} & \textbf{Difference} \\
\midrule
Break-even Rate (\%)        & $75.8$     & $71.9$     & $-3.9$ \\
Average Trial Cost (\$M)    & $577.5$    & $656.1$    & $+78.6$ \\
Average Lifetime Sales (\$M)& $4{,}672.7$ & $4{,}859.1$ & $+186.4$ \\
Average Net Return (\$M)    & $4{,}095.2$ & $4{,}203.0$ & $+107.8$ \\
Average ROI Ratio           & $8.09$     & $7.41$     & $-0.68$ \\
\bottomrule\bottomrule
\end{tabular}
\begin{minipage}{\textwidth}
\vspace{4pt}
\centering
\footnotesize\textbf{Notes:} Statistics computed on the approved-products subsample (4,749 observations) pooled across all strategies. All monetary values in millions of U.S. dollars.
\end{minipage}
\end{table}

\begin{figure}[tp]
    \centering
    \includegraphics[width=1\linewidth]{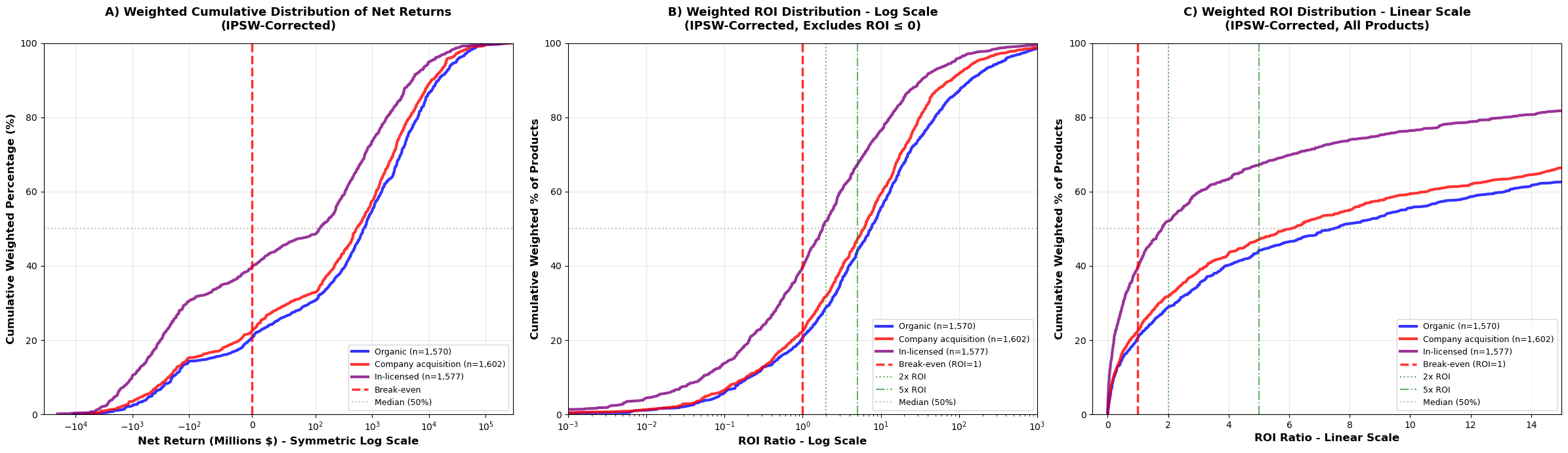}
    \caption{\textbf{Figure A.2} - Weighted Cumulative Distribution Functions}
    \label{fig:ecdf_net_return}
\end{figure}

\begin{figure}[tp]
    \centering
    \includegraphics[width=1\linewidth]{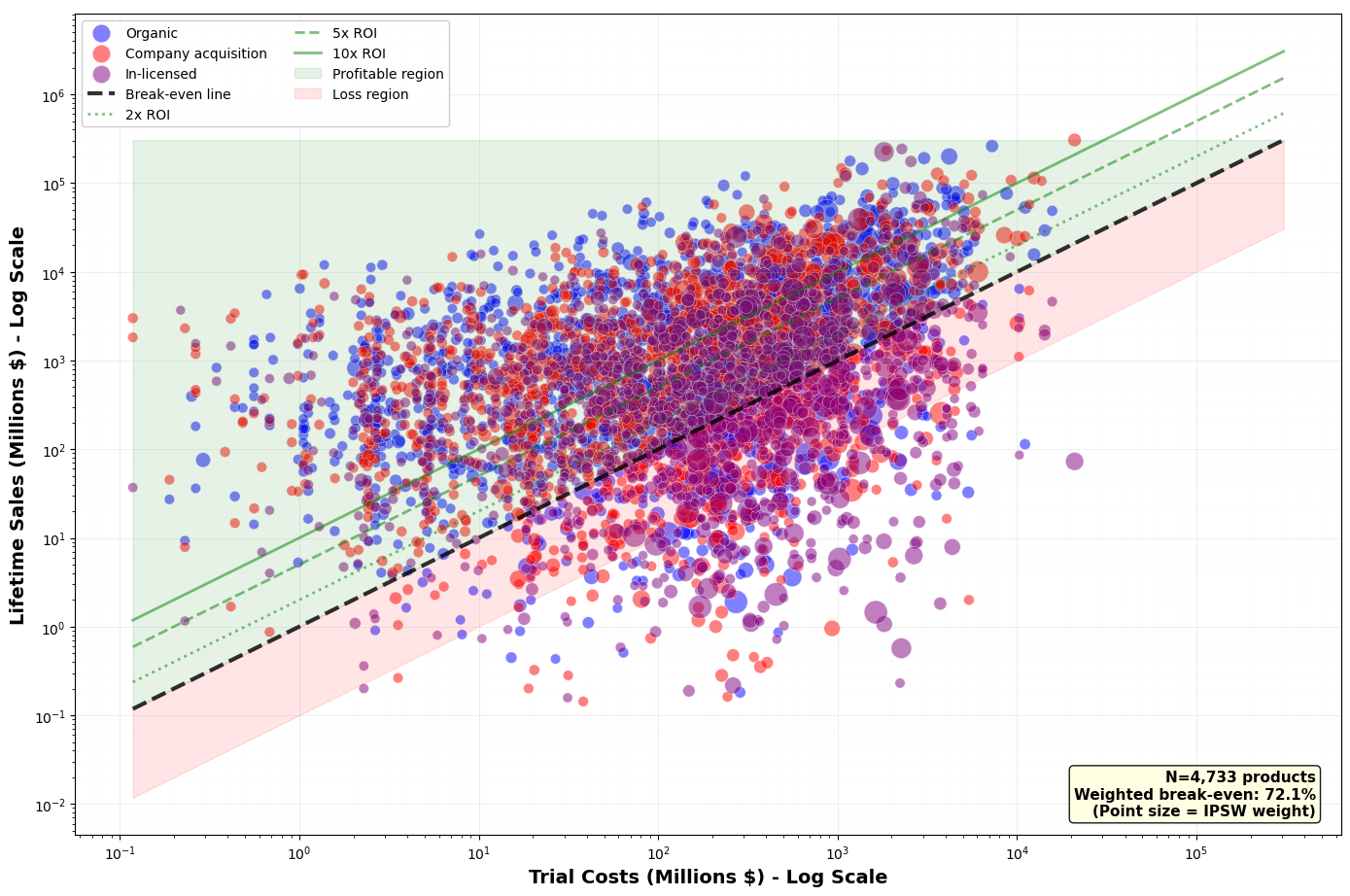}
    \caption{\textbf{Figure A.3} - IPSW-Weighted: Trial Costs vs Lifetime Sales (Log-Log Scale)}
    \label{fig:scatter_costs_sales}
\end{figure}

\begin{figure}[tp]
    \centering
    \includegraphics[width=1\linewidth]{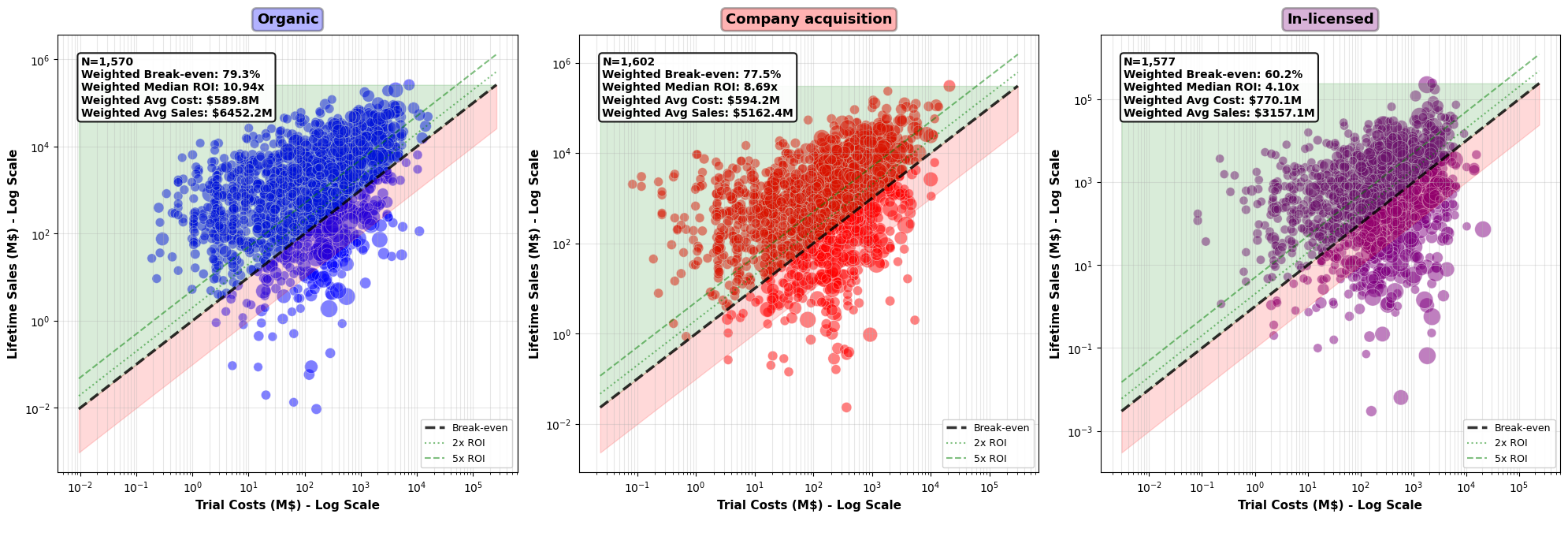}
    \caption{\textbf{Figure A.4} - IPSW-Weighted Break-Even Analysis by Strategy (Log-Log Scale)}
    \label{fig:scatter_costs_sales_strategy}
\end{figure}

This lower commercial performance for in-licensed products is corroborated by Fig.~\ref{fig:ecdf_net_return}: panel A reports the IPSW-weighted cumulative distribution of net returns, where in-licensed products exhibit a clear leftward shift relative to both organic and company-acquisition strategies, indicating a higher probability of lower and negative net returns. Panels B and C focus on the ROI-based measure. On both log and linear scales, the weighted ECDFs show that organic products stochastically dominate in-licensed products: at any given ROI threshold above one, a larger fraction of internally developed projects exceeds that threshold. Median and upper-quantile differences are particularly sizable, suggesting that in-house development not only achieves break-even more frequently but also generates a thicker right tail of high-return outcomes. Company acquisition tends to lie between the two, performing closer to organic than to in-licensed products. Fig.~\ref{fig:scatter_costs_sales} and~\ref{fig:scatter_costs_sales_strategy} complement these results by plotting the joint distribution of trial costs and lifetime sales. Even after IPSW correction, in-licensed projects cluster closer to the break-even frontier, while organic projects display broader dispersion above it. These aggregate patterns should be interpreted in light of the heterogeneity analysis of Section~\ref{subsec:dml_main_results}: the overall return disadvantage of in-licensed products is driven primarily by incremental projects, for which we find statistically significant negative differences relative to organic in-house development, while novel projects - for which estimated differences are statistically indistinguishable from zero - do not contribute to this aggregate pattern.

\begin{figure}[tp]
    \centering
    \includegraphics[width=1\linewidth]{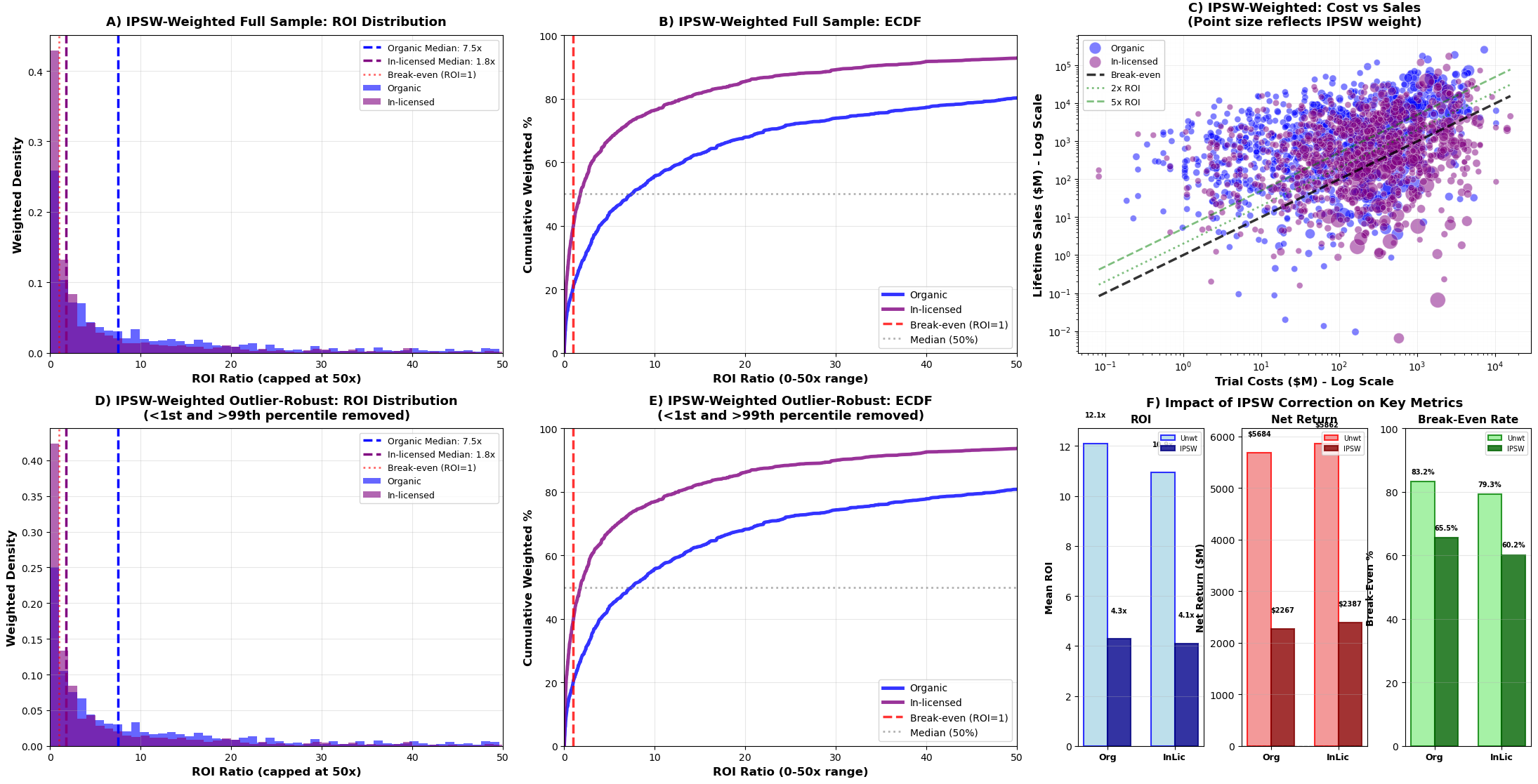}
    \caption{\textbf{Figure A.5} - Focus on Organic vs In-licensed: Full Sample vs Outlier-Robust Analysis}
    \label{fig:roi_robust}
\end{figure}

To ensure robustness, an additional outlier-robust specification trims the weighted sample at the 1st and 99th percentiles of the ROI distribution (Fig.~\ref{fig:roi_robust}). The weighted, outlier-robust ROI distributions and corresponding ECDFs indicate that in-licensed products are more densely concentrated in ranges of modest to intermediate returns (e.g., below 50× ROI), where they exhibit relatively strong performance and a higher concentration of outcomes. By contrast, in-house organic development displays greater dispersion and a heavier right tail, reflecting a higher prevalence of projects achieving very large ROI realizations. As a consequence, while in-house strategy dominates the upper tail of the ROI distribution (at least for projects that successfully reach the market), in-licensed strategy appears comparatively more effective in generating consistent returns in lower-to-medium ROI ranges. This heterogeneity in return distributions across strategies — and potentially across risk categories within strategies — suggests that selection patterns may be more complex than simple positive or negative selection, motivating the more sophisticated DML analysis used in Section~\ref{sec:dml_results}.

\subsection{Results}
\label{app:results}

Here we show DML estimates for the three additional outcomes introduced in Appendix~\ref{app:appendix_outcomes}: trial costs ($Y_3$), lifetime sales ($Y_4$), and the ROI ratio ($Y_5$). Together, the first two decompose the aggregate net return result of Section~\ref{subsec:dml_main_results} into its cost and revenue components, allowing us to identify which channel primarily drives the return differences between licensed and in-house developed products. For $Y_3$ and $Y_4$, the final stage of the DML procedure follows the standard OLS specification described in Section~\ref{subsec:dml_specification}. For $Y_5$, the final stage is modified to employ weighted least squares (WLS), where each observation is weighted by the inverse of its estimated selection probability from the IPSW procedure of Appendix~\ref{app:ipsw_selection_bias}:
\begin{equation*}
\hat{\theta}_{\text{lic}}^{\text{IPSW}} = \left[\sum_{i=1}^{n} w_i \tilde{D}_{\text{lic},i}^2 \right]^{-1} \left[\sum_{i=1}^{n} w_i \tilde{D}_{\text{lic},i} \tilde{Y}_i\right]
\end{equation*}
This modification corrects for the differential probability of observing approved versus failed projects, since $Y_5$ is constructed exclusively on approved products. The WLS formulation preserves the asymptotic properties of the DML estimator while accounting for differential selection probabilities across products \citep{hirano2003efficient, wooldridge2007inverse}. The machine learning algorithms employed for each outcome follow the specification described in Section~\ref{subsec:dml_learners}.

\begin{figure}[tp]
    \centering
    \includegraphics[width=1\linewidth]{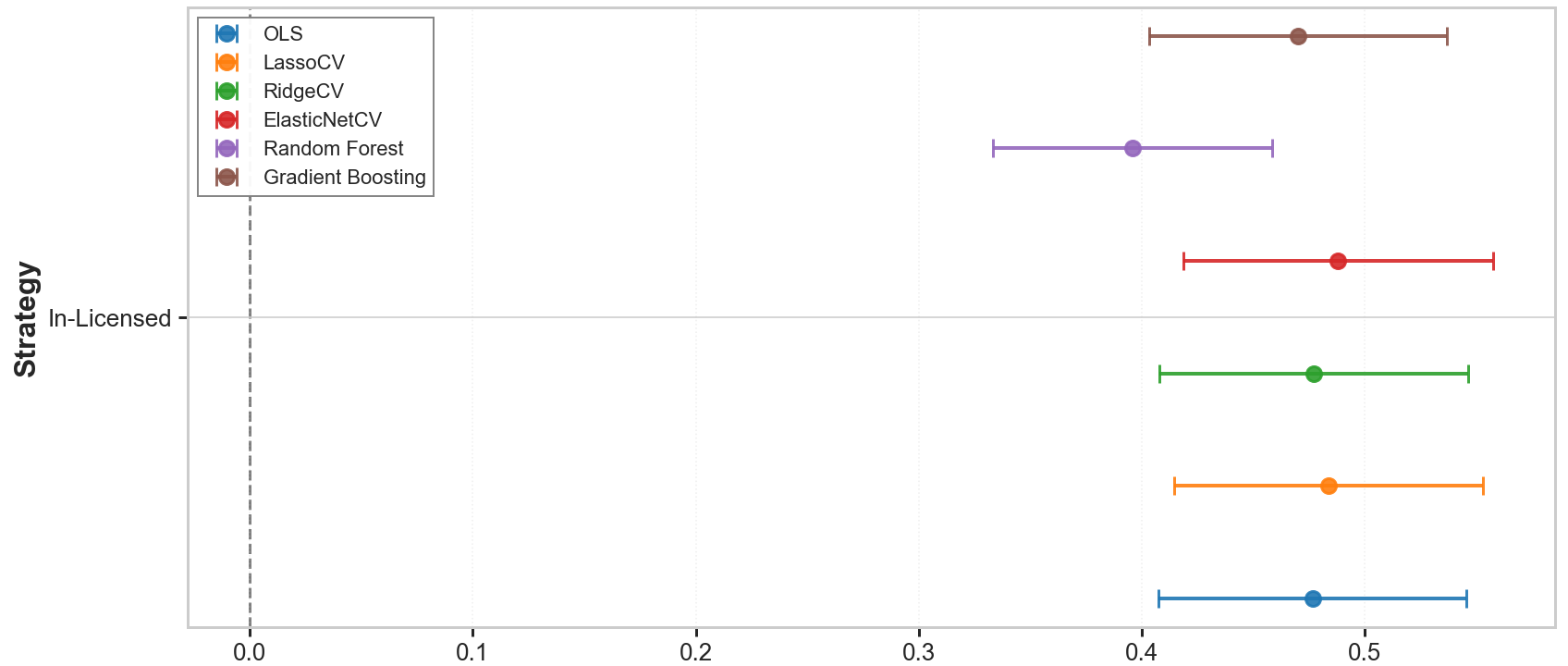}
    \caption{\textbf{Figure A.6} - DML Estimate: Trial Costs ($Y_3$, in logs) with 95\% Confidence Intervals}
    \label{fig:dml_y3}
\end{figure}

\paragraph{Development Costs ($Y_3$).} Trial costs are substantially and significantly higher for in-licensed products (Fig.~\ref{fig:dml_y3}). Across all machine learning estimators, in-licensing exhibits significantly higher trial costs than in-house development: point estimates range from 0.396 (Random Forest) to 0.488 (Elastic Net), corresponding to increases of 48.6\% to 62.9\% in trial costs relative to organic development.\footnote{For log-transformed outcomes, the percentage change is computed as $(\exp(\hat{\theta}_{\text{lic}})-1) \times 100\%$.} Estimates from parametric models (OLS, Lasso, Ridge, Elastic Net) cluster tightly around 0.477--0.488, while tree-based methods yield slightly lower but still substantial estimates (0.396 for Random Forest, 0.470 for Gradient Boosting), all highly statistically significant ($p<0.001$). The qualitative conclusion is robust: licensing substantially increases clinical trial costs relative to in-house development, consistent with \citet{girotra2007valuing}. These findings provide strong empirical support for Proposition 3: licensing increases total social development costs when coordination costs $C_{\text{c}}(\tau)$ are substantial, arising from imperfect knowledge transfer between organizations \citep{szulanski1996exploring}, additional regulatory compliance activities, and the transaction costs of managing complex contractual relationships involving milestone payments and royalty structures \citep{kogut1992knowledge}.

\begin{figure}[tp]
    \centering
    \includegraphics[width=1\linewidth]{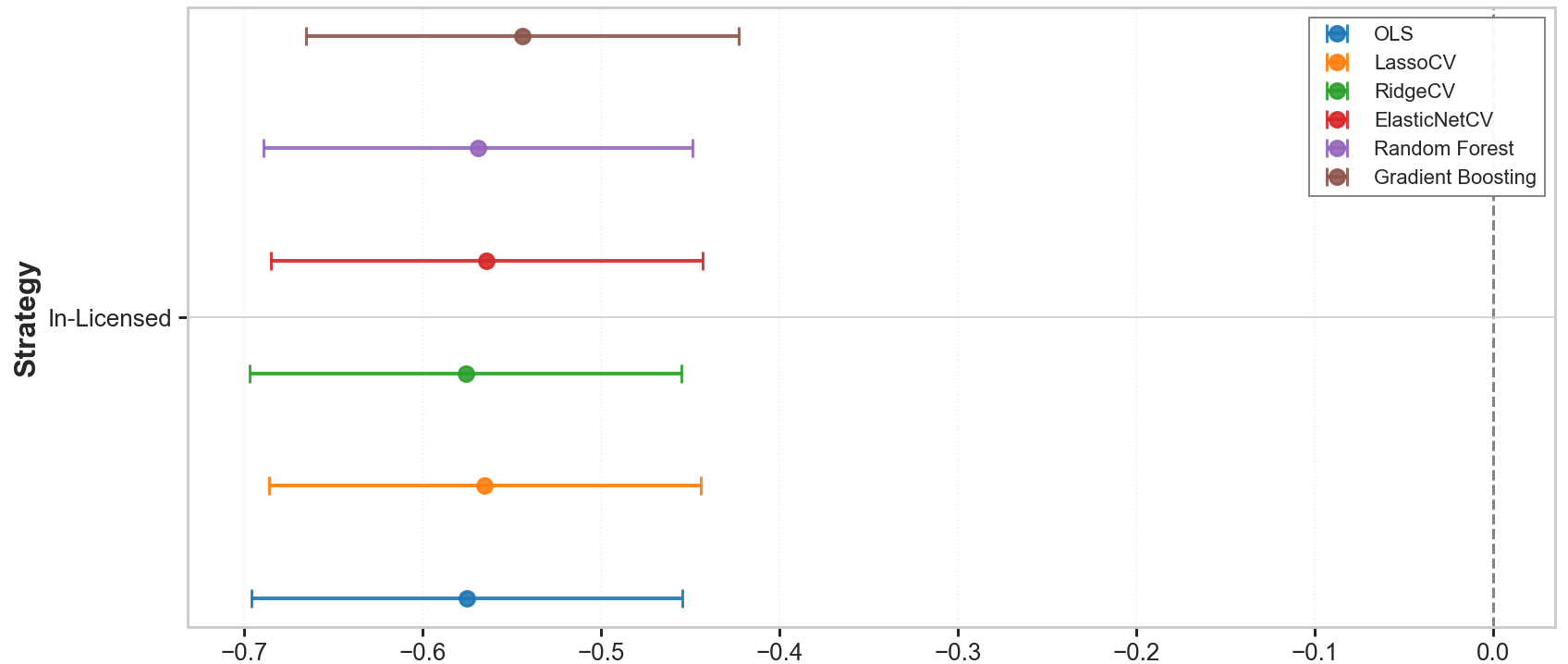}
    \caption{\textbf{Figure A.7} - DML Estimate: Lifetime Sales ($Y_4$, in logs) with 95\% Confidence Intervals}
    \label{fig:dml_y4}
\end{figure}

\paragraph{Lifetime Sales ($Y_4$).} Commercial revenues exhibit the opposite pattern. In-licensed products generate significantly lower commercial revenues (Fig.~\ref{fig:dml_y4}): DML estimates range from $-0.544$ (Gradient Boosting) to $-0.576$ (Ridge), corresponding to 42--44\% lower sales (all $p<0.001$). If licensed products are statistically associated with higher success probabilities, why do approved licensed products generate lower revenues? Several mechanisms may explain this pattern. First, positive selection in terms of success probability may reflect selection of lower-risk, more incremental projects that face greater market competition and limited differentiation, rather than breakthrough innovations commanding premium pricing. Second, selection may occur along dimensions that improve regulatory success without increasing commercial value, including a focus on specialized rather than blockbuster markets \citep{grabowski1994returns} and reduced effective patent life for products licensed late in development \citep{hegde2018patent}.

\begin{figure}[tp]
    \centering
    \includegraphics[width=1\linewidth]{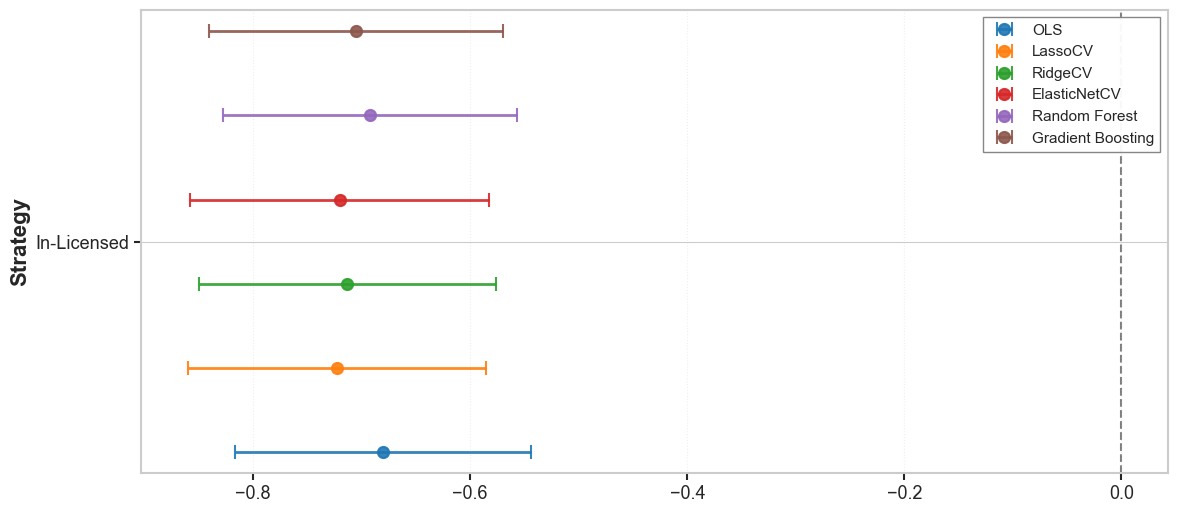}
    \caption{\textbf{Figure A.8} - DML Estimate: ROI ($Y_5$, IPSW-Weighted + Trimmed) with 95\% Confidence Intervals}
    \label{fig:dml_y5}
\end{figure}

\paragraph{Return on Investment ($Y_5$).} The ROI ratio combines both measures, revealing that in-licensed products yield substantially lower returns (Fig.~\ref{fig:dml_y5}). DML estimates cluster tightly around $-0.681$ to $-0.723$, corresponding to 49--52\% lower ROI (all $p<0.001$), and are remarkably stable across different machine learning algorithms, with standard errors around 0.070 and all confidence intervals excluding zero by substantial margins.

\paragraph{Connecting to the Heterogeneity Analysis.} The three outcomes jointly tell a coherent story: licensing is associated with higher clinical trial costs, lower lifetime sales, and lower ROI across all machine learning specifications. These compounding disadvantages on both the cost and revenue sides generate the aggregate net return penalty documented in Section~\ref{sec:dml_results}. Crucially, however, these unconditional estimates mask the heterogeneity uncovered in Section~\ref{subsec:dml_main_results}: the aggregate return disadvantage is driven entirely by incremental projects, for which we find statistically significant negative differences in net returns relative to organic development. Novel projects, for which estimated return differences are statistically indistinguishable from zero, do not contribute to this aggregate pattern. In other words, the unconditional lower returns documented here reflect the competitive equilibrium outcome for the incremental segment of pharmaceutical innovation, in which the success probability advantage of licensing is fully offset by higher costs and lower revenues.

\setcounter{table}{0}
\setcounter{figure}{0}
\section{Interacted DML Robustness: Bootstrap Validation and Outcome Trimming}
\label{app:dml_interaction_validation}

\subsection{Bootstrap Validation of Analytical Standard Errors}
\label{app:bootstrap}

Because $\hat{S}_i$ enters the interaction term $D_{ji} \times \tilde{S}_i$ as a generated regressor in the continuous interaction model of Section~\ref{subsec:dml_specification_interaction}, the analytical standard errors produced by the second-stage DML step treat $\hat{S}_i$ as fixed and therefore ignore additional sampling uncertainty arising from the first-stage estimation of the success probability classifier. We validate these analytical standard errors via a two-stage bootstrap procedure with $B = 500$ replications. Each replication proceeds as follows. In the first stage, a bootstrap sample of size $N_1$ is drawn with replacement from the $Y_1$ dataset, and the Random Forest classifier is retrained on this resample to obtain a new set of predicted success probabilities $\hat{S}^*_i$. In the second stage, a bootstrap sample of size $N_2$ is drawn with replacement from the $Y_2$ dataset; using the newly estimated $\hat{S}^*_i$, the centered interaction term $\tilde{S}^*_i$ is recomputed and the interacted DML is re-estimated. Bootstrap standard errors are the standard deviations of the empirical distributions across the $B = 500$ replications, and confidence intervals are computed as empirical percentiles, requiring no normality assumption.

Table~\ref{tab:bootstrap_ses} reports the analytical and bootstrap standard errors for both strategies, together with the SE ratio and the bootstrap 95\% confidence interval. The SE ratios for the in-licensed coefficients are 1.024 and 0.940 for $\hat{\theta}_{\text{lic}}$ and $\hat{\gamma}_{\text{lic}}$ respectively — both negligibly close to unity and in opposite directions — indicating that the additional uncertainty from the first-stage estimation of $\hat{S}_i$ contributes negligibly to the total sampling variance of the DML estimates. For company acquisition, the SE ratios are similarly close to unity (1.143 and 0.943), and the bootstrap confidence intervals confirm the neutrality result documented in Section~\ref{subsec:dml_main_results}.

\begin{table}[tp]
\centering
\caption{\textbf{Table B.1} - Analytical vs.\ Bootstrap Standard Errors}
\label{tab:bootstrap_ses}
\resizebox{\textwidth}{!}{%
\begin{tabular}{lrrrrrl}
\hline\hline
Parameter & Analyt.\ Est. & Analyt.\ SE & Boot.\ SE & SE Ratio & Boot.\ $p$ & Boot.\ 95\% CI \\
\hline
In-licensed (main effect) & $-746.3$ & $152.8$ & $156.5$ & $1.024$ & $<0.001$ & $[-1{,}023.5,\ -384.8]$ \\
In-licensed (interaction) & $-2{,}443.4$ & $659.0$ & $619.4$ & $0.940$ & $<0.001$ & $[-3{,}671.8,\ -1{,}158.3]$ \\
Company acq.\ (main effect) & $-30.7$ & $183.5$ & $209.8$ & $1.143$ & $0.600$ & $[-509.7,\ 322.7]$ \\
Company acq.\ (interaction) & $-839.1$ & $829.1$ & $782.1$ & $0.943$ & $0.268$ & $[-2{,}190.1,\ 634.2]$ \\
\hline\hline
\end{tabular}}
\begin{minipage}{\textwidth}
\footnotesize\textbf{Notes:} SE Ratio $= \text{SE}_{\text{boot}} / \text{SE}_{\text{analyt}}$. Bootstrap $p$-values and confidence intervals are computed from the empirical percentile distribution. All monetary estimates in millions of U.S.\ dollars.
\end{minipage}
\end{table}

\begin{figure}[tp]
\centering
\includegraphics[width=1\linewidth]{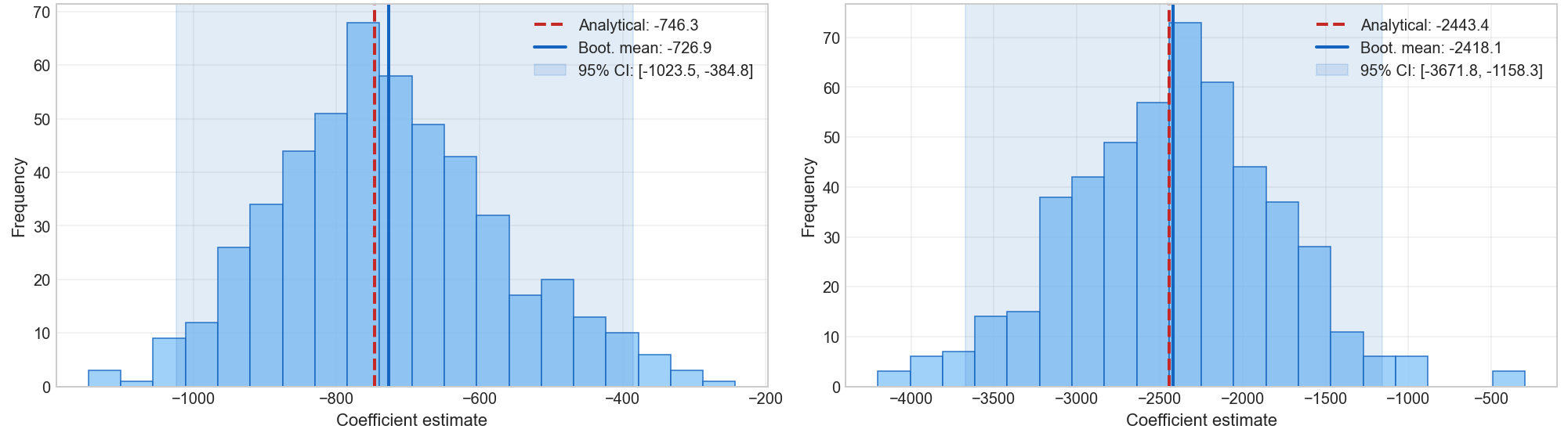}
\caption{\textbf{Figure B.1} - Bootstrap distributions of $\hat{\theta}_{\text{lic}}$ (left panel) and $\hat{\gamma}_{\text{lic}}$ (right panel) for the in-licensed strategy.}
\label{fig:bootstrap_dist}
\end{figure}

\begin{figure}[tp]
\centering
\includegraphics[width=1\linewidth]{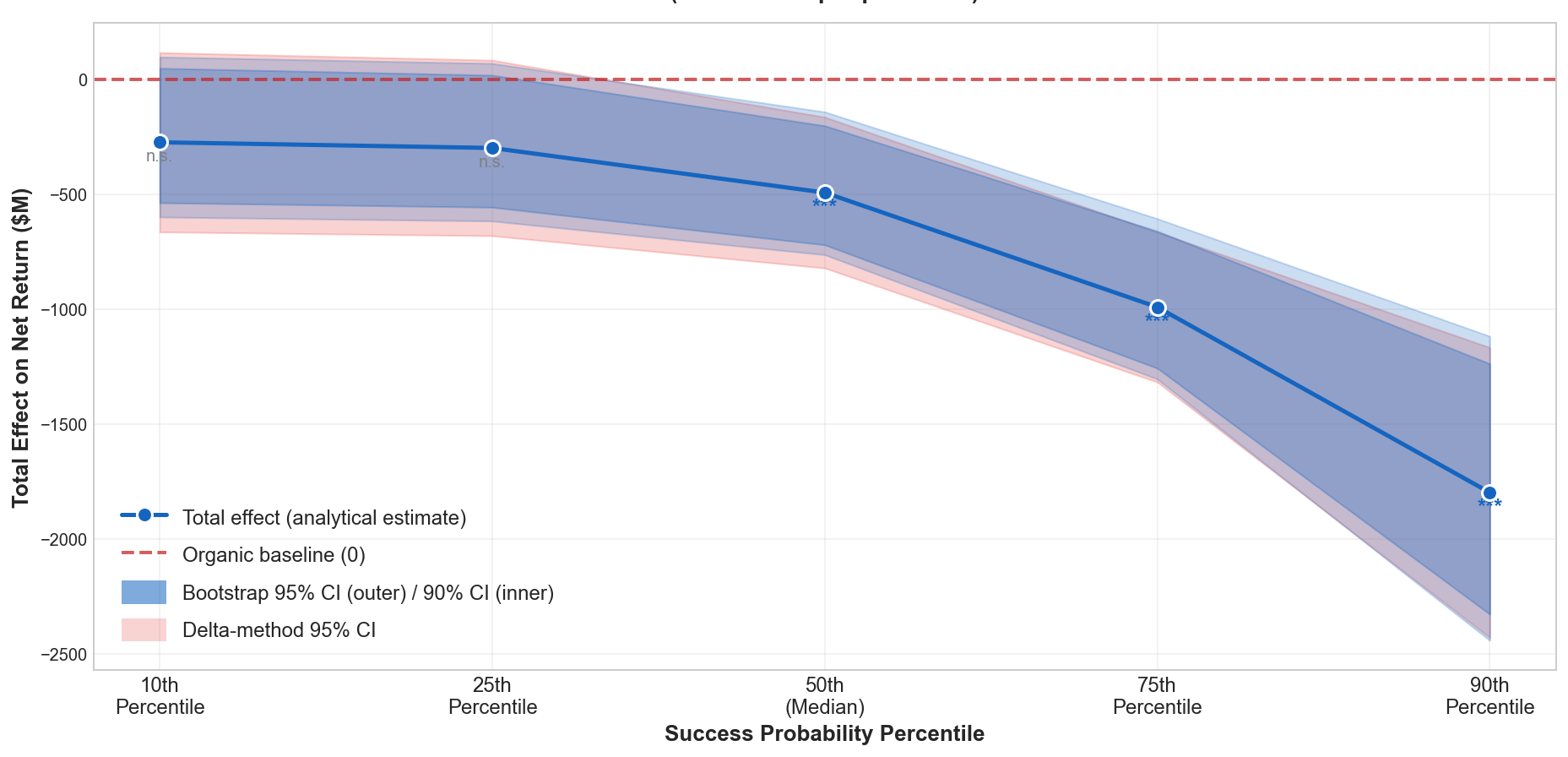}
\caption{\textbf{Figure B.2} - Total effect of in-licensing on net return (\$M) relative to organic development.}
\label{fig:bootstrap_ci}
\end{figure}

Figure~\ref{fig:bootstrap_dist} plots the bootstrap distributions of $\hat{\beta}_1$ (main effect) and $\hat{\beta}_3$ (interaction coefficient) for the in-licensed strategy. Both distributions are approximately symmetric and unimodal around the analytical point estimates, with the bootstrap means of $-726.9$ and $-2{,}418.1$ closely tracking the analytical estimates of $-746.3$ and $-2{,}443.4$ respectively. Figure~\ref{fig:bootstrap_ci} plots the total effect curve together with the bootstrap and delta-method 95\% confidence bands across the full range of the success probability distribution. The near-identical overlap of the two bands confirms that generated-regressor bias is negligible and that the heterogeneity result — statistically insignificant for novel projects at the 10th and 25th percentiles, and negative and significant from the median onwards — is fully robust to first-stage sampling uncertainty.

\subsection{Outcome Trimming}
\label{app:trimming}

As a further robustness check, we re-estimate the interacted DML model on a trimmed sample in which the net return outcome is restricted to the 5th--95th percentile of its distribution for the licensing strategy, removing 1,360 observations with extreme values and retaining 12,235 products. Trimming mainly addresses the concern that the heterogeneity pattern documented in Section~\ref{subsec:dml_main_results} may be driven by a small number of outlier observations with very large positive or negative net returns. Table~\ref{tab:trim_marginal} reports the marginal effects of in-licensing on net return at the usual five percentiles of the predicted success probability distribution. The qualitative pattern is fully preserved: the total effect is statistically indistinguishable from zero for novel projects at the 10th and 25th percentiles ($+5.7$\,\$M, $p = 0.621$ and $+1.2$\,\$M, $p = 0.912$ respectively), turns negative and marginally insignificant at the median ($-15.7$\,\$M, $p = 0.114$), and becomes negative and strongly significant for incremental projects at the 75th ($-74.7$\,\$M, $p < 0.001$) and 90th percentiles ($-177.1$\,\$M, $p < 0.001$). The magnitudes are substantially smaller than in the untrimmed specification by construction, reflecting the compression of the outcome distribution, but the sign pattern and the novel/incremental asymmetry are fully consistent with the main results.

\begin{table}[tp]
\centering
\caption{\textbf{Table B.2} - Heterogeneous Effect on Net Returns ($Y_2$, Trim 5--95th Pct) by Innovation-Risk Percentile: In-Licensing}
\label{tab:trim_marginal}
\small
\renewcommand{\arraystretch}{1.20}
\begin{tabular}{lrrrrc}
\toprule\toprule
 & & & \multicolumn{3}{c}{\textit{Total Effect: $\hat{\theta}_{\text{lic}} + \hat{\gamma}_{\text{lic}} \tilde{S}_i$}} \\
\cmidrule(lr){4-6}
\textbf{Percentile of $\tilde{S}_i$} & \textbf{Main Effect ($\hat{\theta}_{\text{lic}}$)} & \textbf{Interaction ($\hat{\gamma}_{\text{lic}} \tilde{S}_i$)} & \textbf{Estimate} & \textbf{SE} & \textbf{$p$-value} \\
\midrule
10th (most novel)       & $-49.39$ & $+55.05$  & $+5.66$          & $11.44$ & $0.621$ \\
25th                    & $-49.39$ & $+50.62$  & $+1.23$          & $11.09$ & $0.912$ \\
Median (50th)           & $-49.39$ & $+33.70$  & $-15.69$         & $9.92$  & $0.114$ \\
75th                    & $-49.39$ & $-25.27$  & $-74.66^{***}$   & $9.48$  & $<$0.001 \\
90th (most incremental) & $-49.39$ & $-127.71$ & $-177.10^{***}$  & $18.95$ & $<$0.001 \\
\bottomrule\bottomrule
\end{tabular}
\begin{minipage}{\textwidth}
\vspace{4pt}
\centering
\footnotesize\textbf{Notes:} $^{*}p<0.10$, $^{**}p<0.05$, $^{***}p<0.01$. Outcome trimmed at the 5th--95th percentile of the net return distribution ($N = 12{,}235$; bounds: [\$-386.4M, \$3{,}369.1M]). $\hat{\theta}_{\text{lic}} = -49.39$ (SE $8.88$) is the main effect at mean $\tilde{S}$; $\hat{\gamma}_{\text{lic}} = -279.47$ (SE $36.62$) is the interaction coefficient.
\end{minipage}
\end{table}

\clearpage
\newpage

\section*{Acknowledgements}

\subsection*{Data}
Data are from Evaluate Pharma\textsuperscript{\textcircled{R}}, a leading commercial database for pharmaceutical intelligence. The data are available upon subscription to Evaluate Pharma. The authors gratefully acknowledge Evaluate Pharma for access to these data.

\subsection*{Use of Generative Artificial Intelligence}
During the preparation of this work, the authors used Claude for grammar, spelling, and minor clarity edits. After using this tool, the authors reviewed and edited the content as needed and take full responsibility for the content of the published article.

\clearpage
\newpage

\bibliographystyle{apalike}
\bibliography{ref}

\begin{thebibliography}{}

\bibitem[Adams and Brantner, 2010]{adams2010spending}
Adams, C.~P. and Brantner, V.~V. (2010).
\newblock Spending on new drug development 1.
\newblock {\em Health economics}, 19(2):130--141.

\bibitem[Akcigit et~al., 2016]{akcigit2016buy}
Akcigit, U., Celik, M.~A., and Greenwood, J. (2016).
\newblock Buy, keep, or sell: Economic growth and the market for ideas.
\newblock {\em Econometrica}, 84(3):943--984.

\bibitem[Akcigit and Kerr, 2018]{akcigit2018growth}
Akcigit, U. and Kerr, W.~R. (2018).
\newblock Growth through heterogeneous innovations.
\newblock {\em Journal of Political Economy}, 126(4):1374--1443.

\bibitem[Akerlof, 1970]{akerlof1970lemons}
Akerlof, G.~A. (1970).
\newblock The market for "lemons": Quality uncertainty and the market mechanism.
\newblock {\em Quarterly Journal of Economics}, 84(3):488--500.

\bibitem[Arora et~al., 2021]{arora2021knowledge}
Arora, A., Belenzon, S., and Sheer, L. (2021).
\newblock Knowledge spillovers and corporate investment in scientific research.
\newblock {\em American Economic Review}, 111(3):871--898.

\bibitem[Arora et~al., 2022]{arora2022science}
Arora, A., Belenzon, S., and Suh, J. (2022).
\newblock Science and the market for technology.
\newblock {\em Management Science}, 68(10):7176--7201.

\bibitem[Arora et~al., 2001]{arora2001}
Arora, A., Fosfuri, A., and Gambardella, A. (2001).
\newblock Markets for technology and their implications for corporate strategy.
\newblock {\em Industrial and corporate change}, 10(2):419--451.

\bibitem[Arora et~al., 2004]{arora2004markets}
Arora, A., Fosfuri, A., and Gambardella, A. (2004).
\newblock {\em Markets for technology: The economics of innovation and corporate strategy}.
\newblock MIT press.

\bibitem[Arora and Gambardella, 2010a]{arora2010ideas}
Arora, A. and Gambardella, A. (2010a).
\newblock Ideas for rent: an overview of markets for technology.
\newblock {\em Industrial and corporate change}, 19(3):775--803.

\bibitem[Arora and Gambardella, 2010b]{arora2010market}
Arora, A. and Gambardella, A. (2010b).
\newblock The market for technology.
\newblock {\em Handbook of the Economics of Innovation}, 1:641--678.

\bibitem[Arora et~al., 2009]{arora2009breath}
Arora, A., Gambardella, A., Magazzini, L., and Pammolli, F. (2009).
\newblock A breath of fresh air? firm type, scale, scope, and selection effects in drug development.
\newblock {\em Management Science}, 55(10):1638--1653.

\bibitem[Athey and Imbens, 2019]{athey2019machine}
Athey, S. and Imbens, G.~W. (2019).
\newblock Machine learning methods that economists should know about.
\newblock {\em Annual Review of Economics}, 11(1):685--725.

\bibitem[Austin, 2009]{austin2009balance}
Austin, P.~C. (2009).
\newblock Balance diagnostics for comparing the distribution of baseline covariates between treatment groups in propensity-score matched samples.
\newblock {\em Statistics in medicine}, 28(25):3083--3107.

\bibitem[Azoulay et~al., 2019]{azoulay2019public}
Azoulay, P., Graff~Zivin, J.~S., Li, D., and Sampat, B.~N. (2019).
\newblock Public r\&d investments and private-sector patenting: evidence from nih funding rules.
\newblock {\em The Review of economic studies}, 86(1):117--152.

\bibitem[Belloni et~al., 2014]{belloni2014inference}
Belloni, A., Chernozhukov, V., and Hansen, C. (2014).
\newblock Inference on treatment effects after selection among high-dimensional controls.
\newblock {\em Review of Economic Studies}, 81(2):608--650.

\bibitem[Budish et~al., 2015]{budish2015firms}
Budish, E., Roin, B.~N., and Williams, H. (2015).
\newblock Do firms underinvest in long-term research? evidence from cancer clinical trials.
\newblock {\em American Economic Review}, 105(7):2044--2085.

\bibitem[Chernozhukov et~al., 2018]{chernozhukov2018double}
Chernozhukov, V., Chetverikov, D., Demirer, M., Duflo, E., Hansen, C., Newey, W., and Robins, J. (2018).
\newblock Double/debiased machine learning for treatment and structural parameters.

\bibitem[Cunningham et~al., 2021]{cunningham2021killer}
Cunningham, C., Ederer, F., and Ma, S. (2021).
\newblock Killer acquisitions.
\newblock {\em Journal of political economy}, 129(3):649--702.

\bibitem[Danzon et~al., 2005]{danzon2005productivity}
Danzon, P.~M., Nicholson, S., and Pereira, N.~S. (2005).
\newblock Productivity in pharmaceutical--biotechnology r\&d: the role of experience and alliances.
\newblock {\em Journal of Health Economics}, 24(2):317--339.

\bibitem[de~Villemeur et~al., 2022]{de2022biopharmaceutical}
de~Villemeur, E.~B., Scannell, J.~W., and Versaevel, B. (2022).
\newblock Biopharmaceutical r\&d outsourcing: Short-term gain for long-term pain?
\newblock {\em Drug Discovery Today}, 27(11):103333.

\bibitem[DiMasi et~al., 2010]{dimasi2010trends}
DiMasi, J.~A., Feldman, L., Seckler, A., and Wilson, A. (2010).
\newblock Trends in risks associated with new drug development: success rates for investigational drugs.
\newblock {\em Clinical Pharmacology \& Therapeutics}, 87(3):272--277.

\bibitem[DiMasi and Grabowski, 2007]{dimasi2007cost}
DiMasi, J.~A. and Grabowski, H.~G. (2007).
\newblock The cost of biopharmaceutical r\&d: is biotech different?
\newblock {\em Managerial and decision Economics}, 28(4-5):469--479.

\bibitem[DiMasi et~al., 2016]{dimasi2016innovation}
DiMasi, J.~A., Grabowski, H.~G., and Hansen, R.~W. (2016).
\newblock Innovation in the pharmaceutical industry: new estimates of r\&d costs.
\newblock {\em Journal of health economics}, 47:20--33.

\bibitem[DiMasi et~al., 2003]{dimasi2003price}
DiMasi, J.~A., Hansen, R.~W., and Grabowski, H.~G. (2003).
\newblock The price of innovation: new estimates of drug development costs.
\newblock {\em Journal of health economics}, 22(2):151--185.

\bibitem[Gans and Stern, 2003]{gans2003product}
Gans, J.~S. and Stern, S. (2003).
\newblock The product market and the market for “ideas”: commercialization strategies for technology entrepreneurs.
\newblock {\em Research policy}, 32(2):333--350.

\bibitem[Girotra et~al., 2007]{girotra2007valuing}
Girotra, K., Terwiesch, C., and Ulrich, K.~T. (2007).
\newblock Valuing r\&d projects in a portfolio: Evidence from the pharmaceutical industry.
\newblock {\em Management Science}, 53(9):1452--1466.

\bibitem[Grabowski and Vernon, 1994]{grabowski1994returns}
Grabowski, H.~G. and Vernon, J.~M. (1994).
\newblock Returns to r\&d on new drug introductions in the 1980s.
\newblock {\em Journal of health economics}, 13(4):383--406.

\bibitem[Hay et~al., 2014]{hay2014clinical}
Hay, M., Thomas, D.~W., Craighead, J.~L., Economides, C., and Rosenthal, J. (2014).
\newblock Clinical development success rates for investigational drugs.
\newblock {\em Nature biotechnology}, 32(1):40--51.

\bibitem[Hegde and Luo, 2018]{hegde2018patent}
Hegde, D. and Luo, H. (2018).
\newblock Patent publication and the market for ideas.
\newblock {\em Management Science}, 64(2):652--672.

\bibitem[Hermosilla, 2021]{hermosilla2021rushed}
Hermosilla, M. (2021).
\newblock Rushed innovation: Evidence from drug licensing.
\newblock {\em Management Science}, 67(1):257--278.

\bibitem[Hirano et~al., 2003]{hirano2003efficient}
Hirano, K., Imbens, G.~W., and Ridder, G. (2003).
\newblock Efficient estimation of average treatment effects using the estimated propensity score.
\newblock {\em Econometrica}, 71(4):1161--1189.

\bibitem[H{\"u}nermund et~al., 2023]{hunermund2023double}
H{\"u}nermund, P., Louw, B., and Caspi, I. (2023).
\newblock Double machine learning and automated confounder selection: A cautionary tale.
\newblock {\em Journal of Causal Inference}, 11(1):20220078.

\bibitem[Kogan et~al., 2017]{kogan2017technological}
Kogan, L., Papanikolaou, D., Seru, A., and Stoffman, N. (2017).
\newblock Technological innovation, resource allocation, and growth.
\newblock {\em The quarterly journal of economics}, 132(2):665--712.

\bibitem[Kogut and Zander, 1992]{kogut1992knowledge}
Kogut, B. and Zander, U. (1992).
\newblock Knowledge of the firm, combinative capabilities, and the replication of technology.
\newblock {\em Organization science}, 3(3):383--397.

\bibitem[Kola and Landis, 2004]{kola2004can}
Kola, I. and Landis, J. (2004).
\newblock Can the pharmaceutical industry reduce attrition rates?
\newblock {\em Nature reviews Drug discovery}, 3(8):711--716.

\bibitem[Krieger et~al., 2022]{krieger2022missing}
Krieger, J., Li, D., and Papanikolaou, D. (2022).
\newblock Missing novelty in drug development.
\newblock {\em The Review of Financial Studies}, 35(2):636--679.

\bibitem[Liu, 2024]{liu2024pharmaceutical}
Liu, Q. (2024).
\newblock Pharmaceutical innovation collaboration, evaluation, and matching.
\newblock {\em Journal of Health Economics}, 98:102922.

\bibitem[Lo et~al., 2019]{lo2019machine}
Lo, A.~W., Siah, K.~W., and Wong, C.~H. (2019).
\newblock {\em Machine learning with statistical imputation for predicting drug approvals}, volume 60 No. 10.1162.
\newblock SSRN.

\bibitem[Nicholson et~al., 2005]{nicholson2005biotech}
Nicholson, S., Danzon, P.~M., and McCulloch, J. (2005).
\newblock Biotech-pharmaceutical alliances as a signal of asset and firm quality.
\newblock {\em Journal of Business}, 78(4):1433--1464.

\bibitem[Orsenigo et~al., 2001]{orsenigo2001}
Orsenigo, L., Pammolli, F., and Riccaboni, M. (2001).
\newblock Technological change and network dynamics: lessons from the pharmaceutical industry.
\newblock {\em Research policy}, 30(3):485--508.

\bibitem[Palermo et~al., 2019]{palermo2019reliable}
Palermo, V., Higgins, M.~J., and Ceccagnoli, M. (2019).
\newblock How reliable is the market for technology?
\newblock {\em Review of Economics and Statistics}, 101(1):107--120.

\bibitem[Pammolli et~al., 2011]{pammolli2011productivity}
Pammolli, F., Magazzini, L., and Riccaboni, M. (2011).
\newblock The productivity crisis in pharmaceutical r\&d.
\newblock {\em Nature reviews Drug discovery}, 10(6):428--438.

\bibitem[Pisano, 1997]{pisano1997}
Pisano, G.~P. (1997).
\newblock {R\&D Performance, Collaborative Arrangements, and the Market-for-Know-How: A Test of the `Lemons' Hypothesis in Biotechnology}.
\newblock Working Paper 97--105, Harvard Business School.

\bibitem[Reepmeyer et~al., 2011]{reepmeyer2011outlicensing}
Reepmeyer, G., Gassmann, O., and Ruether, F. (2011).
\newblock Out-licensing in markets with asymmetric information: the case of the pharmaceutical industry.
\newblock {\em International Journal of Innovation Management}, 15(4):755--795.

\bibitem[Robinson and Stuart, 2007]{robinson2007financial}
Robinson, D.~T. and Stuart, T.~E. (2007).
\newblock Financial contracting in biotech strategic alliances.
\newblock {\em The Journal of Law and Economics}, 50(3):559--596.

\bibitem[Rosenbaum and Rubin, 1985]{rosenbaum1985constructing}
Rosenbaum, P.~R. and Rubin, D.~B. (1985).
\newblock Constructing a control group using multivariate matched sampling methods that incorporate the propensity score.
\newblock {\em The American Statistician}, 39(1):33--38.

\bibitem[Siah et~al., 2021]{siah2021predicting}
Siah, K.~W., Kelley, N.~W., Ballerstedt, S., Holzhauer, B., Lyu, T., Mettler, D., Sun, S., Wandel, S., Zhong, Y., Zhou, B., et~al. (2021).
\newblock Predicting drug approvals: The novartis data science and artificial intelligence challenge.
\newblock {\em Patterns}, 2(8).

\bibitem[Szulanski, 1996]{szulanski1996exploring}
Szulanski, G. (1996).
\newblock Exploring internal stickiness: Impediments to the transfer of best practice within the firm.
\newblock {\em Strategic management journal}, 17(S2):27--43.

\bibitem[Tyagi et~al., 2025]{tyagi2025does}
Tyagi, H., Hermosilla, M., and Shah, R. (2025).
\newblock Does transparency hinder technological novelty? evidence from large pharmaceutical firms.
\newblock {\em Manufacturing \& Service Operations Management}, 27(5):1415--1432.

\bibitem[Williams, 2013]{williams2013intellectual}
Williams, H.~L. (2013).
\newblock Intellectual property rights and innovation: Evidence from the human genome.
\newblock {\em Journal of Political Economy}, 121(1):1--27.

\bibitem[Wooldridge, 2007]{wooldridge2007inverse}
Wooldridge, J.~M. (2007).
\newblock Inverse probability weighted estimation for general missing data problems.
\newblock {\em Journal of econometrics}, 141(2):1281--1301.

\end{thebibliography}

\end{document}